\documentclass[longauth]{aa}  

\usepackage{graphicx}
\usepackage{txfonts}

\usepackage{multirow}
\usepackage[colorlinks=true,allcolors=blue]{hyperref}

\let\oldAA\AA
\renewcommand{\AA}{\text{\oldAA}\xspace}
\newcommand{\target}{K20-ID7\xspace}

\newcommand{\nii}{[N{\sc{ii}}]}

\newcommand{\oiid}{[O{\sc{ii}}]$\lambda\lambda$3727,29\xspace}
\newcommand{\oii}{[O{\sc{ii}}]\xspace}

\newcommand{\oiii}{[O{\sc{iii}}]}
\newcommand{\ha}{H$\alpha$\xspace}

\newcommand{\hg}{H$\gamma$\xspace}
\newcommand{\hd}{H$\delta$\xspace}
\newcommand{\hb}{H$\beta$\xspace}
\newcommand{\siid}{[S{\sc{ii}}]$\lambda\lambda6716,31$\xspace}
\newcommand{\sii}{[S{\sc{ii}}]\xspace}
\newcommand{\siileft}{[S{\sc{ii}}]$\lambda$6717\xspace}
\newcommand{\siiright}{[S{\sc{ii}}]$\lambda$6731\xspace}

\newcommand{\oiiid}{[O{\sc{ii}}]$\lambda\lambda$5007,4959\xspace}

\newcommand{\hei}{He{\sc{i}}$\lambda$1.083\xspace}

\newcommand{\z}{$z$\xspace}    
\newcommand{\sfr}{M${_\odot}$ yr$^{-1}$\xspace}        
\newcommand{\msun}{M${_\odot}$\xspace}

\newcommand{\kms}{${\rm km~s^{-1}}$\xspace}

\newcommand{\ergscm}{${\rm erg~s^{-1}~cm^{-2}}$\xspace}

\newcommand{\Pag}{Pa$\gamma$\xspace}
\newcommand{\Pab}{Pa$\beta$\xspace}
\newcommand{\siiileft}{[S\textsc{iii}]$\lambda$9069\xspace}
\newcommand{\siiiright}{[S{\sc{iii}}]$\lambda$9533\xspace}
\newcommand{\siii}{[S{\sc{iii}}]\xspace}

\begin{document} 

   \title{Resolving stellar populations, star formation,
and interstellar medium conditions with JWST in a large spiral galaxy at $z\approx2$}

   \authorrunning{E. Parlanti, G. Tozzi et al.}
   \titlerunning{JWST view of K20-ID7}
   
   \author{Eleonora Parlanti\inst{\ref{MPE},\ref{iNorm}}\thanks{These authors contributed equally to this work.}, 
            Giulia Tozzi\inst{\ref{MPE}}$^{\star}$,
            Natascha M. Förster Schreiber\inst{\ref{MPE}},
            Claudia Pulsoni \inst{\ref{MPE}},
            Letizia Scaloni \inst{\ref{MPE}, \ref{unibo},\ref{inafbo}},
            Stavros Pastras \inst{\ref{MPE}, \ref{MPA}},
            Pascal Oesch\inst{\ref{uniGeneva},\ref{DAWN}},
            Capucine Barfety \inst{\ref{MPE}},
            Francesco Belfiore \inst{\ref{ESO}}, 
            Jianhang Chen \inst{\ref{MPE}},
            Giovanni Cresci \inst{\ref{inafflo}},
            Ric Davies \inst{\ref{MPE}},
            Frank Eisenhauer \inst{\ref{MPE}, \ref{TUM}},
            Juan M. Espejo Salcedo \inst{\ref{MPE}},
            Reinhard Genzel \inst{\ref{MPE}, \ref{berkeley}},
            Rodrigo Herrera-Camus \inst{\ref{uni_concepcion}},
            Jean-Baptiste Jolly \inst{\ref{MPE}}.
            Lilian L. Lee \inst{\ref{MPE}},
            Minju M. Lee \inst{\ref{DAWN}, \ref{denmark}},
            Daizhong Liu \inst{\ref{purple}}, 
            Dieter Lutz \inst{\ref{MPE}}, 
            Filippo Mannucci \inst{\ref{inafflo}},
            Giovanni Mazzolari \inst{\ref{MPE}}, 
            Thorsten Naab \inst{\ref{MPA}}, 
             Amit Nestor Shachar \inst{\ref{telaviv}},
            Sedona H. Price \inst{\ref{pitt}}, 
            Alvio Renzini \inst{\ref{padova}}, 
            T. Taro Shimizu \inst{\ref{MPE}}, 
            Amiel Sternberg \inst{\ref{telaviv}},
            Martina Scialpi \inst{\ref{MPE}, \ref{unitrento}, \ref{uniflo}, \ref{inafflo}},
            Eckhard Sturm \inst{\ref{MPE}},
            Linda J. Tacconi \inst{\ref{MPE}},
            Hannah Übler  \inst{\ref{MPE}},
            Stijn Wuyts \inst{\ref{Bath}}
          }
   \institute{
        Max-Planck-Institut f\"ur extraterrestrische Physik (MPE), Gie{\ss}enbachstra{\ss}e 1, 85748 Garching, Germany\label{MPE}
   \and
        Scuola Normale Superiore, Piazza dei Cavalieri 7, I-56126 Pisa, Italy\label{iNorm}
    \and
        Department of Physics and Astronomy “Augusto Righi”, University of Bologna, Via Piero Gobetti 93/2, I-40129 Bologna, Italy \label{unibo}
    \and
        INAF – Astrophysics and Space Science Observatory of Bologna, Via Piero Gobetti 93/3, I-40129 Bologna, Italy \label{inafbo}
    \and 
        Max-Planck-Institut für Astrophysik (MPA), Karl-SchwarzschildStr. 1, D-85748 Garching, Germany \label{MPA}
    \and
        Department of Astronomy, University of Geneva, Chemin Pegasi 51, CH1290 Versoix, Switzerland\label{uniGeneva}
    \and
        Cosmic Dawn Center (DAWN), Niels Bohr Institute, University of Copenhagen, Jagtvej 128, DK-2200 København N, Denmark\label{DAWN}
    \and 
    European Southern Observatory, Karl-Schwarzschild-Str. 2, 85748, Garching bei München, Germany \label{ESO}
    \and 
        INAF - Osservatorio Astrofisico di Arcetri, Largo E. Fermi 5, I-50125, Florence, Italy \label{inafflo}
    \and
    Department of Physics, Technical University of Munich, 85748 Garching, Germany \label{TUM}
    \and
    Departments of Physics and Astronomy, University of California,
Berkeley, CA 94720, USA \label{berkeley}
    \and
    Departamento de Astronomía, Universidad de Concepción, Barrio
Universitario, Concepción, Chile 
\label{uni_concepcion}
\and 
DTU-Space, Technical University of Denmark, Elektrovej 327, DK2800 Kgs. Lyngby, Denmark 
\label{denmark}
\and Purple Mountain Observatory, Chinese Academy of Sciences, 10
Yuanhua Road, Nanjing 210023, China \label{purple}
\and 
School of Physics and Astronomy, Tel Aviv University, Tel Aviv 69978, Israel \label{telaviv}
\and
Space Telescope Science Institute (STScI), 3700 San Martin Drive, Baltimore, MD 21218, USA \label{pitt}
\and
Osservatorio Astronomico di Padova, Vicolo dell’Osservatorio 5, Padova, I-35122, Italy \label{padova}
\and 
University of Trento, Via Sommarive 14, Trento, I-38123, Italy \label{unitrento}
\and 
Dipartimento di Fisica e Astronomia, Università di Firenze, Via G. Sansone 1, I-50019, Sesto F.no (Firenze), Italy\label{uniflo}
\and 
Department of Physics, University of Bath, Claverton Down, Bath, BA2 7AY, UK \label{Bath}\\
    \email{eleonora.parlanti@sns.it, gtozzi@mpe.mpg.de}
             }
   \date{}

  \abstract 
  {Cosmic noon represents the prime epoch where today's massive galaxies assembled most of their stellar masses, and it is an ideal period for observations with both the space-based \textit{James Webb Telescope} (JWST) and ground-based near-IR integral-field unit (IFU) spectrographs. This work analyzes JWST NIRSpec Micro Shutter Array (MSA) and NIRCam Wide Field Slitless Spectroscopy (WFSS) observations of \target, a large spiral, star-forming (SF) galaxy at $z=2.224$ with evidence of radial gas inflows. Leveraging ground-based IFU ERIS observations, we conducted a comprehensive and resolved study of the interstellar medium and stellar properties of this galaxy, covering the rest-frame optical to near-IR. Our analysis -- using several emission-line diagnostics, resolved spectral energy distribution (SED) fitting of high-resolution imaging, and \Pab line detection in NIRCam WFSS data -- reveals massive SF clumps  ($M_{\star}\simeq(0.67-3.5)$$\times$$10^{9}$ M$_{\odot}$)  with star formation rates (SFRs) of $3-24$ \sfr, low dust attenuation ($A_V$$\simeq$0.4), electron densities ($n_{\rm e}$$\lesssim$300 cm$^{-3}$), and ionization parameter (log($U$)$\simeq$$-3.0$). The central bulge is modestly massive ($M_{\star}$=(7$\pm$3)$\times$10$^{9}$ M$_{\odot}$), heavily obscured ($A_V$=6.43$\pm$0.55), and likely formed most of its stellar mass in the past (SFR=82$\pm$42 \sfr over the last $\sim$100 Myr). Yet, it continues to form stars at a lower rate (SFR$=12\pm8$ \sfr over the last $\sim$10 Myr). We infer a relatively low sulfur abundance of log(S/O)$\simeq$-1.9, which may have implications for sulfur production via type I supernova explosions. Moreover, all distinct galaxy regions feature a metallicity of 12+log(O/H)$\approx$8.54, likely due -- along with the enhanced N/O abundance (i.e., log(N/O)$\simeq$$-1.0$) -- to dilution effects from radial inflows of metal-poor gas. Lastly, we find tentative evidence of a negative gradient in stellar age, suggesting possible inside-out growth for \target.}

   \keywords{ISM: abundances -- galaxies: high-redshift -- galaxies: ISM -- galaxies: star formation}

   \maketitle
%

\section{Introduction}\label{sec:intro}

Present-day massive galaxies assembled half their current mass at $z$\,$\sim$\,1\,--\,3 -- the so-called ``cosmic noon'' epoch  -- when the star formation rate (SFR) density of the Universe reached its peak ($z$\,$\approx$\,2; \citealt{Madau:2014}), due to the large amount of available gas \citep{Tacconi:2020, Walter:2020}. At this epoch, gas-rich, turbulent star-forming disks represent the bulk of the massive main-sequence galaxy population \citep{Forster:2020}. These galaxies commonly feature central compact bulges, accreting black holes, galactic winds, and giant star-forming complexes on kiloparsec scales \citep{Genzel:2011,Genzel:2014,Wuyts:2012,Wuyts:2013,Lang:2014,Tacchella:2018,Forster:2018, Forster:2019,Ferreira:2023, Espejo:2025,Kalita:2025}. Star-forming galaxies at $z$\,$\approx$\,2 represent a perfect laboratory for probing how galaxies grow and self-regulate and for investigating the origin of the tight scaling relations between the global properties of star-forming galaxies observed in the local Universe and up to high redshifts \citep[e.g.,][]{Speagle:2014, Maiolino:2019}. 

While \textit{James Webb Space Telescope} (JWST) observations have mostly focused on the $z$\,>\,4 Universe -- yielding extraordinary discoveries and new questions \citep[e.g.,][]{Finkelstein:2023, CurtisLake:2023, Carniani:2024, Naidu:2026} -- cosmic noon remains the key epoch to explore for ultimately addressing the processes driving galaxy evolution. Unlike $z$\,>\,4, $z$\,$\approx$\,2 offers a unique advantage: it is an ideal redshift for both JWST and ground-based near-IR integral-field unit (IFU) observations with adaptive optics (AOs), such as SINFONI (Spectrograph for INtegral Field Observations in the Near Infrared; \citealt{Eisenhauer2003}) and its successor ERIS (Enhanced Resolution Imager and Spectrograph; \citealt{Davies:2023}) at the Very Large Telescope (VLT). While near-IR AO-assisted IFU spectrographs can reach much higher spectral resolutions ($R$\,$\approx$\,11000 with VLT/ERIS), uniquely enabling a thorough study of gas kinematics via line emission, JWST's unprecedented broad wavelength coverage and high sensitivity provide imaging and spectroscopy of $z$\,$\approx$\,2  galaxies across a continuous rest-frame UV to near-IR wavelength range \citep{Ferruit:2022, Rieke:2023}. This gives us access to a wealth of interstellar medium (ISM) and stellar population properties (e.g., metallicity, dust extinction, ionization state, and stellar mass). Moreover, with their high spatial resolution ($\sim$\,0.1$''$), both facilities can probe physical scales down to $\sim$\,1 kpc, comparable to the Toomre scale \citep{Toomre:1964}, the fundamental scale governing gravitational instabilities in gas-rich, rotationally supported galactic disks at $z$\,$\approx$\,1\,--\,3 (e.g., \citealt{Genzel:2011}). Therefore, by leveraging both JWST and ground-based AO IFU facilities, we can now obtain a full picture of galaxies at cosmic noon, including gas kinematics, stellar properties, and ISM conditions \citep{Park:2024,Ju:2025, Nestor:2025,Slob:2025}.

In this work, we present JWST observations of \target (GS4\_29868; RA\,=\,03:32:29.1, Dec\,=\,--\,27:46:28.5),  a star-forming main-sequence galaxy (M$_\star$\,$\approx$\,4\,$\times$\,10$^{10}$ M$_\odot$; \citealt{Forster:2018}) at $z$\,=\,2.224, featuring a large, well-defined spiral morphology. This galaxy is an ideal laboratory for investigating many of the key mechanisms occurring at cosmic noon, from bulge growth and the settling of morphological features to star formation in kiloparsec-scale clumps and radial gas transport. Located in GOODS-South, one of the most targeted extragalactic fields, \target has a plethora of observations from large space surveys and deep dedicated ground-based follow-ups. It was indeed extensively observed from space with the \textit{Hubble Space Telescope} (HST) as part of the Great Observatories Origins Deep Survey (GOODS; \citealt{Giavalisco:2004}), the Cosmic Assembly Near-Infrared Deep Extragalactic
Legacy Survey (CANDELS; \citealt{Grogin:2011,Koekemoer:2011}), and the 3D-HST \citep{Brammer:2012,Skelton:2014} surveys. It was also observed at longer wavelengths with \textit{Spitzer} \citep{Dickinson:2003,Ashby:2013,Whitaker:2014} and \textit{Herschel} \citep{Lutz:2011} and benefits from deep ground-based AO-assisted IFU observations at high spectral and spatial resolutions.
The first deep high-resolution IFU data were obtained with VLT/SINFONI, as part of the AO SINS/zC-SINF survey \citep{Forster:2018}, where the \ha\ kinematics revealed a globally ordered velocity field compatible with disk rotation \citep{Forster:2018,Espejo:2025a} and provided evidence of radial gas inflows along a spiral arm \citep{Genzel:2023}. More recently, \target was observed with VLT/ERIS at superior spectral ($R$\,$\approx$\,10500) and spatial resolutions as part of the Guaranteed Time Observation (GTO) program GALPHYS (PI: N. M. Förster Schreiber). The detailed \ha\ kinematics and morphology from ERIS data have enabled a thorough study of kiloparsec-scale star-forming clumps (Förster Schreiber et al. in prep.) as well as a robust characterization of gas inflows (Pulsoni et al., in prep.).

For galaxies at $z$\,$\sim$\,2, JWST data provide continuous coverage from the rest-frame near-UV to the rest-frame near-IR. \target has a wealth of archival JWST spectroscopy and imaging, including data from the GTO JWST Deep Extragalactic Survey \citep[JADES;][]{Eisenstein:2026} in the \textit{Hubble} Ultra Deep Field (HUDF) and surrounding GOODS-South \citep{Bunker:2024}, as well as from the Cycle 1 General Observer (GO) program First Reionization Epoch Spectroscopically Complete Observations (FRESCO; \citealt{Oesch:2023}).

In this work, we leverage the rich dataset available for \target to present a complete characterization of its ISM conditions and stellar properties on scales down to $\sim$\,1~kpc, thereby gaining insights into its formation and evolution mechanisms. This paper is structured as follows. In Sect. \ref{sec:data} we present all the data analyzed in this work from JWST, HST/ACS, and ALMA and provide details on the data reduction. In Sect. \ref{sec:ism}, we infer local ISM properties for the different galaxy regions covered by distinct MSA shutters. In Sects. \ref{sec:sed_cigale} and \ref{sec:pab}, we focus on resolved SED fitting of NIRCam and HST imaging and discuss spatially mapping star formation via \Pab in NIRCam WFSS data. We then compare the resolved results with integrated SED analyses, including available mid- and far-IR photometry. In Sect. \ref{sec:discussion}, we interpret the inferred results in the context of galaxy growth and evolution scenarios, and in Sect. \ref{sec:conclusions}, we draw our conclusions. 

Throughout this work, we use the following notation to refer to key emission-line diagnostics:
\begin{itemize}
\item R2 = log([O{\sc{ii}}]$\lambda\lambda$3726,29/H$\beta$);
\item R3 = log([O{\sc{iii}}]$\lambda$5007/H$\beta$);
\item O3O2 = R3 -- R2;
\item N2 = log([N{\sc{ii}}]$\lambda$6584/H$\alpha$);
\item O3N2 = R3 -- N2;
\item N2O2 = log([N{\sc{ii}}]$\lambda$6584/\oiid);
\item S2 = log[([S{\sc{ii}}]$\lambda$6716 + [S{\sc{ii}}]$\lambda$6731)/H$\alpha$];
\item S23 = log[([S{\sc{ii}}]$\lambda\lambda$6716,31 + [S{\sc{iii}}]$\lambda\lambda$9069,9533)/H$\beta$];
\item N2S2 = log([N{\sc{ii}}]$\lambda$6584/\siid).
\end{itemize}
We adopted cosmological parameters from \citet{Planck:2015} -- i.e., $H_0
= 67.7$ \kms\,Mpc$^{-1}$, $\Omega_{\rm m}$ = 0.307, and $\Omega_\Lambda$= 0.691 -- and a Chabrier initial mass function \citep{Chabrier:2003}.

\begin{figure*}
    \centering
    \includegraphics[width=\linewidth]{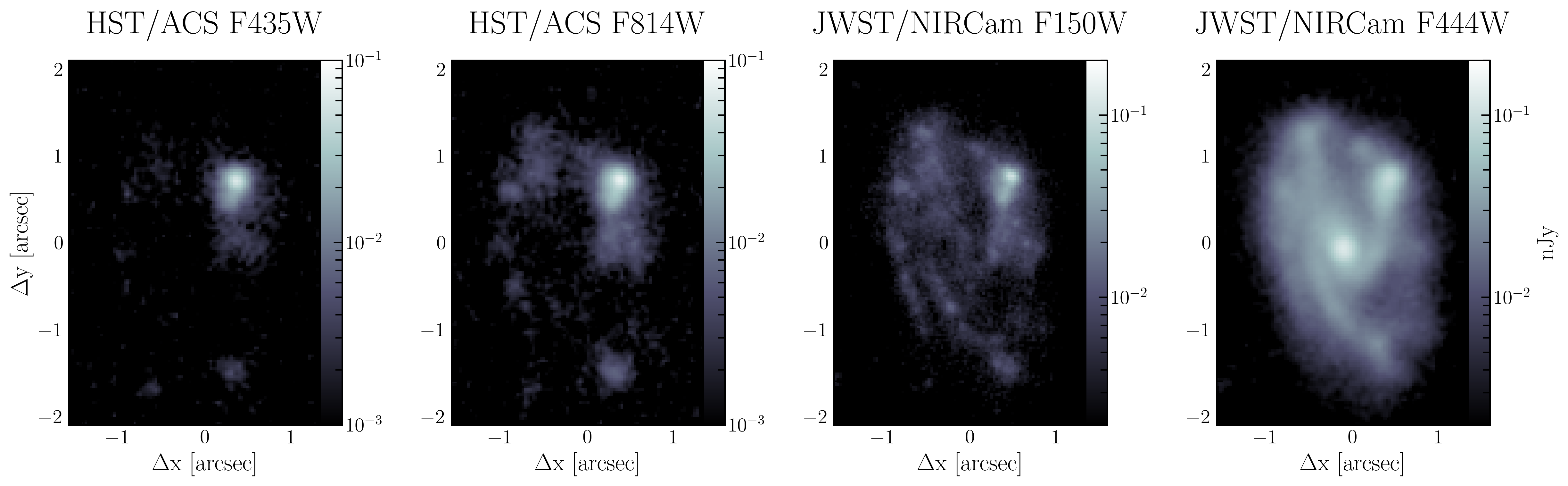}
    \caption{Images of \target in four representative HST/ACS (F435W, F814W) and JWST/NIRCam (F150W, F444W) filters, tracing galaxy emission from the rest-frame UV to near-IR wavelengths. Redder NIRCam imaging unambiguously reveals a regular, unperturbed galaxy morphology with well-defined spiral arms and a compact core, which is brightest in the F444W filter. All images are expressed in nanojansky.
    }
    \label{fig:HST_JWST_photo}
\end{figure*}

\section{Data}\label{sec:data}
In this section,  we present all data employed in this paper, mostly from JWST, but also from HST (ACS images in five filters) and ALMA (Band 6 continuum at 1.2\,mm), along with the data reduction. The JWST data consist of: NIRCam imaging in 11 filters; NIRSpec MSA spectra at $R$\,$\sim$\,100, $R$\,$\sim$\,1000, and $R$\,$\sim$\,2700; and NIRCam WFSS F444W data ($R$\,$\sim$\,1600).

\subsection{JWST/NIRCam and HST/ACS imaging}\label{sec:data_imaging}
\target was recently observed as part of the JADES survey \citep{Eisenstein:2026} in the Hubble Ultra Deep Field \citep{Bunker:2024}, which acquired NIRCam imaging in 11 different broad- and medium-band filters (F090W, F115W, F150W, F182M, F200W, F210M, F277W, F335M, F356W, F410M, and F444W), and NIRSpec MSA spectroscopy (presented in Sect. \ref{sec:data_msa}). Additional NIRCam data come from the Cycle 1 FRESCO survey (PI: P. Oesch, program ID: 1895; \citealt{Oesch:2023}) -- a NIRCam WFSS program, which obtained grism spectroscopy (see Sect. \ref{sec:data_grism}) and F444W imaging, along with parallel medium-band imaging in F182M and F210M.
Among the HST imaging available, we only used publicly released HST/ACS imaging (F435W, F606W, F775W, F814W, and F850LP) from the GOODS and CANDELS surveys \citep{Giavalisco:2004,Grogin:2011,Koekemoer:2011} and discarded the WFC3/IR images, since they cover the same wavelengths as the bluer NIRCam filters but at a lower sensitivity and spatial resolution.

For five NIRCam filters (F182M, F210M, F277W, F356W, and F444W), we adopted the high-quality reductions publicly released on the DAWN JWST Archive\footnote{\url{ https://dawn-cph.github.io/dja/}} (DJA). These were obtained using the \texttt{grizli}\footnote{\url{https://github.com/gbrammer/grizli}} reduction software \citep{Brammer:2019,grizli}, which performs standard calibrations, astrometric alignment with \texttt{DrizzlePac}\footnote{\url{https://www.stsci.edu/scientific-community/software/drizzlepac}}, and background subtraction optimized to preserve faint emission and minimize noise \citep[see details in][]{Valentino:2023}. Since the reduced NIRCam images in the other six filters (F090W, F115W, F150W, F200W, F335M, F410M) have not yet been publicly released on DJA, we retrieved uncalibrated NIRCam data from the Mikulski Archive for Space Telescopes (MAST) portal\footnote{\url{https://archive.stsci.edu/}}, and reduced them using a custom-made reduction pipeline (see \citealt{Espejo:2025} for a detailed description). This reduction leverages the CrabToolkit\footnote{\url{https://github.com/1054/Crab.Toolkit.JWST}}, which follows standard JWST image processing steps\footnote{\url{https://jwst-docs.stsci.edu/jwst-science-calibration-pipeline/stages-of-jwst-data-processing}}, and applies additional steps to improve the removal of artifacts through a combination of manual masking and publicly available templates
\citep{2025jwst.rept.9225S}. The reduced images are then astrometrically corrected, improved with an additional flat background subtraction, and finally produced with a 0.025$''$ pixel scale aligned with a north up and east left orientation, consistent with DJA reductions. Similarly, we also produced HST images with the same northeast orientation and 0.025$''$ pixel sampling.

Figure \ref{fig:HST_JWST_photo} displays four representative HST/ACS (F435W and F814W) and JWST/NIRCam (F150W and F444W) images of \target, each showing the morphology of the galaxy emission at different wavelengths, from the rest-frame UV to near-IR. The comparison between JWST/NIRCam and HST/ACS images immediately highlights how crucial it is to cover redder wavelengths with JWST/NIRCam to get a complete, unambiguous view of this galaxy. At redder wavelengths, a well-defined spiral structure clearly emerges, featuring three spiral arms, bright clumps, and a compact red core that becomes particularly bright in the reddest F444W filter. A first composite-color image of \target, including the NIRCam F444W band, was presented in \citet{Genzel:2023}, revealing the galaxy's large outstanding spiral and the central compact core for the first time.

\subsection{NIRSpec MSA spectroscopy}\label{sec:data_msa}

\begin{figure*}
    \centering
    \includegraphics[width=\linewidth]{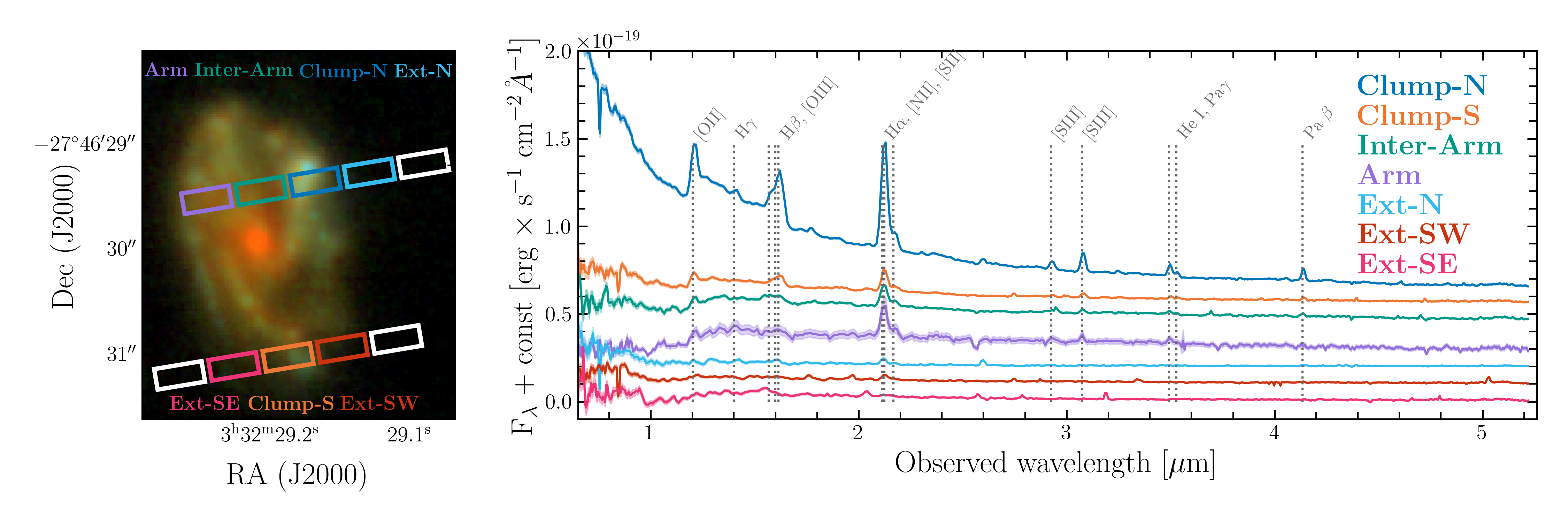}
    \caption{NIRSpec MSA spectroscopy of \target.
    \textit{Left}: Location of the two distinct MSA mask arrays, with open shutters overlaid on a NIRCam RGB image of \target composed of F356W+F444W (red), F182M+F200W+F277W (green), and F090W+F115W+F150W (blue) filters. \textit{Right}: PRISM/CLEAR spectra extracted from the seven colored MSA shutters (showing galaxy emission, color-coded as in the left panel), using our custom reduction pipeline (described in Sect. \ref{sec:data_msa}). The white shutters contain no galaxy emission. The spectra from different regions exhibit different continuum shapes and line intensities across the rest-frame near-UV to near-IR wavelength range. For clarity, each spectrum is vertically shifted by an arbitrary constant.
    }
    \label{fig:Nircam_RGB_and_prism}
\end{figure*}

In addition to NIRCam imaging, the JADES survey also acquired deep NIRSpec MSA spectroscopy of targets in the HUDF and surrounding GOODS-South field at low ($R$\,$\sim$\,100, with PRISM/CLEAR), medium ($R$\,$\sim$\,1000, with G140M/F070LP, G235M/F170LP, and G395M/F290LP), and high resolution ($R$\,$\sim$\,2700, with G395H/F290LP). In particular, \target's NIRSpec/MSA data were obtained as part of the JADES program 1210 (PI: N. Lützgendorf) and consist of two distinct datasets associated with different pointings (Fig. \ref{fig:Nircam_RGB_and_prism}, left panel). The two pointings were centered on the two UV-bright blue clumps visible in HST/ACS imaging (see Fig. \ref{fig:HST_JWST_photo}), which were classified as two separate entries (i.e., two distinct galaxies) in the CANDELS and 3D-HST catalogs \citep{Skelton:2014}.

\target's NIRSpec/MSA data were publicly released as part of the first JADES data release \citep{Bunker:2024}, along with a public reduced version of the spectra obtained using the NIRSpec MSA GTO pipeline \citep{Scholtz:2025}. This pipeline applies a point-source path-loss correction and performs a local background subtraction by measuring the background in lateral shutters and subtracting it from the "science" spectrum of the central shutter. While this standard procedure has widely proven to work well for compact sources (e.g., more distant galaxies at \z$>$\,4; \citealt{Bunker:2024}), it is not suitable for extended sources such as \target, where the side shutters are also placed on emitting regions of the galaxy. Therefore, if used for background measurement, they result in line and continuum emission self-subtraction. Moreover, an accurate path-loss correction is essential to properly account for both geometric losses (from galaxy light partially falling outside the micro-shutter mask) and diffraction losses (from light lost along the NIRSpec optical path). The standard pipeline performs reliable path-loss correction for either point-like sources or uniformly illuminated micro-shutters. Since \target's MSA data do not match any of these options, the standard path-loss correction in the public reduced data likely produced incorrect fluxes and altered the global spectral shape due to the wavelength dependence of the correction.

For all these reasons, we implemented a custom reduction of \target's NIRSpec/MSA spectra, which i) performs improved background subtraction and path-loss correction and ii) extracts science spectra from all MSA shutters showing galaxy emission. In the following, we describe the main steps of our adopted reduction procedure. We retrieved the 
count rate data from the MAST archive to extract the 2D trace of each spectrum for each visit, apply the wavelength calibration and flat-field correction, and skip the path-loss correction and background subtraction steps. After rectifying the 2D traces, we obtained 2D rectified spectra relative to each grating and filter configuration, nod, and exposure.

By visually inspecting the resulting 2D rectified spectra, we detected galaxy emission in most of the MSA shutters in the PRISM/CLEAR configuration (seven in total; the colored shutters in the right panel of Fig. \ref{fig:Nircam_RGB_and_prism}). We therefore created a master background, which we subtracted from each shutter spectrum using the following procedure. Using the MSA pipeline, we reduced the data from all the shutters left open in the MSA mask for the same visit, focusing on specifically selected source-free regions, and created the corresponding rectified background 2D spectra. For each visit, we then generated a unique 2D master background -- computed as the median of the 2D rectified background spectra -- and removed it from the 2D rectified science spectra previously obtained for \target. By combining all 2D background-subtracted spectra of \target, we obtained the final 2D PRISM spectra for each visit. From these, we extracted the 1D spectra for each shutter using a 5-pixel wide slit. We also adopted the same procedure for the "error" extension of the observed data. For medium- and high-resolution data, we followed a different procedure for background subtraction, since galaxy emission is detected in fewer shutters (i.e., five) than in the PRISM/CLEAR configurations. In particular, since we do not detect galaxy emission in the light blue and pink shutters shown in the left-hand panel of Fig. \ref{fig:Nircam_RGB_and_prism}, unlike in the PRISM data, we used these to separately measure the background for each pointing. From these two background shutters, we therefore extracted 1D background spectra using a 5-pixel-wide slit and subtracted them from the 1D science spectra extracted from the other shutters of the same visit. This yielded medium- and high-resolution spectra for four distinct regions of \target.
We note that, with this approach, the expected self-subtraction effect on the \ha\ flux is at most $\sim$\,10\%, while it is negligible for other emission lines.

Finally, to correct all MSA spectra for path losses, we measured the flux in the NIRCam images of \target for the region encompassed by each shutter. We then used these flux values to rescale the NIRSpec/MSA spectra via a linear fit to the ratio of the observed NIRCam fluxes to the observed NIRSpec/MSA counterparts in each shutter.

\renewcommand{\arraystretch}{1.2}
\begin{table}[ht]
    \centering
    \footnotesize
        \caption{NIRSpec/MSA observations of \target used in this paper.}
    \begin{tabular}{cccc}
            \hline
            \hline

        Dataset & $\Delta\lambda$ [$\mu$m]& $R$ & $t_{\rm exp}$ [ks]\\
        \hline
        PRISM/CLEAR&  0.6 -- 5.3& 30 -- 300&  33.6\\
        G140M/F070LP& 0.7 -- 1.9$^{(\star)}$ & 500 -- 890 & 8.4\\
        G235M/F170LP& 1.7 -- 3.2 & 720 -- 1340 & 8.4\\
        G395M/F290LP& 2.9 -- 5.2 & 730 -- 1315 & 8.4\\
        G395H/F290LP& 2.9 -- 5.2 &  1930 -- 3615 & 8.4\\
        \hline
    \end{tabular}
    \tablefoot{The columns list the wavelength coverage, nominal spectral resolution \citep{Ferruit:2022}, and exposure time of the observations in each distinct grating and filter comnfigurations. 
    The $t_{\rm exp}$ values reported here refer to the central shutter for each visit. 
    $^{(\star)}$The nominal wavelength range is 0.7 -- 1.3 $\mu$m. However, as in JADES \citep{Bunker:2023}, we recover the spectrum up to 1.9$\mu$m, effectively acting as G140M/F070LP+F100LP.}
    \label{tab:obstime}
\end{table}

In summary, with our custom MSA reduction, we obtain in total seven PRISM/CLEAR spectra (see Fig. \ref{fig:Nircam_RGB_and_prism}) and five spectra for each medium- and high-resolution grating and filter configuration. All together these spectra probe seven distinct galaxy regions: clumps (labeled as Clump-N and Clump-S), spiral arm (Arm), inter-arm (Inter-Arm), and more external diffuse regions (Ext-N, Ext-SW, and Ext-SE). Both low-resolution PRISM ($R$\,$\sim$\,100) and all medium-resolution spectra ($R$\,$\sim$\,1000) cover a total spectral range from 0.6\,$\mu$m to 5.2\,$\mu$m, corresponding to rest-frame near-UV to near-IR wavelengths at $z$\,$\sim$\,2. In contrast, high-resolution observations ($R$\,$\sim$\,2700) were only obtained with the reddest grating (i.e., G395H/F290LP) and are, therefore, limited to the  2.9 -- 5.2 $\mu$m wavelength range encompassing \target's rest-frame near-IR emission. Details on the covered wavelength range, spectral resolution, and exposure time relative to each MSA gratingand filter  configuration are summarized in Table \ref{tab:obstime}. 
The exposure time reported refers to the central shutter of each array; the exposure time decreases to 2/3 and 1/3 with increasing distance from the central shutter.

\subsection{NIRCam WFSS data}\label{sec:data_grism}
In this work, we used JWST/NIRCam WFSS F444W observations of \target from FRESCO \citep{Oesch:2023}, obtained with an exposure time of 7 ks. These consist of grism spectra covering the range 3.8 -- 5.0 $\mu$m at a spectral resolution of $R$\,$\sim$\,1600, which probe rest-frame near-IR wavelengths at $z$\,$\sim$\,2. Slitless spectra of \target were acquired with the grism dispersion axis at a 90$^{\circ}$ position angle  (i.e., aligned WE, with increasing wavelengths toward E). All FRESCO NIRCam WFSS data, including those of \target, were reduced using the publicly available \texttt{grizli} code (\citealt{Brammer:2012,grizli}). Continuum-subtracted spectra were then produced by running a median filter along each row of the 2D grism spectra, adopting a 12-pixel central gap in the filtering to minimize line self-subtraction \citep{Kashino:2023}. For further details on the observational design and reduction of FRESCO observations, we refer the reader to the main papers of the survey (e.g., \citealt{Oesch:2023,Nelson:2024,Neufeld:2024}). 

\begin{figure}
    \centering
    \includegraphics[width=1\linewidth]{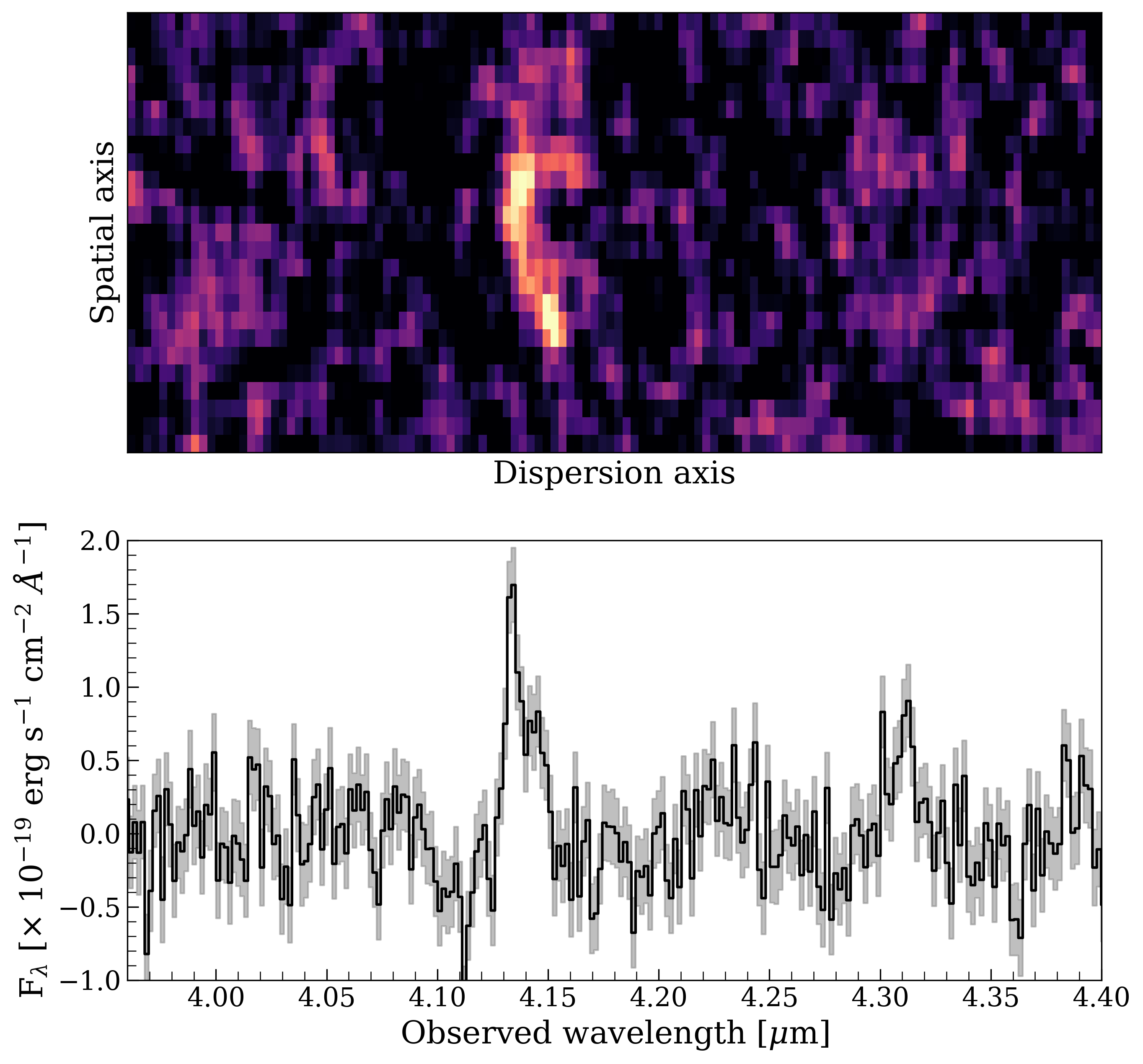}
    \caption{Continuum-subtracted NIRCam F444W WFSS data of \target, zoomed-in over 3.95 -- 4.40 $\mu$m, encompassing \Pab line emission. In the 2D slitless spectra (upper panel), the \Pab line emission is clearly spatially resolved perpendicular to the dispersion axis and extended along the parallel direction. This results in a double-peaked \Pab line profile in the total 1D spectrum (lower panel), a consequence of resolved spatial offsets combined with velocity gradients along the grism dispersion axis (discussed in Sect. \ref{sec:pab}).}
    \label{fig:Pa_beta_observed}
\end{figure}

As shown in Fig. \ref{fig:Pa_beta_observed}, the continuum-subtracted spectra of \target clearly exhibit Pa$\beta$ line emission.
The Pa$\beta$ line is detected with a total S/N of 7, computed as the ratio between the integrated line flux and the noise level estimated from the error spectrum. We derived the total flux by fitting the spectrum with a model consisting of a linear continuum and two Gaussian components. The error was estimated using two Gaussian components with the same line widths and amplitudes equal to the error level.
No additional emission lines are detected. Compared to optical Balmer emission lines (e.g., \ha\ and \hb), near-IR Paschen lines are less affected by dust attenuation and, therefore, can probe recent ($\sim$\,10 Myr) star formation in a more obscured regime.
In the 2D slitless spectra (upper panel), \Pab appears spatially resolved along the spatial axis (i.e., perpendicular to the grism dispersion axis) and extended along the spectral axis. This results in a double-peaked \Pab line profile in the total 1D spectrum (lower panel). As later addressed in Sect. \ref{sec:pab}, this is a consequence of real spatial offsets combined with velocity gradients along the grism dispersion axis.

\subsection{ALMA data}\label{sec:data_alma}
We reduce the available ALMA Band 6 1.2\,mm continuum data (PI: J. Scholtz, program ID 2018.1.01044.S) using \verb|CASA|. We retrieve the raw visibilities from the ALMA archive, and calibrate them by executing the script delivered with the data. We then clean the data by using the task \verb|tclean| in \verb|CASA|, and image them using natural weighting and a pixel scale of 0.1'' to obtain the best sensitivity. The final continuum image has an angular resolution of 0.5\arcsec$\times$0.6\arcsec and a sensitivity of 28$\mu$Jy beam$^{-1}$. 
We compute the continuum flux by summing all pixels within a circular aperture of radius 2\arcsec. The uncertainty on the observed flux is estimated via a Monte Carlo approach: we randomly placed 5000 apertures of the same size in regions free of galaxy emission and measured the flux in each, hence we assumed as an error the standard deviation of the resulting distribution.
Following this analysis we obtain a 1.2\,mm continuum flux of 0.36\,$\pm$\,0.11 mJy.

\section{Local ISM properties from spectroscopy}\label{sec:ism}

The wavelength range covered by the NIRSpec MSA data ($\sim$\,0.7 -- 5.2 $\mu$m) encompasses several emission lines, which we used to infer the properties of the galactic ISM via various emission-line diagnostics. In particular, we derived nebular dust attenuation, the ionization parameter, electron density ($A_{V,{\rm neb}}$, log($U$), $n_{\rm e}$), and chemical abundances (O, N, S) for the galaxy regions where the key emission lines are detected at S/N\,$>$\,3. In Appendix \ref{apx:spectral_fitting}, we describe the spectral fitting of the MSA spectra and summarize the best-fit results. For shutters where the same emission lines were detected with more than one spectral setup, we adopted the line fluxes from the medium-resolution data as fiducial. These have a higher S/N than the high-resolution data and resolve close spaced emission lines (e.g., \hb+\,\oiii, H$\alpha$\,+\,\nii, \sii line doublet) well, unlike the PRISM data. Therefore, when available, we only used the medium-resolution line fluxes to derive ISM properties, thereby also avoiding issues related to possible inaccuracies in the absolute flux calibration \citep{Deugenio:2025}.

\subsection{Dust attenuation, ionization, and electron density}\label{sec:extinction}

The first important physical property to derive is the (nebular) dust attenuation ($A_{V,{\rm neb}}$), which allows us to correct all observed line fluxes and then measure the other ISM properties. Within the broad spectral range covered by the NIRSpec MSA data of \target, there are several hydrogen emission lines (i.e., H$\delta$, H$\gamma$, \hb, \ha, \Pag, \Pab), which can be used to infer multiple measurements of $A_{V,{\rm neb}}$ from distinct line ratios. In Table \ref{tab:ism_table}, we report our derived $A_{V,{\rm neb}}$ measurements, obtained by combining the highest S/N hydrogen lines (i.e., \hb, \ha, \Pab, when detected). We assumed the \citet{Calzetti:2000} dust attenuation law and adopted theoretical line ratios of 2.86, 0.0569, 0.163 -- as predicted for warm ionized gas ($T=10^4$ K; \citealt{Osterbrock:2006}) -- for \ha/\hb, \Pab/\ha, and \Pab/\hb, respectively. For the inter-arm region, we derived $A_{V,{\rm neb}}$ from the \Pab/\ha ratio only, since no measurements for the \hb\ flux are available in this shutter\footnote{For the inter-arm region, \hb\ is only detected in the PRISM spectrum but appears blended with \oiiid. Therefore, it cannot be disentangled.}. 
For the clump regions (Clump-N and Clump-S), where all three hydrogen lines are detected, we obtain consistent measurements of $A_{V,{\rm neb}}$ from different hydrogen line ratios within 2$\sigma$.
Small differences in the derived $A_V$ values using different line ratios may be due to the different shapes of attenuation curves at high redshift \citep{Markov:2025} compared to local ones. To verify this, we also computed $A_{V,{\rm neb}}$ by adopting the high-$z$ attenuation curve by \citealt{Reddy:2026}, finding values comparable within the uncertainty with our estimates reported in Table \ref{tab:ism_table}. 
In the rest of this section, we adopt 
as fiducial values the weighted averages of the available measurements for each region, as reported in Table \ref{tab:ism_table}\footnote{For Clump-S, we consider only the two inferred $A_{V,{\rm neb}}$ estimates, excluding the upper limit derived from \ha/\hb.}. For the Arm region, \ha\ is the only hydrogen line detected, so we cannot estimate $A_{V,{\rm neb}}$. However, this is not crucial, since \nii\ and \ha\ are the only emission lines detected in this region. Thus, we can only use the \nii/\ha\ emission-line diagnostic (discussed in Sect. \ref{sec:chemical_abundances} to derive metallicity), for which dust extinction is negligible due to the proximity of these lines  in wavelength.
We reiterate that the $A_{V,{\rm neb}}$ values inferred in this section from hydrogen line ratios refer to nebular dust attenuation, which might be higher than the attenuation affecting stellar continuum ($A_{V,{\rm star}}$ inferred from SED modeling in Sect. \ref{sec:sed_cigale}), since HII regions are typically more embedded in dust.

\renewcommand{\arraystretch}{1.5}  
\begin{table}
    \centering
    \footnotesize
\caption{ISM properties inferred from NIRSpec MSA spectra for the three higher-S/N shutters of \target.}
\label{tab:ism_table}
\resizebox{0.95\columnwidth}{!}{
    \begin{tabular}{c|c|c|c|c}
    \hline \hline
    Diagnostic & Arm & Inter-Arm & Clump-N & Clump-S\\  \hline

    \multicolumn{5}{c}{$A_{V,{\rm neb}}$ [mag]}  
        
    \\ \hline

    \ha/\hb        &- & -                  & $0.6^{+0.1}_{-0.1}$ & $<$0.9 \\ 
    Pa$\beta$/\ha &-& $1.3^{+0.4}_{-0.5}$ & $0.2^{+0.1}_{-0.1}$ & $0.6^{+0.4}_{-0.6}$  \\ 
    Pa$\beta$/\hb &-& -                  & $0.4^{+0.1}_{-0.1}$ & $0.3^{+0.3}_{-0.4}$  \\ 

    Fiducial &-&  $1.3^{+0.4}_{-0.5}$  & $0.4 ^{+0.1}_{-0.1}$ & $0.4 ^{+0.3}_{-0.3}$   \\

    \hline

        \multicolumn{5}{c}{$n_{\rm e}$ [cm$^{-3}$]}  
        
    \\ \hline

    \sii        &-&    $<1900$             & $260^{+120}_{-139}$ & $<310$ \\ 
    \hline

    \multicolumn{5}{c}{log($U$)}                                                                                                                                                         \\ \hline

    \sii, \siii  &-&  $-3.3^{+0.3}_{-0.2}$               & $-3.05 \pm 0.05$  & $-3.0 \pm 0.2$ \\ 
    \oii, \oiii &-& $-3.1^{+0.3}_{-0.5}$& $-2.97 \pm 0.02$ &  $ -3.1 \pm 0.1$ \\ 
    Fiducial &-&  $-3.3 \pm 0.3$  &  $-3.01 \pm 0.03$ & $-3.1 \pm 0.2$ \\
   
    \hline

    \multicolumn{5}{c}{12+log(O/H)}                                                                                                                                                         \\ 
    \hline
    Fiducial       & $8.57^{+0.20}_{-0.21}$ & $8.58^{+0.06}_{-0.06}$    & $8.49^{+0.06}_{-0.06}$   & $8.51^{+0.08}_{-0.08}$    \\

    \hline

    \multicolumn{5}{c}{log(N/O)}                                                                                                                                                         \\ \hline
    N2O2       &-& $-0.8 \pm 0.2$  & $-0.95 \pm 0.05 $ & $-1.1 \pm 0.1 $  \\ 
    N2S2       &-& $-0.9 \pm 0.2$  & $-0.95 \pm 0.10 $ &  $-1.2 \pm 0.2 $   \\ 
   Fiducial &-& $-0.9 \pm 0.2$   & $-0.95 \pm 0.08$  &  $-1.2 \pm 0.2$  \\
 \hline

     \multicolumn{5}{c}{12+log(S/H)}                                                                                                                                                         \\ \hline
    S23       &-& $6.6^{+0.2}_{-0.2}$ & $6.6^{+0.2}_{-0.2}$ &  $6.7^{+0.3}_{-0.2}$  \\ 
    \hline

 \multicolumn{5}{c}{log(S/O)}        \\                                                                                                                                                 
 \hline
        &-& $-2.0 \pm 0.2$  & $-1.9 \pm 0.2 $ &  $-1.8 \pm 0.3 $  \\ 
    \hline
    \end{tabular}}
    \tablefoot{Top to bottom: Nebular attenuation $A_{V,{\rm neb}}$, derived from different hydrogen line ratios \citep{Calzetti:2000}; electron density $n_{\rm e}$ derived from the \sii line ratio \citep{Sanders:2016}; ionization parameter log($U$) using sulfur- and oxygen based diagnostics \citep{Diaz:2000}; gas-phase metallicity (12+log(O/H)) derived by combining various strong-line diagnostics \citep{Sanders:2025}; log(N/O) abundance \citep{Hayden-Pawson:2022}; 12+log(S/H) abundance \citep{Diaz:2022}; log(S/O) abundance, derived by combining the 12+log(S/H) and the fiducial 12+log(O/H) values. All upper limits correspond to $2\sigma$.
    }

\end{table}

Following \citet{Sanders:2016}, we then measured $n_{\rm e}$ from the dust-corrected \siid line doublet. For Clump-N, we find an electron density of $n_{\rm e}$\,$\simeq$\,260 cm$^{-3}$, consistent with the typical values inferred for $z$\,$\approx$\,2 star-forming regions \citep{Sanders:2016, Davies:2021, Topping:2025}. Unfortunately, the \sii$\lambda$6717/$\lambda$6731 ratio for the other two regions (i.e., Clump-S and Inter-Arm) yields nonphysical (negative) values of $n_{\rm e}$; therefore, we adopted a conservative 2$\sigma$ upper limit.

Given the high sensitivity and the wide wavelength coverage of \target's NIRSpec/MSA spectra, we derived two independent measurements of the ionization parameter from the \oii/\oiii\ and \sii/\siii line ratios, respectively, for each high-S/N galaxy region. 
The \siiileft emission line is detected in the medium-resolution grating only in the Clump-N spectra. 
However, we can estimate it for the inter-arm and Clump-S regions using the dust-corrected flux of the \siiiright line, assuming a theoretical ratio of \siiiright/\siiileft = 2.61 \citep{Osterbrock:2006}.
We note that for Clump-S, we have a measurement of the \siiileft from the PRISM spectrum. However, most of the PRISM data are affected by a bad feature near this emission line (see Fig. \ref{fig:bestfit_prism}).
Finally, we obtain a dust-corrected flux of \siiileft of $\sim$\,$2.7\times 10^{-19}$ erg s$^{-1}$ cm$^{-2}$ and $3.2 \times 10^{-19}$ erg s$^{-1}$ for the inter-arm and Clump-S regions, respectively. 
Using the \oii/\oiii- and \sii/\siii-based calibrators from \citealt{Diaz:2000}, we thus infer two independent values of log($U$) for each region. All inferred measurements indicate log($U$)\,$\simeq$\,--\,3.0 for both Clump-N and Clump-S, with a similar value -- albeit with larger uncertainties -- for the inter-arm region. This value is low, yet consistent with findings for other high-$z$ galaxies \citep{Reddy:2023}.

\subsection{Gas phase metallicity}\label{sec:metallicity}
In this work, we used metallicity calibrations from \citet{Sanders:2025}, based on six distinct diagnostics, namely R2, R3, O32, N2, O3N2, and S2 (see the end of Sect. \ref{sec:intro} for all definitions). For the inter-arm region, where direct flux measurement of the \hb emission lines needed for four diagnostics were unavailable, we estimated the \hb\ flux from our fiducial dust-extinction-corrected \ha\ flux (i.e., measured from medium-resolution data and corrected for $A_{V,{\rm neb}}=1.3^{+0.4}_{-0.5}$; see Table \ref{tab:ism_table}). We assumed a theoretical \ha/\hb\ line ratio of 2.86 \citep{Osterbrock:2006}, yielding an \hb flux of 2.75$\times 10^{-17}$ erg s$^{-1}$ cm$^{-2}$ for the inter-arm region.
In the Arm region, we only used N2, since \ha and \nii\ are the only detected emission lines.

Following an approach similar to that described in \citealt{Curti:2020klever}, we derived the metallicity that simultaneously minimizes the $\chi^2$ between the observed line ratios and the expected value according to each calibrator using the \texttt{python} library \verb|lmfit|.  The resulting best-fit metallicity values, including their errors (which account for the intrinsic scatter of the calibrators), are reported in Table \ref{tab:ism_table}.
We find no statistically significant difference in metallicity between the various regions. The values are consistent within the uncertainties and indicate slightly subsolar metallicities, with 12\,+\,log(O/H)\,=\,$8.49-8.58$ (i.e., $\sim$\,$0.60-0.80$\,$Z_{\odot}$\footnote{We assumed a solar metallicity 12\,+\,log(O/H)$_{\odot}$\,=\,8.69 \citep{Asplund:2021}.}). Nonetheless, it is interesting that the star-forming clumps, where newborn stars are expected to inject metals into the ISM, do not exhibit higher metallicity than the other regions. We speculate that any more prominent intrinsic differences (if present) may be smoothed out by the spatial extent of the shutters, which encompass emission from physically distinct regions. Indeed, the Clump-N shutter is not perfectly centered on the clump, and the Inter-Arm shutter may be contaminated by emission from the clump close by (see Fig. \ref{fig:Nircam_RGB_and_prism}). Interestingly, all the derived metallicities are consistent with that expected from the fundamental metallicity relation (FMR; \citealt{Mannucci:2010}) for \target, given its stellar mass and SFR ($M_{\star}$\,$\approx$\,4\,$\times$\,10$^{10}$ \msun and SFR\,$\approx$\,120 \sfr, inferred later in Sects. \ref{sec:sed_cigale} and \ref{sec:pab}). Using the parameterization by \citet{Curti:2020}, the FMR predicts a gas-phase metallicity of 12\,+\,log(O/H)\,$\approx$\,8.57. As discussed in Sect. \ref{sec:discussion_inflow}, this is connected to the presence of gas inflows (\citealt{Genzel:2023}; Pulsoni et al., in prep.).

\subsection{Chemical abundances: N/O, S/H, S/O}
\label{sec:chemical_abundances}
Among chemical elements, nitrogen is one of the most extensively studied, from the local Universe up to high redshifts \citep{Pilyugin:2012,Shapley:2015,Masters:2016, Strom2018,Stiavelli:2025, Sanders:2025}. In star-forming galaxies, its abundance can be derived from the relatively bright \nii\, and \oii emission lines, via the N2O2 diagnostic (e.g., \citealt{Steidel2016,Strom2018,Hayden-Pawson:2022}). Since these lines originate from nearly the same ionization zones and have similar ionization potentials, their ratio minimally depends on ISM ionization conditions. Alternatively, one can infer N/O using the N2S2 diagnostic, which assumes a constant S/O abundance. This method is less affected by dust extinction, since the lines involved are closer in wavelength than [N{\sc{ii}}] and \oii in N2O2. Yet, the \sii emission lines have a lower ionization potential (10.4 eV) and, therefore, may be emitted from regions with softer radiation fields. Moreover, the assumption of a constant S/O abundance, which is necessary to use \sii as a proxy for oxygen, may not always be true. For all these reasons, we provide two independent measurements of N/O in the various galaxy regions, adopting the calibrations from \citet{Hayden-Pawson:2022} (see Table \ref{tab:ism_table}). For each region,  N2O2- and N2S2 diagnostics deliver consistent, values-based N/O ratios. We adopted their average as the fiducial value. The N/O abundance in Clump-S is 0.3\,dex higher than in the Inter-Arm, with the Clump-N lying between the two.
We obtain similar results by also using the calibrations for N2O2 and N2S2 presented in \citet{Strom2018}.

Similar to oxygen, sulfur is produced by massive stars; therefore, the relative S/O abundance is expected to remain constant at about the solar value (i.e., log(S/O)$_{\odot}$\,=\,$-1.57$; \citealt{Asplund:2021}). A sulfur abundance of 12\,+\,log(S/H) has been studied only for a handful of galaxies at $z$\,>\,1 \citep[e.g. ][]{Sanders:2020, Rogers:2024, Rogers:2026} and can be derived from the optical \sii and [S{\sc{iii}}] emission lines. Compared to the \oii and [O{\sc{iii}}] lines used to infer 12\,+\,log(O/H), the sulfur lines have a milder (albeit exponential) dependence on electron temperature, due to their lower ionization potential, which makes them detectable even at solar or supersolar abundances \citep{Diaz2007}. For \target, we estimated 12\,+\,log(S/H) from the S23 
diagnostic (\citealt{Diaz:2022}; see Table \ref{tab:ism_table}). Interestingly, all three examined galaxy regions consistently feature a subsolar S/H abundance (12\,+\,log(S/H)$_{\odot}$\,=\,7.12; \citealt{Asplund:2009}), that is 12\,+\,log(S/H)\,$\approx$\,$6.6-6.8$. By finally combining these measurements of 12\,+\,log(S/H) with our derived fiducial values of 12\,+\,log(O/H), we estimate a relative log(S/O) abundance of about $-1.9$, consistent for all regions, which is systematically lower than the expected nearly constant (solar) value of -1.57 $\pm$ 0.05 \citep{Asplund:2021}.
In Sect. \ref{sec:sulfur}, we investigate the implications of this result, discussing possible formation and depletion channels for sulfur.  

\section{Resolved SED fitting}\label{sec:sed_cigale}

In this section, we present a spatially resolved SED model of high-resolution JWST/NIRCam and HST/ACS imaging to derive 2D maps of the main galaxy's physical properties (e.g., stellar mass, age, dust attenuation, and SFR). We then compare the resolved results with those obtained from integrated SED fitting, which includes low-resolution photometry at longer wavelengths, as well as those inferred from NIRSpec MSA in Sects. \ref{sec:ism}.

\subsection{PSF matching and SED modeling with CIGALE}\label{sec:sed_analysis}
We matched all reduced 6$''$\,$\times$\,6$''$ cutouts to the spatial resolution of the F444W band (FWHM\,$=$\,0.15$''$). For JWST/NIRCam and HST/ACS data, we used \texttt{WebbPSF} models and empirical PSFs from 3DHST \citep{Skelton:2014} to create F444W matching kernels using the \texttt{python} package \texttt{photutils.psf.matching} and thus convolved each image to the F444W PSF. To estimate the background noise to adopt as the error on pixel fluxes while accounting for spatial correlations between pixels introduced by the data reduction, PSF matching, and instrumental features, we created a segmentation map for each cutout and performed flux measurements in 100 randomly placed "empty" apertures. These measurements were taken for a suite of aperture sizes with variable linear sizes of $N$ pixels (up to $N$\,=\,10; see also \citealt{Tacchella:2015sins}). We derived the effective rms from the flux distribution as a function of aperture size and quantified the rms within one pixel. Since the effective rms is expected to be spatially constant at a fixed aperture size, we associated this error with the flux of all pixels in a given cutout. An additional contribution comes from Poisson noise, which might be significant in bright regions. From the reduced HST and JWST images of \target (in counts), we find that Poisson noise varies across the galaxy and with spectral band, ranging from negligible to 30\%\,--\,60\% above the background noise. This translates to relative Poisson uncertainties in pixel flux from approximately 0.1\% to a few percent. To account for this additional variable noise contribution, we conservatively included an additional 10\% error on all pixel fluxes in our SED model.

\begin{figure*}
    \centering
    \includegraphics[width=0.90\linewidth]{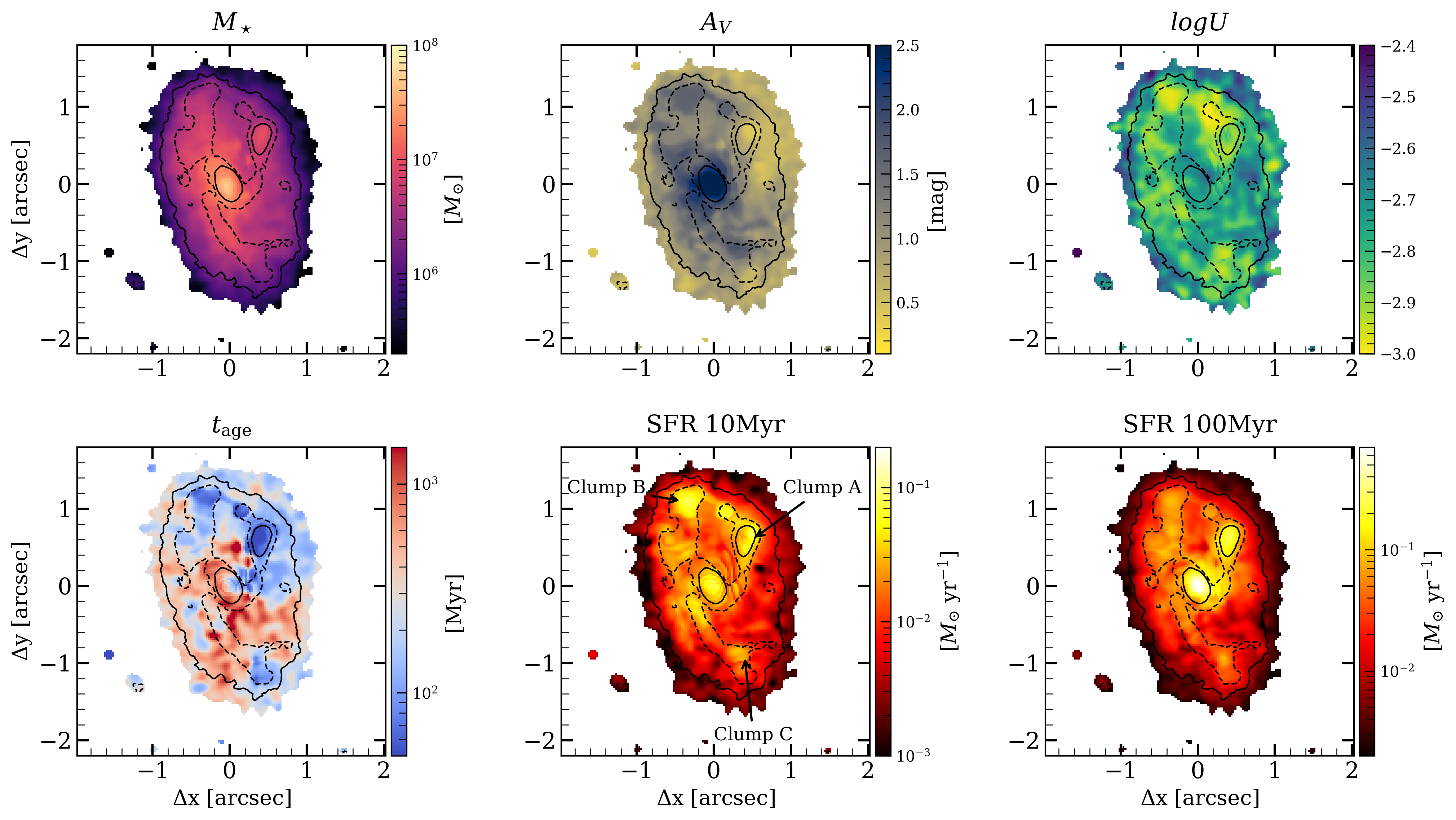}
    \caption{2D maps at a 25 mas pixel scale of the main galaxy properties derived from our resolved SED fitting with \texttt{CIGALE} of HST/ACS and JWST/NIRCam images. Left to right, top: Maps of stellar mass and SFRs averaged over the last 10 Myr and 100 Myr. Bottom: $A_V,{\rm star}$, stellar age, and $\chi^2_{\rm red}$. The $\chi^2_{\rm red}$ map shows the goodness of our pixel-by-pixel SED model, with values $\sim$\,1 across the entire galaxy. 
    The solid contours correspond to 10 and 100$\sigma$ levels of F444W continuum emission, highlighting the locations of the bright Clump A and of the bulge.
    The dashed lines trace the spiral arms and represent the 3$\sigma$ contours of the residual F444W emission, obtained by subtracting the F444W image from a Gaussian-smoothed version with a kernel of 10 pixels. The location of the three main star-forming clumps is highlighted in the SFR10 map.
    }
    \label{fig:SED_fitting_maps}
\end{figure*}

To model the SED of individual pixels, we used the SED fitting code \texttt{CIGALE} \citep{Boquien:2019} and only fit the SED of pixels with S/N\,>\,7 on the F444W band, based on the noise estimate inferred above. This adopted S/N threshold enables continuous mapping of the whole galaxy and guarantees a S/N\,$\gtrsim$\,2 in the bluer NIRCam filters, which indeed yields good reduced $\chi^2$ values ($\chi^2_{{\rm red}}$\,$\approx$\,1) in individual pixel fits (see Fig. \ref{fig:SED_fitting_maps}, bottom right panel). The SED of each S/N(F444W)\,>\,7 pixel was then modeled independently from the other pixels under the same set of assumptions. For all pixels, we fixed the redshift to $z$\,=\,2.224, as measured from \ha\ \citep{Forster:2018}, and assumed a decaying exponential star formation history (SFH). We allowed ages since the onset of star formation to range from 50 Myr and 2900 Myr (i.e., the age of the Universe at $z$\,=\,2.224), with $e$-folding timescales $\tau$ spanning 10 Myr to 8000 Myr (i.e., approximately corresponding to a constant SFH). We adopted the \citet{Bruzual:2003} stellar population synthesis models, assuming a \citet{Chabrier:2003} initial mass function, and fixing the metallicity to $Z$\,=\,0.008\footnote{\texttt{CIGALE} only provides limited metallicity values (i.e., $Z$\,=\,0.008 and $Z$\,=\,0.02), with no intermediate value in between.}, to reduce the number of free parameters and avoid degeneracy with other physical quantities. This metallicity value (corresponding to 0.4\,$Z_{\odot}$) is slightly lower than that inferred in Sect. \ref{sec:metallicity}. Yet, we find it delivers better results compared to fixing the metallicity to the solar value ($Z$\,=\,0.02). In addition, we assumed the \citet{Calzetti:2000} attenuation law and included nebular emission. We adopted suitable priors consistent with the ISM properties inferred in Sect. \ref{sec:ism}, while allowing a sufficiently large parameter space to also reproduce the properties of regions across the full system, including those that may differ from the regions for which we have measurements from the MSA spectra. Among these, we set a two-value grid for $n_{\rm e}$ (100 cm$^{-3}$ and 1000 cm$^{-3}$) and allowed log($U$) to vary from --\,3.4 and --\,2.0. 
With all these ingredients, our resolved SED model  reproduces the variety of single-pixel SED shapes well, which reflect the different intrinsic natures and properties of distinct galaxy regions, such as clumps, spiral-arm regions, and inter-arm regions (see Appendix \ref{sec:sed_fitting_models} for representative examples of best-fit SEDs).

\subsection{2D maps of host galaxy properties}\label{sec:sed_results}

Figure \ref{fig:SED_fitting_maps} displays the resulting spatially resolved maps of relevant galaxy parameters, computed in \texttt{CIGALE} as the likelihood-weighted mean of the following priors: stellar mass ($M_{\star}$), average SFRs computed over the last 10 Myr and 100 Myr (SFR10 and SFR100, respectively); stellar dust attenuation ($A_{V,{\rm star}}$); stellar age ($t_{\rm age}$); and reduced $\chi^2$. Table \ref{tab:sed_properties} summarizes the main physical properties for the most relevant regions (i.e., bulge and clumps), extracted using dedicated apertures (Fig. \ref{fig:Pa_beta_reconstructed}), and for the whole galaxy.
In particular, for the central red compact core, we adopted a 0.25\arcsec-diameter aperture, corresponding to twice its effective radius. This was measured by modeling the F444W image with GALFIT \citep{galfit}, using a double S\'ersic profile \citep{SersicProfile} (Pulsoni et al., in prep.). One of the two components was used to reproduce the extended disk (S\'ersic index $n_{\rm disk} = 0.35\pm0.01$, $R_{\rm e,disk} = 6.5\pm\,0.2$ kpc), while the other was used to reproduce the central mass concentration ($n_{\rm core} = 1.16\pm0.06$, $R_{\rm e,core} = 1.01\pm\,0.14$ kpc). With $n_{\rm core} = 1.16\pm0.06$, the central core is best classified as either a disky- or pseudo-bulge (\citealt{KormendyKennicutt2004}; detailed discussion will be presented in Pulsoni et al., in prep.). In the remainder of the paper, we simply refer to this central component as ``bulge,'' with no attempt to interpret its dynamical nature, which is beyond the purpose of this paper.

The most prominent feature in the stellar mass map is the central bulge ($M_{\rm \star,Bulge}$\,=\,(7\,$\pm$\,3)\,$\times$\,10$^{9}$ \msun), which is then followed by the brightest and largest star-forming clump, Clump A ($M_{\rm \star,A}$\,=\,(3.5\,$\pm$\,0.7)\,$\times$\,10$^{9}$ \msun). With a total galaxy stellar mass of ($M_{\star}$\,=\,(3.7\,$\pm$\,1.0)\,$\times$\,10$^{10}$ \msun), we find that the bulge and Clump A contribute by 19\% and 9\%, respectively. The SFR10 map follows the clumps and the spiral arms well, whose locations are visually highlighted by the dashed contours and reveal recent active star formation in the central bulge (SFR10$_{\rm Bulge}$\,=\,12\,$\pm$\,8 \sfr). Averaged over a longer time frame, the SFR100 map displays a different
morphology: Clump A has been forming stars intensely for the last 100 Myr, while Clump-B and the spiral arms show weaker past activity than the bulge and Clump A, except for the spiral arm region between them. Interestingly, the bulge and its surroundings appear to have experienced enhanced star formation over the last 100 Myr (SFR100$_{\rm Bulge}$\,=\,82\,$\pm$\,42 \sfr), suggesting that the bulge formed its stars mostly in the past.

In terms of dust extinction, the stellar attenuation map ($A_{V,{\rm star}})$ indicates the highly obscured nature of the nuclear region ($A_{V,{\rm star}}$\,$=$\,2.83\,$\pm$\,0.24).
If we account for extra nebular line emission attenuation (i.e., $A_{V,{\rm neb}}$\,$\approx$\,$A_{V,{\rm star}}$\,/\,0.44; \citealt{Calzetti:2000}), we obtain $A_{V,{\rm neb}}$\,=\,6.43\,$\pm$\,0.55. Significant extinction ($A_{V,{\rm star}}$\,=\,1.5\,$\simeq$\,2.0) is also found in regions of intense star formation, 
such as along the spiral arms, in Clump B, and around Clump A. The core of Clump A, in contrast, appears less attenuated ($A_{V,{\rm star}}$\,$\simeq$\,0.5), consistent with the fact that it is also detected in the rest-frame UV.

Finally, the stellar age map suggests a trend toward the inner galaxy regions: younger stars ($t_{\rm age}$\,=\,50\,--\,100 Myr) preferentially lie in the clumps, while older populations dominate the bulge and the inter-arm regions ($t_{\rm age}$\,$\approx$\,$400-2000$ Myr). 
The relatively old ages inferred for the bulge, combined with its elevated SFR100, furthermore indicate that it formed most of its stellar mass in the past, while continuing to maintain low star formation activity.
However, stellar age is more uncertain than the other inferred parameters, with typical uncertainties of 50\%\,--\,70\% on single pixels.

An alternative scenario could involve an unattenuated, quiescent bulge with extremely old stars and no recent star formation, which would still explain its observed red color. To test this hypothesis, we extracted the bulgeś integrated SED using our adopted 0.25$''$ aperture and fit it with \texttt{CIGALE}, 
following the same prescriptions as in Sect. \ref{sec:sed_analysis} but with $A_{V,{\rm star}}$\,=\,0. With this assumption, we obtain a poor best-fit ($\chi^2_{\rm red}$\,=\,12), as the model fails to reproduce the observed SED at redder wavelengths. The observed slope is much steeper than the best-fit $A_{V,{\rm star}}$\,=\,0 model, which flattens and does not reproduce most of the NIRCam photometry.

In Appendix \ref{sec:sed_fitting_check}, we validate the results from our resolved SED fitting, by comparing the 2D \ha distribution inferred from the dust corrected SFR with the observed \ha flux observed with ERIS.
Figure \ref{fig:ha_comparison} show that the \ha flux expected from the SED fitting is overall in good agreement with the observed ERIS \ha line map at similar angular resolutions (Pulsoni et al., in prep.).

\renewcommand{\arraystretch}{1.5} 
\begin{table*}
    \centering
    \footnotesize
\caption{Physical properties of the bulge, clumps, and whole galaxy, derived from our SED fitting and analysis of the \Pab line emission.}
\label{tab:sed_properties}
    \resizebox{0.75\textwidth}{!}
{\begin{tabular}{l|ccccc|cc}
    \hline
    \hline
     \multirow{2}{*}{ } & \multicolumn{5}{c|}{Resolved} & \multicolumn{2}{c}{Integrated}\\
     \cline{2-8}
     & Bulge & Clump A & Clump B & Clump C & Galaxy & Galaxy$_{\rm 4.4\mu m}$ & Galaxy$_{\rm 1.2 mm}$\\
     \hline
    $d$ [$''$] & 0.25 & 0.8 & 0.5 & 0.3 & 3 & 3 & 3\\
    $M_{\star}$ [10$^9$\,M${_\odot}$] &  7 $\pm$ 3 $^{(\star)}$ & 3.5 $\pm$ 0.7 & 1.5 $\pm$ 0.3 & 0.67 $\pm$ 0.19 & 37 $\pm$ 10 & 23 $\pm$ 5 & 36 $\pm$ 8\\
    $t_{\rm age}$ [Myr] &  <\,850 & 123 $\pm$ 75 & 136 $\pm$ 103 & <\,483 & 288 $\pm$ 178 & <\,223 & 413 $\pm$ 294\\
    $A_{V,{\rm star}}$ [mag] & 2.83 $\pm$ 0.24 & 0.77 $\pm$ 0.11 & 1.50 $\pm$ 0.16 & 1.58 $\pm$ 0.21 & 1.01 $\pm$ 0.22 & 1.19 $\pm$ 0.07 & 1.01 $\pm$ 0.09\\
    SFR10 [M${_\odot}$\,yr$^{-1}$] &  12 $\pm$ 8 $^{(\star)}$ & 21 $\pm$ 6 & 13 $\pm$ 4 & 2.8 $\pm$ 1.3 & 109 $\pm$ 66 & 93 $\pm$ 20 & 71 $\pm$ 19\\
    SFR100 [M${_\odot}$\,yr$^{-1}$] & 82 $\pm$ 42 $^{(\star)}$ & 54 $\pm$ 17 & 18 $\pm$ 6 & 6 $\pm$ 3 & 288 $\pm$ 178 & 384 $\pm$ 143 & 146 $\pm$ 130\\
    SFR$_{\rm UV+IR}$ [M${_\odot}$\,yr$^{-1}$] & - & - & - & - & - & - & 105 $\pm$ 17\\
    \hline
    SFR$^0_{\rm Pa\beta}$ [M${_\odot}$\,yr$^{-1}$] &  1.2\,$\pm$\,1.0 $^{(\star)}$ & $21 \pm 5$ & $10 \pm 2$ & $3 \pm 1$ & - & \multicolumn{2}{c}{$66 \pm 22$} \\
    SFR$_{\rm Pa\beta}$ [M${_\odot}$\,yr$^{-1}$] & 7.6$^{+8.8}_{-6.6}$ $^{(\star)}$ &  24$^{+6}_{-7}$ &  11$^{+2}_{-2}$ & 3.4$^{+1.6}_{-1.0}$ & - & \multicolumn{2}{c}{ 120$^{+70}_{-70}$}\\
    $A_{V,{\rm neb}}$ & 6.43 $\pm$ 0.55 &  0.4$^{+0.1}_{-0.1}$ &  0.4$^{+0.1}_{-0.1}$ &  0.4$^{+0.3}_{-0.3}$ & - & \multicolumn{2}{c}{ 2.0 $\pm$ 0.6}\\
    \hline

    \end{tabular}
    }
    \tablefoot{For the galaxy, we provide SED-based results from resolved maps (Galaxy) and integrated models, including photometry up to 4.4\,$\mu$m (Galaxy$_{4.4\mu{\rm m}}$) and 1.2\,mm (Galaxy$_{1.2{\rm mm}}$). From top to bottom, we list: the diameter ($d$) of the apertures (shown in Fig. \ref{fig:Pa_beta_reconstructed}) used to sum or average pixel values in resolved maps, and for the integrated analysis; the SED-based physical parameters (Sect. \ref{sec:sed_cigale}), i.e., stellar mass ($M_{\star}$), age ($t_{\rm age}$), and stellar dust attenuation ($A_{V,{\rm star}}$); SFRs averaged over the last 10 Myr and 100 Myr (SFR10 and SFR100), and those derived from UV plus IR luminosity (SFR$_{\rm UV+IR}$); \Pab-based SFRs, inferred by assuming no dust attenuation (SFR$^0_{\rm Pa\beta}$) and corrected for nebular attenuation (SFR$_{\rm Pa\beta}$), using the $A_{V,{\rm neb}}$ values reported in the last row (see Sect. \ref{sec:pab_sfr_map} for details). $^{(\star)}$ $M_{\star}$ and SFRs of the bulge were computed by summing over the pixels within the bulge aperture and multiplying the resulting sum by two (see Sect. \ref{sec:sed_results}).}
    
\end{table*}

\subsection{Integrated SED fitting}\label{sec:sed_integ_results}

In this section, we discuss our SED modeling of integrated HST/ACS and JWST/NIRCam photometry, using the same assumptions as in Sect. \ref{sec:sed_analysis}, to compare the resolved and integrated results. To extract the total HST and JWST photometry from the reduced cutouts, we summed the flux contained in all pixels within the S/N(F444W)\,>\,7 region defined in Sect. \ref{sec:sed_analysis} and estimated the corresponding error by summing pixel uncertainties in quadrature.
Results from this integrated SED fitting, including up to NIRCam F444W photometry, are summarized in Table \ref{tab:sed_properties} (see the column labeled Galaxy$_{4.4{\rm \mu m}}$; the best-fit model is shown in Appendix \ref{sec:sed_fitting_models}).

Among all physical parameters, $M_{\star}$ is the most robustly constrained based on the available rest-frame optical HST/ACS and near-IR JWST/NIRCam photometry, whereas the other quantities would require coverage at longer mid- and far-IR wavelengths to constrain dust emission. As reported in Sect. \ref{sec:sed_results}, the total stellar mass of \target, computed as a sum of all pixels, is $M_{\star}$\,=\,(3.7\,$\pm$\,1.1)\,$\times$\,10$^{10}$ \msun. This total value is slightly higher than the integrated result ($M_{\star}$\,=\,(2.3\,$\pm$\,0.5)\,$\times$\,10$^{10}$ \msun) by about a factor of 1.6, yet the two agree within the uncertainties. 
This difference likely reflects outshining effects from a young stellar population, which lead to underestimated mass-to-light ratios in integrated SED modeling (e.g., \citealt{Maraston:2010,Wuyts:2012,GimenezArteaga:2023}). Previous studies (e.g., \citealt{Sorba:2018,GimenezArteaga:2023,GimenezArteaga:2024}) indicate that stellar masses can indeed be underestimated by up to a factor of 5 in integrated SED fitting, with larger offsets for higher specific SFRs. In the case of \target, outshining seems to play a minor role, given the smaller discrepancy between the resolved and integrated result. This discrepancy is more consistent with that found for star-forming galaxies of similar mass at $z$\,$\simeq$\,2 \citep{Shen:2024} and at $z$\,$\simeq$\,4\,--\,6 \citep{Li:2024,Lines:2025}.

To better constrain SFRs, stellar age, and dust attenuation, we performed an additional integrated SED fitting,. This included the available photometry at longer wavelengths from \textit{Spitzer} IRAC (5.8\,$\mu$m and 8\,$\mu$m) and MIPS 24\,$\mu$m, using the values reported by \citealt{Dickinson:2003} and \citealt{Whitaker:2014}, as well as mid-IR data from \textit{Herschel}/PACS 160\,$\mu$m, for which we adopted the integrated photometry from \citealt{Lutz:2011}. we also included ALMA Band 6 data at 1.2\,mm, introduced in Sect. \ref{sec:data_alma}.
In \texttt{CIGALE}, we modeled dust emission using dust templates from \citet{Dale:2014} (see Appendix \ref{sec:sed_fitting_models} for more details on the best-fit model).

As shown in Table \ref{tab:sed_properties} (see the column labeled Galaxy$_{1.2{\rm mm}}$), including mid- and far-IR photometry yields excellent agreement between the integrated $M_{\star}$ and its sum over pixels, as well as consistent $A_{V,{\rm star}}$ values. By contrast, the inferred SFR10 and SFR100 values are systematically lower than those from both resolved SED fitting and integrated modeling with photometry up to 4.4\,$\mu$m. In particular, the resulting SFR10 is lower than the other two inferred total SFR10 values and the most accurate values previously published in \citet{Forster:2009} (SFR\,$=$\,101 -- 190 \sfr; see Sect. \ref{sec:pab} for details on previously published SFRs). Yet, the wide coverage of this photometric dataset, extending up to far-IR, allowed us to estimate the total SFR from the UV and IR luminosities of the SED best-fit model \citep{Kennicutt:1998}.
Doing so, we obtained a higher SFR (SFR$_{{\rm UV+IR}}$\,=\,105\,$\pm$\,17 \sfr), consistent with the sum over pixels. Investigating the causes of the systematic lower SFR10 derived from the integrated SED analysis with photometry up to 1.2\,mm is beyond the scope of this paper.

\section{2D spatial distribution of \Pab line emission}\label{sec:pab}

\begin{figure*}
    \centering
    \includegraphics[width=0.95\linewidth]{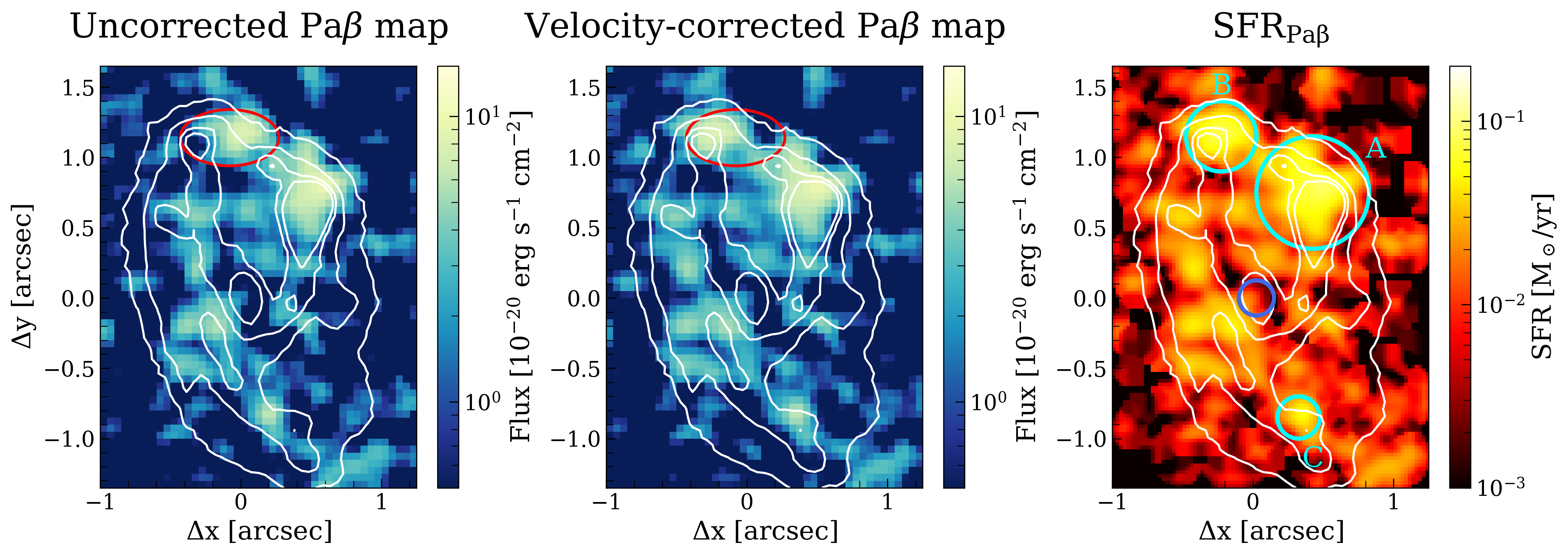}
    \caption{2D spatial distribution  of \Pab in \target from NIRCam grism F444W observations. \textit{Left:} \Pab line map, as reconstructed by \texttt{grizli} \citep{grizli}, with white contours tracing the F444W continuum emission. The \Pab emission appears offset, with respect to the continuum emission, along the \textit{x}-axis (i.e., the grism dispersion direction), due to velocity gradients. This is particularly evident for the northernmost clump (circled in red). \textit{Middle:} \Pab line map after correcting for the \ha\ velocity-field model inferred from VLT/ERIS AO data (Pulsoni et al., in prep.). 
    Here, the \Pab and continuum emission align more closely, as highlighted by their good spatial correspondence with the bright star-forming clumps detected at S/N = 5.
    \textit{Right:} \Pab-traced SFR map, obtained by assuming negligible dust attenuation. Cyan circular apertures (diameters of 0.8$''$, 0.5$''$, and 0.3$''$) and a central blue aperture (0.25$''$) are used to infer the \Pab-based SFRs for the three brightest clumps and the bulge, respectively. The two \Pab maps have a 0.05$''$ pixel scale, while the SFR map is rebinned to a 0.025$''$ pixel scale and smoothed with a Gaussian kernel of $\sigma = 1$ pix.}
    \label{fig:Pa_beta_reconstructed}
\end{figure*}

\subsection{Correction of \Pab line map for velocity gradients}\label{sec:vel_corr}

Compared to long-slit spectroscopy, slitless grism observations
 allow us to recover the overall 2D emission line distribution after removing continuum emission (e.g., \citealt{Nelson:2016, Nelson:2024,Matharu:2022}).
For high-resolution grism data of extended sources, such as NIRCam WFSS ($R$\,$\sim$\,1600, FWHM $\approx$\,190 \kms), the emission-line morphology along the dispersion axis combines spatial and velocity displacements. In contrast, low-resolution spectral data (e.g., HST/WFC3 or JWST/NIRISS, $R$\,$\sim$\,100, FWHM $\approx$\,2000 \kms) contain only spatial information along this dispersion axis. Consequently, accounting for this velocity-position degeneracy is a crucial, necessary step to recover the intrinsic 2D emission-line distribution or 1D line velocity profile from NIRCam WFSS data.

The left panel of Fig. \ref{fig:Pa_beta_reconstructed} shows the continuum-subtracted \Pab emission line map of \target, reconstructed by the FRESCO team using \verb|grizli| (\citealt{grizli}; see \citealt{Oesch:2023,Neufeld:2024} for more details). This map was matched to the spatial resolution of the VLT/ERIS AO-assisted data (FWHM\,=\,0.16$''$\footnote{The \ha\ line map used in this work was smoothed to a spatial resolution of FWHM\,=\,0.16$''$ (Pulsoni et al., in prep.), whereas the nominal resolution of the ERIS AO-assisted data is FWHM\,=\,0.105$''$ (Förster Schreiber et al., in prep.).}, compared to FWHM\,=\,0.13$''$ for NIRCam F444W), using the \texttt{python} package \texttt{photutils.psf.matching}.
The comparison with the continuum emission in the direct F444W image (white contours) clearly reveals a relative spatial offset with respect to \Pab line emission along the horizontal WE direction. Such offsets are fairly systematic and consistent with expectations, given the observed kinematics and orientation of the grism dispersion axis (with increasing wavelengths toward E). \target indeed features a kinematic position angle of 25 deg (i.e., roughly aligned with the NE--SW direction) and a substantial velocity gradient \citep[ranging from -300 to +300 km/s; ][Pulsoni et al. in prep.]{Forster:2018,Genzel:2023}, as seen in the \ha kinematics from high-resolution IFU data. It follows that the largest offsets are observed at greater distances from the galaxy center along the kinematic major axis, where the velocity gradient is strongest.
This can be well appreciated for the northernmost clump (i.e., Clump B, circled in red in the left and middle panels of Fig. \ref{fig:Pa_beta_reconstructed}).
The fact that these offsets between the continuum and line emission are not observed in VLT/ERIS \ha\ data furthermore supports their nonphysical origin.

To break the spatial and kinematic degeneracy and recover the intrinsic \Pab spatial distribution, we relied on the (smoothed) \ha\ velocity map obtained from ERIS AO observations (Pulsoni et al., in prep.). Since \Pab and \ha\ emission originate from the same gas phase, we can reasonably assume the same kinematics for \Pab as for \ha. The mapping in NIRCam grism data is only weakly depends on wavelength and position, with a scale of 10.0\,\AA  per NIRCam F444W pixel (i.e., 10.0\,\AA/0.063$''$). Given the 50 mas pixel scale of the 2D reconstructed \Pab map, a one-pixel shift results in a 7.9\,\AA  wavelength-shift, which corresponds to 57 \kms at $\lambda$\,$\sim$\,4.13\,$\mu$m (i.e., the observed \Pab wavelength for \target). To a first approximation, the velocity-corrected, intrinsic position along the spectral axis (i.e., $x_{\rm int}$) of a given \Pab emitting pixel at a position $x_{\rm obs}$ in the reconstructed image can be recovered as
\begin{equation}
x_{\rm int} = x_{\rm obs} - v(x_{\rm int})/(57~ \rm km~s^{-1}),
\label{eq:vel_corr}
\end{equation}
where $v(x_{\rm int})$ is the velocity for each pixel from the rotating disk model based on the ERIS \ha\ data (Pulsoni et al., in prep.). 

The middle panel of Fig. \ref{fig:Pa_beta_reconstructed} displays the reconstructed \Pab emission line map, after shifting each pixel along the horizontal (dispersion) axis according to Eq. \ref{eq:vel_corr}. Although approximate, our velocity correction provides a much better spatial correspondence between the \Pab and continuum emission: the \Pab emission of the northernmost clump and the galaxy spiral arms now align well with the continuum emission seen in the F444W direct image. The \Pab line distribution reveals that most of the emission arises from the clumps and spiral arms.
The S/N = 7 detection of the integrated \Pab line allows us to recover its spatial distribution with a significance of $\sim$5$\sigma$ in the clumps and $\sim$3$\sigma$ along the spiral arms, while the bulge and the other regions of the galaxy remain undetected in \Pab emission.

\subsection{Star formation traced via \Pab line emission}\label{sec:pab_sfr_map}

We used the velocity-corrected \Pab map to spatially map recent star formation (with a timescale of $\sim$\,10 Myr), probing a more obscured regime compared to optical Balmer emission lines (e.g., \ha\ or \hb).
Following \citet{Reddy:2023a}, the SFR can be inferred from the dust-corrected \Pab luminosity ($L_{{\rm Pa\beta}}$) using their Eq. 2 (assuming $Z$\,=\,0.02, the closest metallicity to our measurements):
\begin{equation}
    \rm SFR_{Pa\beta} [{\rm M_\odot/yr}] = 7.67 \times 10^{-41} ~ L_{Pa\beta} [erg/s].
\label{eq:pab_sfr}
\end{equation}

Since we lack a full 2D map of nebular line attenuation and using $A_{V,{\rm star}}$ may introduce per-pixel correction errors (due to astrometric misalignments), in a first approximation we assumed negligible attenuation at the 1.28 $\mu$m \Pab wavelength and derived a SFR map from the observed (i.e., without dust correction) \Pab luminosity, through a pixel-by-pixel application of the equation above. The resulting \Pab-based SFR map is shown in the rightmost panel of Fig. \ref{fig:Pa_beta_reconstructed}. It has been rebinned to a 0.025$''$ pixel scale (to enable direct comparison with the SFR maps in Fig. \ref{fig:SED_fitting_maps}) and smoothed with a Gaussian kernel of $\sigma$\,=1 pixel. Table \ref{tab:sed_properties} reports the resulting total \Pab-based, not dust-corrected, SFRs (i.e., SFR$^0_{{\rm Pa\beta}}$) obtained by summing the \Pab flux within dedicated circular apertures for Clumps A, B, and C (in cyan; diameters of 0.8$''$, 0.5$''$, 0.3$''$, respectively), as well as for the bulge (in red; diameter of 0.25$''$). A comparison of the SFR10 values derived from our resolved SED fitting (see Fig. \ref{fig:SED_fitting_maps} with those reported in Table \ref{tab:sed_properties}) indicates that \Pab traces recent star formation in the clumps and spiral arm well, where extinction is low and the star formation is indeed recent. A major difference, however, arises for the central bulge, where the \Pab line emission is very weak. Consistent with the picture outlined in Sect. \ref{sec:sed_results}, this results from recent, low-level star formation and heavy extinction in the inner bulge regions, which remain significant even at the \Pab wavelength. 

To place a robust lower limit on the \Pab flux in the bulge, we computed its expected value within our adopted bulge aperture, based on the SFR10 value from our SED modeling. We adopted two distinct values for dust attenuation: 1) stellar attenuation only, with $A_{V,{\rm star}}$\,=\,2.83\,$\pm$0.24, and 2) extra nebular attenuation computed as $A_{V,{\rm neb}}$\,=\,$A_{V,{\rm star}}$\,/\,0.44\,=\,6.43\,$\pm$\,0.55 \citep{Calzetti:2000}. We thus obtain lower limits for the expected observed \Pab flux, which translate into lower limits of 1.78 \sfr and 0.076 \sfr for the intrinsic \Pab-based SFR, both consistent with our SFR$^0_{{\rm Pa\beta}}$ estimate obtained for the bulge aperture.

Finally, by fitting two Gaussian components to the 1D NIRCam WFSS spectrum (shown in Fig. \ref{fig:Pa_beta_observed}), we infer a total \Pab flux of $F_{\rm Pa\beta}=(2.2\,\pm\,0.7)\,\times\,10^{-17}$ \ergscm for the whole galaxy, which yields a global SFR$^0_{{\rm Pa\beta}}=(66\pm22)$ M${_\odot}$\,yr$^{-1}$.
This total 1D \Pab flux is more reliable than the sum of the \Pab fluxes over all pixels of the (velocity-corrected) \Pab line map, which adds noise from more diffuse galaxy regions. Compared to previous published estimates \citep{Forster:2009}, our inferred total SFR$^0_{{\rm Pa\beta}}$ is systematically lower than the values derived from IR luminosity and integrated HST-based SED modeling (SFR$_{{\rm IR}}\simeq101$ M${_\odot}$\,yr$^{-1}$ and SFR$_{{\rm SED}}\simeq110$ M${_\odot}$\,yr$^{-1}$, respectively). Moreover, it is closer to the total \ha-traced SFR, which is corrected for stellar attenuation (SFR$_{{\rm H\alpha}}\simeq73$ M${_\odot}$\,yr$^{-1}$). This latter estimate is known to underestimate the true nebular attenuation (e.g., \citealt{Calzetti:2000,Price:2014,Tacchella:2018}). By accounting for extra attenuation of nebular line emission, \citet{Forster:2009} found SFR$_{{\rm H\alpha}}\simeq190$ M${_\odot}$\,yr$^{-1}$.
This result furthermore indicates that \Pab still suffers from substantial attenuation in more deeply obscured regions, such as the bulge. Therefore, neglecting extinction or improperly accounting for extra nebular attenuation at the 1.28\,$\mu$m \Pab wavelength underestimates the intrinsic SFR of more obscured regions and, consequently, of the entire galaxy.

In Table \ref{tab:sed_properties}, we report the local and total \Pab-based SFRs (SFR$_{{\rm Pa\beta}}$) obtained by correcting for nebular attenuation derived from the hydrogen line ratios. Specifically, for Clumps A and B, and separately Clump C, we report the values respectively inferred for Clump-N and Clump-S from MSA spectra (see Table \ref{tab:ism_table}); for the bulge, we report the $A_{V,{\rm neb}}$ value inferred from $A_{V,{\rm star}}$;
and for the whole galaxy, we report an average estimate based on medium- and broadband imaging (see Appendix \ref{sec:av_imaging}). As expected, the SFR$_{{\rm Pa\beta}}$ of the clumps have not changed much with respect to the SFR$^0_{{\rm Pa\beta}}$ values, due to their low extinction, whereas for the bulge and the whole galaxy, we obtain values consistent with those for SFR10. In particular, we note that the total SFR$_{{\rm Pa\beta}}$ of the galaxy now also matches the global SFR$_{{\rm UV+IR}}$ estimate better.

\section{Implications on the evolution of the galaxy disk}\label{sec:discussion}

\subsection{Hints for inside-out galaxy growth}
\label{sec:stellar_age}

\begin{figure*}[htb]
    \centering
    \includegraphics[width=0.99\linewidth]{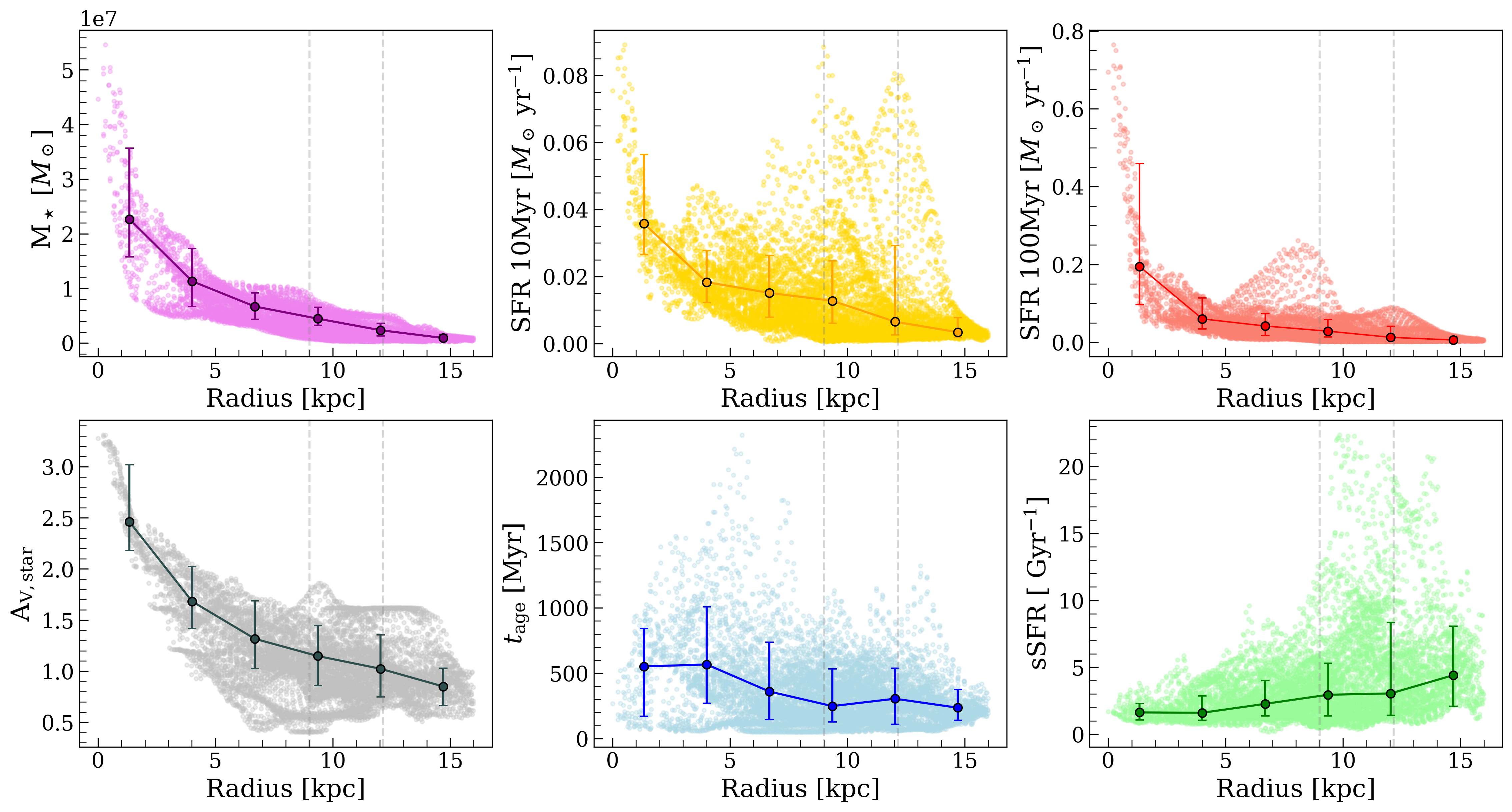}
    \caption{Radial profiles of galaxy properties, as derived from our 2D SED-based maps. Left to right, top: Radial profiles of $M_{\star}$, SFR10, and SFR100. Bottom: $A_{V,{\rm star}}$, $t_{\rm age}$, and specific SFR (sSFR\,=\,SFR10/M$_\star$). The big markers show the median value computed for each ring, with the error bars showing the $\pm 1\sigma$ scatter. In the background, we plot the distribution of individual pixel values. The vertical dashed gray lines correspond to the de-projected radii of Clump A  ($\sim$\,9 kpc) as well as Clumps B and C ($\sim$\,12 kpc).
    }
    \label{fig:Radial}
\end{figure*}

As indicated in Sect. \ref{sec:sed_results}, the 2D stellar age map inferred from our resolved SED fitting reveals a relatively older stellar population in the bulge and the nearby regions and younger stellar populations in the clumps. 
Despite the large non-axisymmetry, we derived radial profiles for each physical parameter (Fig. \ref{fig:Radial}) to better investigate radial trends of the various SED-inferred properties. These profiles are based on the resolved SED maps shown in Fig. \ref{fig:SED_fitting_maps}, using six elliptical annuli at the galaxy's center and computed from the F444W image. Each annulus has a width of 0.25\arcsec and is characterized by a semiminor to semimajor axis ratio of $b/a$\,=\,0.72 \citep{Genzel:2023} and a position angle of 25$^{\circ}$ (counterclockwise from N to E; \citealt{Forster:2018}). As shown in Fig. \ref{fig:Radial}, we computed the median value of the examined property in each elliptical annulus and associated it with an error equal to a $\pm 1\sigma$ scatter. To also better appreciate azimuthal scatter, for each property we show the values of individual pixels as a function of their de-projected distance from the galaxy center, by correcting for the galaxy inclination \citep[122$^{\circ}$, ][]{Genzel:2023}.

The radial profiles overall highlight the presence of a massive, highly obscured yet star-forming bulge as also seen in other cosmic noon galaxies \citep{Arriagada:2025}. Except for the specific SFR (sSFR\,=\,SFR10\,/\,$M_{\star}$) and stellar age, all properties exhibit a declining median radial trend, with the clumps (vertical dashed lines) breaking down the azimuthal symmetry of the 2D SED maps and creating peaks in the scatter distribution. On the contrary, the sSFR features a slightly increasing median profile, reaching peaks in the scatter at the location of the clumps. The stellar age radial profile appears globally flat within the $1\sigma$ scatter, with typical uncertainties on pixel values of 50-70\%. Yet, the  stellar age radial profile tentatively exhibits a negative gradient, suggesting an older stellar population in the inner galaxy regions. We also notice that the stellar age map shown in Fig. \ref{fig:SED_fitting_maps} displays asymmetry between the northern and southern parts of the galaxy, which could furthermore contribute to washing out any possible age gradient. If this negative gradient is real, this would imply an inside-out growth for \target.

\begin{figure}[htb]
    \centering
    \includegraphics[width=0.95\linewidth]{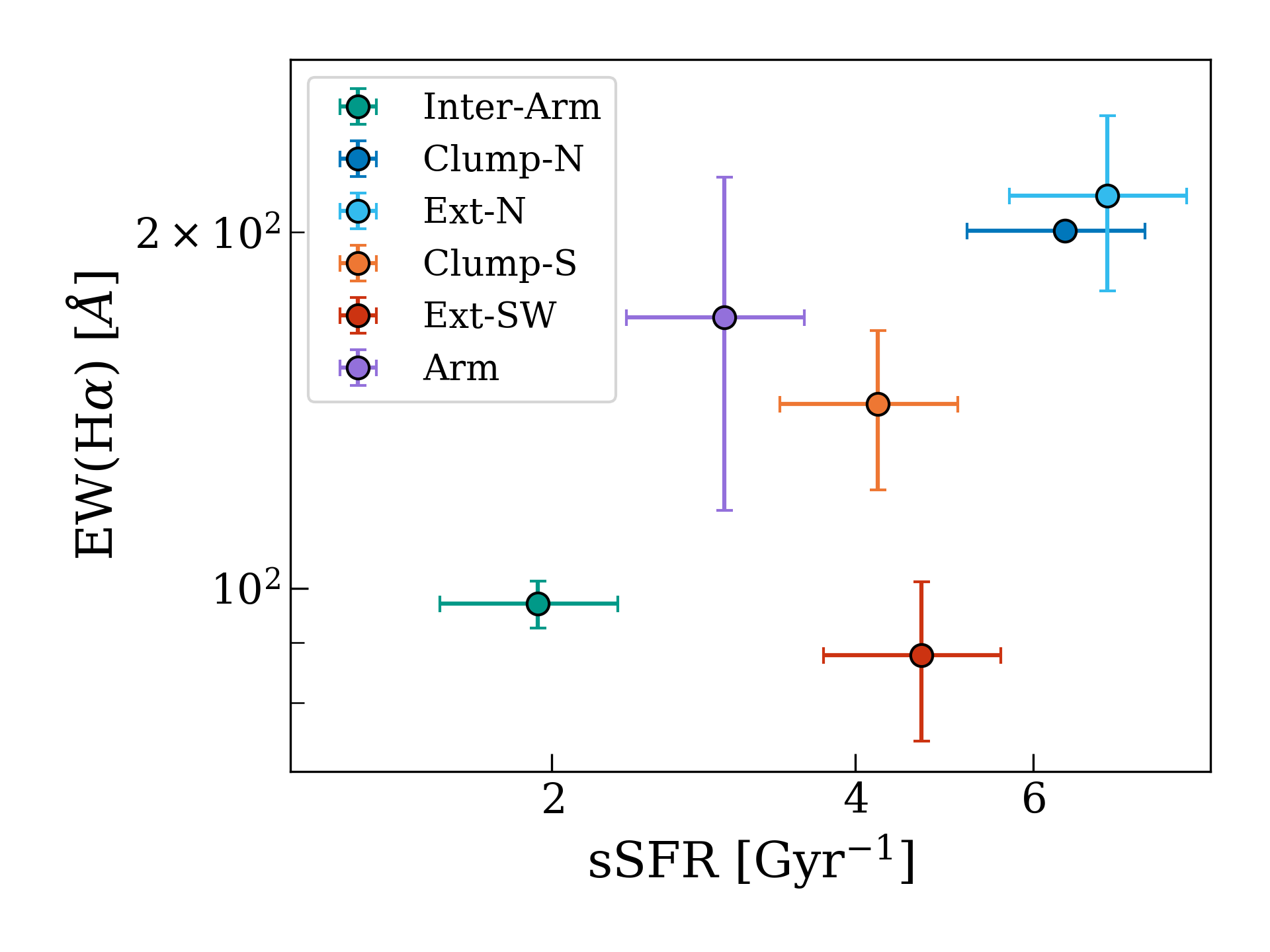}
    \caption{Rest-frame EW(\ha) as a function of sSFR for the various galaxy regions probed by distinct MSA shutters. The EW(\ha) measurements are derived from PRISM/CLEAR data, while the sSFRs are derived from our SED-based SFR10 and $M_\star$ maps. The plot highlights a correlation between the two quantities, with higher EW(\ha) values scaling with higher sSFRs, indicative of younger stellar populations.}
    \label{fig:EW_vs_sSFR}
\end{figure}

Even though our SED-based results cannot prove the inside-out growth of this galaxy, the available NIRSpec MSA spectroscopy independently supports this possible scenario. The detection of both \ha\ and the adjacent continuum emission in the PRISM/CLEAR data indeed allows us to locally estimate the rest-frame \ha\ equivalent width (EW), which is indicative of stellar population age \citep[e.g.,][]{Levesque:2013}. In particular, higher EWs indicate younger populations \citep{Forster:2009,Forster:2011, Reddy:2018}, since EW(\ha) depends on the ratio between the \ha\ line luminosity (a proxy of the current SFR) and the continuum luminosity (a proxy of the past star formation activity). Therefore, we can use the EW(\ha) as an independent age diagnostic to compare the stellar age trend resulting from our SED-based $t_{\rm age}$ map and radial gradient. In Fig. \ref{fig:EW_vs_sSFR}, we display the EW(\ha) as a function of the sSFR for each MSA shutter (same color code as in Fig. \ref{fig:Nircam_RGB_and_prism}). 
In Fig. \ref{fig:EW_vs_sSFR} we note a correlation between EW(\ha) and sSFR, such that regions with larger sSFRs also have higher EWs, as previously observed in individual galaxies at $z$\,$\approx$\,$1-3$ \citep{Reddy:2018}.

In terms of stellar age, the Clump-N and Ext-N regions exhibit higher EW(\ha) values (EW(\ha)\,$\approx$\,200\,\AA), hinting at the presence of a younger stellar population, compared to the inter-arm and Ext-SW shutters (EW(\ha)\,$\approx$\,90 -- 100\,\AA). The Clump-S and Arm regions display intermediate values (despite the large EW(\ha) uncertainty for the latter) of EW(\ha)\,$\approx$\,140 -- 170\,\AA. This relative trend is consistent with the tentative trend showed in our $t_{\rm age}$ map from resolved SED fitting (see Fig. \ref{fig:SED_fitting_maps}). The older stellar population of the inter-arm region is also confirmed by the strong Balmer break in the corresponding PRISM/CLEAR spectrum (see Fig. \ref{fig:Nircam_RGB_and_prism}). Although the distinct shutters probe galaxy regions of different natures (e.g., clumps, spiral arms, and inter-arm regions), considering their distance from the galaxy center, we find higher EW(\ha) values with increasing radii. Specifically,  EW(\ha)$_{\rm Inter-Arm}$\,<\,EW(\ha)$_{\rm Arm}$\,<\,EW(\ha)$_{\rm Clump-N}$. This supports an ``inside-out'' formation scenario for \target \citep{Tacchella:2015b,Nelson:2016}.

\subsection{Connection with gas inflows} \label{sec:discussion_inflow}

Exploiting high-resolution \ha\ SINFONI AO-assisted observations of \target, \citet{Genzel:2023} find kinematic residuals indicative of rapid gas inflows at speeds of 87 km/s. More recent ERIS AO observations at higher spectral and spatial resolutions have confirmed the fast radial inflows and enabled a more accurate association with the spiral arms seen in NIRCam imaging (Pulsoni et al., in prep.). The detection of such radial gas inflows, along with the results inferred in this work, supports the scenario of inflowing gas funneled from the galaxy outskirts toward the central bulge, which dilutes the metal content of the galaxy.
Radial gas inflows detected along spiral arms may be fueled by gas transport from the very outskirts of the disk and by external accretion from the circumgalactic medium (CGM). If the inflowing gas is pristine or metal-poor, this process is expected to leave imprints on the galaxy properties, as discussed in the following subsections.

\subsubsection{“Diluted" gas phase metallicity}\label{sec:n_enrichment}
As mentioned in Sect. \ref{sec:metallicity}, the FMR predicts a metallicity of 12\,+\,log(O/H)\,$\approx$\, 8.57 \citep{Curti:2020} for a galaxy with the same stellar mass and SFR as \target (i.e., $M_{\star}$\,$\approx$\,4\,$\times$\,10$^{10}$ \msun, SFR$_{{\rm Pa\beta}}$\,$\simeq$\,120 \sfr). At such a high SFR, the relation between metallicity and $M_{\star}$ steepens. This has been interpreted to result from prominent dilution effects caused by large inflows of metal-poor gas \citep{Curti:2020}.

Additional evidence supporting dilution effects due to gas inflows comes from nitrogen abundance. The N/O derived in Sect. \ref{sec:chemical_abundances} (Table \ref{tab:ism_table}) is interestingly 0.3\,dex higher at fixed metallicity than the best-fit relation measured for the local galaxies reported in \citet{Hayden-Pawson:2022} and consistent with the large scatter observed a $z$\,$\sim$\,2 \citep{Shapley:2015,Hayden-Pawson:2022}.
In contrast to oxygen and sulfur -- $\alpha$ elements mainly produced on shorter timescales as core collapse supernovae (CCSNe) explode -- nitrogen has two different creation channels \citep[see][for a review]{Kobayashi:2023}: a primary, metallicity-independent channel, in which nitrogen is expelled by CCSNe \citep{Maiolino:2019}, and a secondary channel, in which nitrogen originates from intermediate-mass, metal-rich stars in the asymptotic giant branch phase, which dominate nitrogen production at high metallicity \citep[e.g.,][]{Vincenzo:2016}. 
Given the different formation mechanisms of nitrogen compared to oxygen, measuring its abundance provides insights into the star formation and chemical enrichment histories of the galaxy \citep{Sanders:2023}.

With the advent of JWST, nitrogen enhancement is found to be common in very metal-poor ($Z$\,$\sim$\,0.1\,$Z_{\odot}$) and high-$z$ ($z$\,>\,4) star-forming galaxies \citep[e.g.,][]{Bunker:2023, Cameron:2023,Isobe:2023, Topping:2024}, though it is not ubiquitous \citep{Rogers:2024}.
Various mechanisms were proposed to explain such nitrogen enhancement, including an excess of Wolf-Rayet stars \citep{Flury:2025, Welch:2025}, varying IMFs \citep{Arellanocordova:2025}, and enrichment by population III stars \citep{Nandal:2024}. Considering the inflow detection in \target \citep[][Pulsoni et al. in prep.]{Genzel:2023}, we may be witnessing an overall gas--metallicity dilution, as traced by O/H, rather than a nitrogen enhancement \citep[e.g.,][]{Stiavelli:2025}. The overabundance of N/O relative to subsolar O/H may arise from lower-metallicity gas from the disk outskirts or external accretion, which dilutes the gas-phase oxygen abundance in the disk while preserving the relative N/O abundance.

\subsubsection{Low-metallicity clumps}\label{sec:low_metallicity_clump}

Cosmological simulations show that gas inflows from the CGM are often associated with off-center star-forming clumps, bright in \ha\ and poor in metals \citep{Ceverino:2016}. 
As reported in Sect. \ref{sec:metallicity}, we find no significant difference in the metallicity of the clumps compared to that in the other probed galaxy regions. Yet, we note that prominent metallicity gradients are smoothed due to the spatial extent of the MSA shutters. In particular, we expect a large amount of metals to be ejected during the explosion of massive stars in Clump-N due to intense star formation ($\sim$\,20 \sfr), contrary to observations. Indeed, Clump-N is found in the same region as the strongest inflow detection (\citealt{Genzel:2023}; Pulsoni et al., in prep.). This suggests the clump may have an intrinsic metallicity lower than that inferred from the Clump-N MSA data, which reduces the O/H abundance due to pristine gas accretion, while replenishing the star-forming region with gas.

Estimating the amount of pristine gas needed to produce a given observed decrease in metallicity involves assuming an infall model, where the pristine inflowing gas is responsible for diluting metallicity and boosting star formation. By using Eq. 8 from \citet{Mannucci:2010} and assuming a decreased metallicity of only 0.1\,dex for Clump-N, we find that the accreted pristine gas in the clump should account for 20\% of the mass of the gas. This would explain such a decrease in metallicity. Assuming a gas fraction of $M_{\rm gas}/M_{\star}$\,=\,1.8 \citep{Genzel:2023}, and considering our inferred stellar mass ($M_{\star}$\,$\approx$\,4\,$\times$\,10$^{10}$ \msun), we estimate that the pristine gas accreted on the clump is on the order of $10^9 M_\odot$. Clump-S is also located in a region where kinematic residuals indicate the presence of gas inflows. Applying similar reasoning for Clump-N, accretion of $\sim 10^8 M_{\odot}$ of pristine gas would be required to account for a 0.1\,dex dilution.

\subsection{Sulfur abundance at $z$\,=\,2.2}\label{sec:sulfur}

The detection of the \sii and [S{\sc{iii}}] emission lines in the NIRSpec MSA spectra allowed us to locally measure the sulfur abundance of \target (in Sect. \ref{sec:chemical_abundances}), which had previously been determined only for a handful of galaxies at $z$\,>\,1 (e.g., \citealt{Sanders:2020,Rogers:2024, Rogers:2026}). Before the advent of JWST, detecting the \siii emission lines at $z$\,$\gtrsim$\,1.5 was indeed challenging, because they are emitted at the boundary between optical and near-IR wavelengths; hence, they are redshifted outside the atmospheric transmission windows accessible from the ground. 
Sulfur is an $\alpha$ element (similar to oxygen), mainly produced inside the core of massive stars and returned to the ISM on short timescales through the explosion of CCSNe. Contrary to nitrogen, S/O typically does not depend on galactic metallicity or SFH, but rather on supernova yields only. Hence, it is usually assumed to be constant throughout cosmic time and close to the solar value (log(S/O)$_\odot$\,$\sim$\, 
$-1.7$; \citealt{Asplund:2009}).

In Sect. \ref{sec:chemical_abundances}, we reported a value of log(S/O)\,\,--\,2.0 $\pm$ 0.2 and \,--\,1.9 $\pm$ 0.2 for the inter-arm and Clump-N regions of \target. The measured value is systematically lower than the solar value of log(S/O)$_\odot$\,$\sim$\, 
$-1.6$ \citep{Asplund:2021}. For Clump-S, we still obtain a subsolar S/O abundance compatible with the solar value within the large uncertainties.
A low S/O abundance is also observed in other $z$\,>\,2 galaxies \citep{Rogers:2024, Rogers:2026} at both low- and high-metallicities. 
The inferred abundance pattern may result from a secondary (subdominant) sulfur production channel through type Ia SNe \citep{Kobayashi:2020}, which produces different oxygen and sulfur yields, thereby altering the relative S/O abundance. The subsolar nature of these high-z sources, hence, might indicate that type Ia SNe have not had time to explode yet.  
Another possible explanation for the low S/O values is that sulfur may be captured by ice and dust grains \citep{hilyBlant:2022, PerezDiaz:2024}.
Studies of local galaxies and individual HII regions have found large scatters for S/O abundance \citep[e.g.,][]{Diaz:2022,PerezDiaz:2024}, with some studies revealing tentative evidence of lower S/O at increasing 12\,+\,log(O/H) \citep{Diaz:2022}, while others find no such trend \citep{Izotov:2006}. However, a similar decreasing trend with metallicity is also observed for other $\alpha$ elements, which may be ascribed to the depletion of oxygen into dust grains \citep{Izotov:2006}.
In light of this, our measurements of abundance patterns in \target highlight the need for caution when using nitrogen or sulfur to infer 12\,+\,log(O/H) gas-phase metallicities \citep{PerezDiaz:2024}. In particular, larger samples of $z$\,>\,1 galaxies with available S/O measurements are needed to validate the assumption of constant S/O abundance at high redshifts.

\section{Conclusions}\label{sec:conclusions}

Relying on a novel set of JWST observations of \target, including both NIRSpec and NIRCam data, this work presents a comprehensive resolved study of the ISM and stellar population properties of \target, a large star-forming galaxy at $z$\,=\,2.224 with an outstanding spiral morphology. In doing so, our analysis leverages the synergy between JWST and ground-based IFU AO observations and introduces two key novelties in the data reduction of NIRSpec MSA spectra and in the post-processing of NIRCam WFSS data: 1) the extraction of science spectra from multiple MSA shutters -- not just the central one (as in the standard reduction procedure) -- thereby probing seven distinct galaxy regions in total; the recovery of the intrinsic 2D spatial distribution of the \Pab line emission, achieved by correcting the \Pab line map reconstructed from NIRCam WFSS data for velocity gradients. This correction uses the \ha\ velocity-field model derived from IFU ERIS AO data at a high spectral resolution ($R$\,$\sim$\,10500) and comparable spatial resolution (Pulsoni et al., in prep.). Below, we summarize the main results of this work:

\begin{itemize}

    \item  By means of different emission-line diagnostics, we investigated the ISM properties in three distinct galaxy regions, i.e., two clumps and an inter-arm region. Overall, we found the clumps to exhibit lower dust attenuation ($A_{V,{\rm neb}}$\,$\simeq$\,0.4), and feature relatively low electron densities ($n_{\rm e}$\,$\lesssim$\,300 cm$^{-3}$) and ionization parameters (log($U$)\,$\simeq$\,$-3.0$). As inferred from our resolved SED analysis, the clumps are found to be massive ($M_{\star}$\,=\,0.67\,--\,3.5\,$\times$\,10$^{9}$ M$_{\odot}$) and have SFRs ranging from $3-20$ \sfr. The inter-arm region is more dust-attenuated ($A_{V,{\rm neb}}$\,$\simeq$\,1.3) and has similar ionization conditions to those of the clumps. For this region, we derived a 2$\sigma$ upper limit on the electron density of $n_{\rm e}$\,<\,1900 cm$^{-3}$.\\

    \item In the same regions, we also explored the gas-phase metallicity 12\,+\,log(O/H), by using strong-line diagnostics as well as nitrogen and sulfur chemical abundances. We also obtained a metallicity measurement for a fourth region, corresponding to a spiral arm. We find no statistically significant difference in metallicity between the various regions, featuring 12\,+\,log(O/H) values in the $8.49-8.58$ range and uncertainties of about $0.1-0.2$\,dex overall.\\

    \item We find evidence of metallicity patterns that are likely imprints of the known gas inflows detected in \target (\citealt{Genzel:2023}; Pulsoni et al., in prep.). Our inferred metallicity values are compatible with the FMR-expected metallicity (i.e., 12\,+\,log(O/H)\,$\approx$\,8.57) for a galaxy with the same stellar mass and SFR as those derived for \target (i.e., $M_{\star}$\,$\approx$\,4\,$\times$\,10$^{10}$ \msun, SFR$_{{\rm Pa\beta}}$\,$\simeq$\,120 \sfr). At such high SFRs, the steepening of the $Z-M_{\star}$ relation indicates dilution effects due to gas inflows of metal-poor gas. Similarly, such diluted metallicity could also explain the observed 0.3\,dex enhancement in the log(N/O) abundance (i.e., $\simeq$\,-1) in the log(N/O) -- 12\,+\,log(O/H) plane.\\

    \item Given the \sii and [S{\sc iii}] emission lines detected in the NIRSpec MSA data, we measured the sulfur abundance in the inter-arm and clump regions (12\,+\,log(S/H) of 6.6$^{+0.2}_{-0.2}$ for the first two, and 6.7$^{+0.3}_{-0.2}$ for the third one). When combined with the 12\,+\,log(O/H) values, these yield a relative log(S/O)=-1.9, which is lower than the expected constant (solar) value (i.e., log(S/O)\,$\approx$\,$-1.7$), yet consistent within the uncertainties. This might indicate a secondary production channel for sulfur via the explosion of type I SNe on longer timescales.\\

    \item The 2D map of stellar age, derived from our resolved SED fitting, and the extracted radial profile tentatively show a negative gradient, with older stellar populations  on average closer to the center ($t_{\rm age}$\,$\approx$\,$400-2000$ Myr) and younger ones in the outer regions (particularly in the clumps, $t_{\rm age}$\,=\,$50-100$ Myr). This trend is also supported by an inverse increasing trend in sSFR, and the \ha\ EWs inferred from the NIRSpec MSA spectra for distinct galaxy regions.\\

    \item Our resolved SED analysis, along with the velocity-corrected \Pab line map, reveals a central moderately massive ($M_{\star}$\,=\,(7\,$\pm$\,3)\,$\times$\,10$^{9}$ \msun) bulge, which is highly obscured ($A_{V,{\rm neb}}$\,=\,6.43\,$\pm$\,0.55) and undetected in line emission at the 1.28\,$\mu$m \Pab wavelength. The bulge exhibits recent ($\sim$\,10 Myr) star formation (SFR10$_{\rm Bulge}$\,=\,12\,$\pm$\,8 \sfr), and even a higher activity in the past (SFR100$_{\rm Bulge}$\,=\,82\,$\pm$\,42 \sfr). Together with the relatively older stellar ages in the inner galaxy regions ($t_{\rm age}$\,$\approx$\,$400-2000$ Myr), the enhanced SFR100 of the bulge supports the scenario that it assembled most of its stellar mass in the past, while maintaining some level of active star formation.

\end{itemize}

Taking advantage of the synergy between deep ground-based AO IFU observations, this work showcases the crucial role that JWST can play in investigating key galaxy formation and evolution processes at cosmic noon. Moreover, the successful analysis of \target presented in this work paves the way for similar studies of larger samples of galaxies at cosmic noon, leveraging available spectroscopy from JWST and ground-based IFU facilities. This will enable an accurate, complete, and resolved characterization of the ISM and stellar population properties.
Moreover, the discovery of the deeply obscured yet star-forming bulge in \target underscores the importance of future high-resolution millimeter continuum observations with ALMA. Such observations would detect obscured star formation activity in the bulge and pin down its current growth rate.

\begin{acknowledgements}
The authors are grateful to S. Carniani for valuable comments and discussion.
G.T., N.M.F.S., C.B., J.C. and J.M.E.S. acknowledge funding by the European Union
(ERC Advanced Grant GALPHYS, 101055023), H.Ü. and G.M. acknowledge funding by the European Union (ERC APEX, 101164796). Views and opinions expressed are, however, those of the authors
only and do not necessarily reflect those of the European Union or
the European Research Council. Neither the European Union nor the
granting authority can be held responsible for them.
L.S. acknowledges financial support from the PhD grant funded on PNRR Funds Notice No. 3264 28-12-2021 PNRR M4C2 Reference IR0000034 STILES Investment 3.1 CUP C33C22000640006.
This work has been financed by the European Union with the Next Generation EU plan, Mission 4, through PRIN-MUR project “PROMETEUS” (202223XPZM), CUP C53D2300080-006.
This work is based on observations made with the NASA/ESA/CSA James Webb Space Telescope. Some of the data were obtained from the Mikulski Archive for Space Telescopes at the Space Telescope Science Institute, which is operated by the Association of Universities for Research in Astronomy, Inc., under NASA contract NAS 5-03127 for JWST. These observations are associated with programs: 1210, 1895, 1180, 6541.
Some of the data products presented herein were retrieved from the Dawn JWST Archive (DJA). DJA is an initiative of the Cosmic Dawn Center (DAWN), which is funded by the Danish National Research Foundation under grant DNRF140.

\end{acknowledgements}

\bibliographystyle{aa}
\bibliography{aa}

@ARTICLE{Deugenio:2025,
       author = {{D'Eugenio}, Francesco and {Cameron}, Alex J. and {Scholtz}, Jan and {Carniani}, Stefano and {Willott}, Chris J. and {Curtis-Lake}, Emma and {Bunker}, Andrew J. and {Parlanti}, Eleonora and {Maiolino}, Roberto and {Willmer}, Christopher N.~A. and {Jakobsen}, Peter and {Robertson}, Brant E. and {Johnson}, Benjamin D. and {Tacchella}, Sandro and {Cargile}, Phillip A. and {Rawle}, Tim and {Arribas}, Santiago and {Chevallard}, Jacopo and {Curti}, Mirko and {Egami}, Eiichi and {Eisenstein}, Daniel J. and {Kumari}, Nimisha and {Looser}, Tobias J. and {Rieke}, Marcia J. and {Rodr{\'\i}guez Del Pino}, Bruno and {Saxena}, Aayush and {{\"U}bler}, Hannah and {Venturi}, Giacomo and {Witstok}, Joris and {Baker}, William M. and {Bhatawdekar}, Rachana and {Bonaventura}, Nina and {Boyett}, Kristan and {Charlot}, Stephane and {Danhaive}, A. Lola and {Hainline}, Kevin N. and {Hausen}, Ryan and {Helton}, Jakob M. and {Ji}, Xihan and {Ji}, Zhiyuan and {Jones}, Gareth C. and {Juod{\v{z}}balis}, Ignas and {Maseda}, Michael V. and {P{\'e}rez-Gonz{\'a}lez}, Pablo G. and {Perna}, Michele and {Pusk{\'a}s}, D{\'a}vid and {Shivaei}, Irene and {Silcock}, Maddie S. and {Simmonds}, Charlotte and {Smit}, Renske and {Sun}, Fengwu and {Villanueva}, Natalia C. and {Williams}, Christina C. and {Zhu}, Yongda},
        title = "{JADES Data Release 3: NIRSpec/Microshutter Assembly Spectroscopy for 4000 Galaxies in the GOODS Fields}",
      journal = {\apjs},
         year = 2025,
        month = mar,
       volume = {277},
       number = {1},
          eid = {4},
        pages = {4},
          doi = {10.3847/1538-4365/ada148},
archivePrefix = {arXiv},
       eprint = {2404.06531},
 primaryClass = {astro-ph.GA},
       adsurl = {https://ui.adsabs.harvard.edu/abs/2025ApJS..277....4D}
}

@ARTICLE{Scholtz:2025,
       author = {{Scholtz}, J. and {Carniani}, S. and {Parlanti}, E. and {D'Eugenio}, F. and {Curtis-Lake}, E. and {Jakobsen}, P. and {Bunker}, A.~J. and {Cameron}, A.~J. and {Arribas}, S. and {Baker}, W.~M. and {Charlot}, S. and {Chevellard}, J. and {Circosta}, C. and {Curti}, M. and {Duan}, Q. and {Eisenstein}, D.~J. and {Hainline}, K. and {Ji}, Z. and {Johnson}, B.~D. and {Jones}, G.~C. and {Kumari}, N. and {Maiolino}, R. and {Maseda}, M.~V. and {Perna}, M. and {P{\'e}rez-Gonz{\'a}lez}, P.~G. and {Rawle}, T. and {Rieke}, M. and {Rinaldi}, P. and {Robertson}, B. and {Saxena}, A. and {Shivaei}, I. and {Silcock}, M.~S. and {Sun}, Y. and {Rodr{\'\i}guez Del Pino}, B. and {Tacchella}, S. and {{\"U}bler}, H. and {Venturi}, G. and {Williams}, C.~C. and {Willmer}, C.~N.~A. and {Willott}, C. and {Witstok}, J.},
        title = "{JADES Data Release 4 -- Paper II: Data reduction, analysis and emission-line fluxes of the complete spectroscopic sample}",
      journal = {arXiv e-prints},
         year = 2025,
        month = oct,
          eid = {arXiv:2510.01034},
        pages = {arXiv:2510.01034},
          doi = {10.48550/arXiv.2510.01034},
archivePrefix = {arXiv},
       eprint = {2510.01034},
 primaryClass = {astro-ph.GA},
       adsurl = {https://ui.adsabs.harvard.edu/abs/2025arXiv251001034S}
}

@ARTICLE{Forster:2011,
       author = {{F{\"o}rster Schreiber}, N.~M. and {Shapley}, A.~E. and {Genzel}, R. and {Bouch{\'e}}, N. and {Cresci}, G. and {Davies}, R. and {Erb}, D.~K. and {Genel}, S. and {Lutz}, D. and {Newman}, S. and {Shapiro}, K.~L. and {Steidel}, C.~C. and {Sternberg}, A. and {Tacconi}, L.~J.},
        title = "{Constraints on the Assembly and Dynamics of Galaxies. II. Properties of Kiloparsec-scale Clumps in Rest-frame Optical Emission of z \raisebox{-0.5ex}\textasciitilde 2 Star-forming Galaxies}",
      journal = {\apj},
         year = 2011,
        month = sep,
       volume = {739},
       number = {1},
          eid = {45},
        pages = {45},
          doi = {10.1088/0004-637X/739/1/45},
archivePrefix = {arXiv},
       eprint = {1104.0248},
 primaryClass = {astro-ph.CO},
       adsurl = {https://ui.adsabs.harvard.edu/abs/2011ApJ...739...45F}
}

@ARTICLE{Tacconi:2020,
       author = {{Tacconi}, Linda J. and {Genzel}, Reinhard and {Sternberg}, Amiel},
        title = "{The Evolution of the Star-Forming Interstellar Medium Across Cosmic Time}",
      journal = {\araa},
         year = 2020,
        month = aug,
       volume = {58},
        pages = {157-203},
          doi = {10.1146/annurev-astro-082812-141034},
archivePrefix = {arXiv},
       eprint = {2003.06245},
 primaryClass = {astro-ph.GA},
       adsurl = {https://ui.adsabs.harvard.edu/abs/2020ARA&A..58..157T}
}

@ARTICLE{Speagle:2014,
       author = {{Speagle}, J.~S. and {Steinhardt}, C.~L. and {Capak}, P.~L. and {Silverman}, J.~D.},
        title = "{A Highly Consistent Framework for the Evolution of the Star-Forming ``Main Sequence'' from z \raisebox{-0.5ex}\textasciitilde 0-6}",
      journal = {\apjs},
         year = 2014,
        month = oct,
       volume = {214},
       number = {2},
          eid = {15},
        pages = {15},
          doi = {10.1088/0067-0049/214/2/15},
archivePrefix = {arXiv},
       eprint = {1405.2041},
 primaryClass = {astro-ph.GA},
       adsurl = {https://ui.adsabs.harvard.edu/abs/2014ApJS..214...15S}
}

@ARTICLE{Forster:2020,
       author = {{F{\"o}rster Schreiber}, Natascha M. and {Wuyts}, Stijn},
        title = "{Star-Forming Galaxies at Cosmic Noon}",
      journal = {\araa},
         year = 2020,
        month = aug,
       volume = {58},
        pages = {661-725},
          doi = {10.1146/annurev-astro-032620-021910},
archivePrefix = {arXiv},
       eprint = {2010.10171},
 primaryClass = {astro-ph.GA},
       adsurl = {https://ui.adsabs.harvard.edu/abs/2020ARA&A..58..661F}
}

@ARTICLE{Nandal:2024,
       author = {{Nandal}, Devesh and {Regan}, John A. and {Woods}, Tyrone E. and {Farrell}, Eoin and {Ekstr{\"o}m}, Sylvia and {Meynet}, Georges},
        title = "{Explaining the high nitrogen abundances observed in high-z galaxies via population III stars of a few thousand solar masses}",
      journal = {\aap},
         year = 2024,
        month = mar,
       volume = {683},
          eid = {A156},
        pages = {A156},
          doi = {10.1051/0004-6361/202348035},
archivePrefix = {arXiv},
       eprint = {2402.03428},
 primaryClass = {astro-ph.GA},
       adsurl = {https://ui.adsabs.harvard.edu/abs/2024A&A...683A.156N}
}

@ARTICLE{Ceverino:2016,
       author = {{Ceverino}, Daniel and {S{\'a}nchez Almeida}, Jorge and {Mu{\~n}oz Tu{\~n}{\'o}n}, Casiana and {Dekel}, Avishai and {Elmegreen}, Bruce G. and {Elmegreen}, Debra M. and {Primack}, Joel},
        title = "{Gas inflow and metallicity drops in star-forming galaxies}",
      journal = {\mnras},
         year = 2016,
        month = apr,
       volume = {457},
       number = {3},
        pages = {2605-2612},
          doi = {10.1093/mnras/stw064},
archivePrefix = {arXiv},
       eprint = {1509.02051},
 primaryClass = {astro-ph.GA},
       adsurl = {https://ui.adsabs.harvard.edu/abs/2016MNRAS.457.2605C}
}

@ARTICLE{Maiolino:2019,
       author = {{Maiolino}, R. and {Mannucci}, F.},
        title = "{De re metallica: the cosmic chemical evolution of galaxies}",
      journal = {\aapr},
         year = 2019,
        month = feb,
       volume = {27},
       number = {1},
          eid = {3},
        pages = {3},
          doi = {10.1007/s00159-018-0112-2},
archivePrefix = {arXiv},
       eprint = {1811.09642},
 primaryClass = {astro-ph.GA},
       adsurl = {https://ui.adsabs.harvard.edu/abs/2019A&ARv..27....3M}
}

@ARTICLE{Arellanocordova:2025,
       author = {{Arellano-C{\'o}rdova}, K.~Z. and {Cullen}, F. and {Carnall}, A.~C. and {Scholte}, D. and {Stanton}, T.~M. and {Kobayashi}, C. and {Martinez}, Z. and {Berg}, D.~A. and {Barrufet}, L. and {Begley}, R. and {Donnan}, C.~T. and {Dunlop}, J.~S. and {Hamadouche}, M.~L. and {McLeod}, D.~J. and {McLure}, R.~J. and {Rowlands}, K. and {Shapley}, A.~E.},
        title = "{The JWST EXCELS survey: direct estimates of C, N, and O abundances in two relatively metal-rich galaxies at z ≃ 5}",
      journal = {\mnras},
         year = 2025,
        month = jul,
       volume = {540},
       number = {4},
        pages = {2991-3007},
          doi = {10.1093/mnras/staf855},
archivePrefix = {arXiv},
       eprint = {2412.10557},
 primaryClass = {astro-ph.GA},
       adsurl = {https://ui.adsabs.harvard.edu/abs/2025MNRAS.540.2991A}
}

@ARTICLE{Cappellari:2023,
    author = {{Cappellari}, M.},
    title = "{Full spectrum fitting with photometry in PPXF: stellar population
        versus dynamical masses, non-parametric star formation history and
        metallicity for 3200 LEGA-C galaxies at redshift $z\approx0.8$}",
    journal = {MNRAS},
    eprint = {2208.14974},
    year = 2023,
    volume = 526,
    pages = {3273-3300},
    doi = {10.1093/mnras/stad2597}
}

@ARTICLE{Diaz:2000,
       author = {{D{\'\i}az}, Angeles I. and {Castellanos}, Marcelo and {Terlevich}, Elena and {Luisa Garc{\'\i}a-Vargas}, Mar{\'\i}a},
        title = "{Chemical abundances and ionizing clusters of Hii regions in the LINER galaxy NGC 4258}",
      journal = {\mnras},
         year = 2000,
        month = oct,
       volume = {318},
       number = {2},
        pages = {462-474},
          doi = {10.1046/j.1365-8711.2000.03737.x},
archivePrefix = {arXiv},
       eprint = {astro-ph/0006193},
 primaryClass = {astro-ph},
       adsurl = {https://ui.adsabs.harvard.edu/abs/2000MNRAS.318..462D}
}

@ARTICLE{Welch:2025,
       author = {{Welch}, Brian and {Rivera-Thorsen}, T. Emil and {Rigby}, Jane R. and {Hutchison}, Taylor A. and {Olivier}, Grace M. and {Berg}, Danielle A. and {Sharon}, Keren and {Dahle}, H{\r{a}}kon and {Owens}, M. Riley and {Bayliss}, Matthew B. and {Khullar}, Gourav and {Chisholm}, John and {Hayes}, Matthew and {Kim}, Keunho J.},
        title = "{The Sunburst Arc with JWST. III. An Abundance of Direct Chemical Abundances}",
      journal = {\apj},
         year = 2025,
        month = feb,
       volume = {980},
       number = {1},
          eid = {33},
        pages = {33},
          doi = {10.3847/1538-4357/ada76c},
archivePrefix = {arXiv},
       eprint = {2405.06631},
 primaryClass = {astro-ph.GA},
       adsurl = {https://ui.adsabs.harvard.edu/abs/2025ApJ...980...33W}
}

@ARTICLE{Wuyts:2012,
       author = {{Wuyts}, Stijn and {F{\"o}rster Schreiber}, Natascha M. and {Genzel}, Reinhard and {Guo}, Yicheng and {Barro}, Guillermo and {Bell}, Eric F. and {Dekel}, Avishai and {Faber}, Sandra M. and {Ferguson}, Henry C. and {Giavalisco}, Mauro and {Grogin}, Norman A. and {Hathi}, Nimish P. and {Huang}, Kuang-Han and {Kocevski}, Dale D. and {Koekemoer}, Anton M. and {Koo}, David C. and {Lotz}, Jennifer and {Lutz}, Dieter and {McGrath}, Elizabeth and {Newman}, Jeffrey A. and {Rosario}, David and {Saintonge}, Amelie and {Tacconi}, Linda J. and {Weiner}, Benjamin J. and {van der Wel}, Arjen},
        title = "{Smooth(er) Stellar Mass Maps in CANDELS: Constraints on the Longevity of Clumps in High-redshift Star-forming Galaxies}",
      journal = {\apj},
         year = 2012,
        month = jul,
       volume = {753},
       number = {2},
          eid = {114},
        pages = {114},
          doi = {10.1088/0004-637X/753/2/114},
archivePrefix = {arXiv},
       eprint = {1203.2611},
 primaryClass = {astro-ph.CO},
       adsurl = {https://ui.adsabs.harvard.edu/abs/2012ApJ...753..114W}
}

@ARTICLE{Wuyts:2013,
       author = {{Wuyts}, Stijn and {F{\"o}rster Schreiber}, Natascha M. and {Nelson}, Erica J. and {van Dokkum}, Pieter G. and {Brammer}, Gabe and {Chang}, Yu-Yen and {Faber}, Sandra M. and {Ferguson}, Henry C. and {Franx}, Marijn and {Fumagalli}, Mattia and {Genzel}, Reinhard and {Grogin}, Norman A. and {Kocevski}, Dale D. and {Koekemoer}, Anton M. and {Lundgren}, Britt and {Lutz}, Dieter and {McGrath}, Elizabeth J. and {Momcheva}, Ivelina and {Rosario}, David and {Skelton}, Rosalind E. and {Tacconi}, Linda J. and {van der Wel}, Arjen and {Whitaker}, Katherine E.},
        title = "{A CANDELS-3D-HST synergy: Resolved Star Formation Patterns at 0.7 < z < 1.5}",
      journal = {\apj},
         year = 2013,
        month = dec,
       volume = {779},
       number = {2},
          eid = {135},
        pages = {135},
          doi = {10.1088/0004-637X/779/2/135},
archivePrefix = {arXiv},
       eprint = {1310.5702},
 primaryClass = {astro-ph.CO},
       adsurl = {https://ui.adsabs.harvard.edu/abs/2013ApJ...779..135W}
}

@ARTICLE{Flury:2025,
       author = {{Flury}, Sophia R. and {Arellano-C{\'o}rdova}, Karla Z. and {Moran}, Edward C. and {Einsig}, Alaina},
        title = "{New ionization models and the shocking nitrogen excess at z > 5}",
      journal = {\mnras},
         year = 2025,
        month = nov,
       volume = {543},
       number = {4},
        pages = {3367-3381},
          doi = {10.1093/mnras/staf1615},
archivePrefix = {arXiv},
       eprint = {2412.06763},
 primaryClass = {astro-ph.GA},
       adsurl = {https://ui.adsabs.harvard.edu/abs/2025MNRAS.543.3367F}
}

@ARTICLE{Masters:2016,
       author = {{Masters}, Daniel and {Faisst}, Andreas and {Capak}, Peter},
        title = "{A Tight Relation between N/O Ratio and Galaxy Stellar Mass Can Explain the Evolution of Strong Emission Line Ratios with Redshift}",
      journal = {\apj},
         year = 2016,
        month = sep,
       volume = {828},
       number = {1},
          eid = {18},
        pages = {18},
          doi = {10.3847/0004-637X/828/1/18},
archivePrefix = {arXiv},
       eprint = {1605.04314},
 primaryClass = {astro-ph.GA},
       adsurl = {https://ui.adsabs.harvard.edu/abs/2016ApJ...828...18M}
}

@ARTICLE{Steidel2016,
       author = {{Steidel}, Charles C. and {Strom}, Allison L. and {Pettini}, Max and {Rudie}, Gwen C. and {Reddy}, Naveen A. and {Trainor}, Ryan F.},
        title = "{Reconciling the Stellar and Nebular Spectra of High-redshift Galaxies}",
      journal = {\apj},
         year = 2016,
        month = aug,
       volume = {826},
       number = {2},
          eid = {159},
        pages = {159},
          doi = {10.3847/0004-637X/826/2/159},
archivePrefix = {arXiv},
       eprint = {1605.07186},
 primaryClass = {astro-ph.GA},
       adsurl = {https://ui.adsabs.harvard.edu/abs/2016ApJ...826..159S}
}

@ARTICLE{Strom2018,
       author = {{Strom}, Allison L. and {Steidel}, Charles C. and {Rudie}, Gwen C. and {Trainor}, Ryan F. and {Pettini}, Max},
        title = "{Measuring the Physical Conditions in High-redshift Star-forming Galaxies: Insights from KBSS-MOSFIRE}",
      journal = {\apj},
         year = 2018,
        month = dec,
       volume = {868},
       number = {2},
          eid = {117},
        pages = {117},
          doi = {10.3847/1538-4357/aae1a5},
archivePrefix = {arXiv},
       eprint = {1711.08820},
 primaryClass = {astro-ph.GA},
       adsurl = {https://ui.adsabs.harvard.edu/abs/2018ApJ...868..117S}
}

@ARTICLE{Vincenzo:2016,
       author = {{Vincenzo}, F. and {Belfiore}, F. and {Maiolino}, R. and {Matteucci}, F. and {Ventura}, P.},
        title = "{Nitrogen and oxygen abundances in the Local Universe}",
      journal = {\mnras},
         year = 2016,
        month = jun,
       volume = {458},
       number = {4},
        pages = {3466-3477},
          doi = {10.1093/mnras/stw532},
archivePrefix = {arXiv},
       eprint = {1603.00460},
 primaryClass = {astro-ph.GA},
       adsurl = {https://ui.adsabs.harvard.edu/abs/2016MNRAS.458.3466V}
}

@ARTICLE{Isobe:2023,
       author = {{Isobe}, Yuki and {Ouchi}, Masami and {Tominaga}, Nozomu and {Watanabe}, Kuria and {Nakajima}, Kimihiko and {Umeda}, Hiroya and {Yajima}, Hidenobu and {Harikane}, Yuichi and {Fukushima}, Hajime and {Xu}, Yi and {Ono}, Yoshiaki and {Zhang}, Yechi},
        title = "{JWST Identification of Extremely Low C/N Galaxies with [N/O] {\ensuremath{\gtrsim}} 0.5 at z 6-10 Evidencing the Early CNO-cycle Enrichment and a Connection with Globular Cluster Formation}",
      journal = {\apj},
         year = 2023,
        month = dec,
       volume = {959},
       number = {2},
          eid = {100},
        pages = {100},
          doi = {10.3847/1538-4357/ad09be},
archivePrefix = {arXiv},
       eprint = {2307.00710},
 primaryClass = {astro-ph.GA},
       adsurl = {https://ui.adsabs.harvard.edu/abs/2023ApJ...959..100I}
}

@ARTICLE{Topping:2024,
       author = {{Topping}, Michael W. and {Stark}, Daniel P. and {Senchyna}, Peter and {Plat}, Adele and {Zitrin}, Adi and {Endsley}, Ryan and {Charlot}, St{\'e}phane and {Furtak}, Lukas J. and {Maseda}, Michael V. and {Smit}, Renske and {Mainali}, Ramesh and {Chevallard}, Jacopo and {Molyneux}, Stephen and {Rigby}, Jane R.},
        title = "{Metal-poor star formation at z > 6 with JWST: new insight into hard radiation fields and nitrogen enrichment on 20 pc scales}",
      journal = {\mnras},
         year = 2024,
        month = apr,
       volume = {529},
       number = {4},
        pages = {3301-3322},
          doi = {10.1093/mnras/stae682},
archivePrefix = {arXiv},
       eprint = {2401.08764},
 primaryClass = {astro-ph.GA},
       adsurl = {https://ui.adsabs.harvard.edu/abs/2024MNRAS.529.3301T}
}

@ARTICLE{Kashino:2023,
       author = {{Kashino}, Daichi and {Lilly}, Simon J. and {Matthee}, Jorryt and {Eilers}, Anna-Christina and {Mackenzie}, Ruari and {Bordoloi}, Rongmon and {Simcoe}, Robert A.},
        title = "{EIGER. I. A Large Sample of [O III]-emitting Galaxies at 5.3 < z < 6.9 and Direct Evidence for Local Reionization by Galaxies}",
      journal = {\apj},
         year = 2023,
        month = jun,
       volume = {950},
       number = {1},
          eid = {66},
        pages = {66},
          doi = {10.3847/1538-4357/acc588},
archivePrefix = {arXiv},
       eprint = {2211.08254},
 primaryClass = {astro-ph.GA},
       adsurl = {https://ui.adsabs.harvard.edu/abs/2023ApJ...950...66K}
}

@ARTICLE{Kobayashi:2020,
       author = {{Kobayashi}, Chiaki and {Karakas}, Amanda I. and {Lugaro}, Maria},
        title = "{The Origin of Elements from Carbon to Uranium}",
      journal = {\apj},
         year = 2020,
        month = sep,
       volume = {900},
       number = {2},
          eid = {179},
        pages = {179},
          doi = {10.3847/1538-4357/abae65},
archivePrefix = {arXiv},
       eprint = {2008.04660},
 primaryClass = {astro-ph.GA},
       adsurl = {https://ui.adsabs.harvard.edu/abs/2020ApJ...900..179K}
}

@ARTICLE{Kobayashi:2023,
       author = {{Kobayashi}, Chiaki and {Taylor}, Philip},
        title = "{Chemo-Dynamical Evolution of Galaxies}",
      journal = {arXiv e-prints},
         year = 2023,
        month = feb,
          eid = {arXiv:2302.07255},
        pages = {arXiv:2302.07255},
          doi = {10.48550/arXiv.2302.07255},
archivePrefix = {arXiv},
       eprint = {2302.07255},
 primaryClass = {astro-ph.GA},
       adsurl = {https://ui.adsabs.harvard.edu/abs/2023arXiv230207255K}
}

@ARTICLE{Cameron:2023,
       author = {{Cameron}, Alex J. and {Katz}, Harley and {Rey}, Martin P. and {Saxena}, Aayush},
        title = "{Nitrogen enhancements 440 Myr after the big bang: supersolar N/O, a tidal disruption event, or a dense stellar cluster in GN-z11?}",
      journal = {\mnras},
         year = 2023,
        month = aug,
       volume = {523},
       number = {3},
        pages = {3516-3525},
          doi = {10.1093/mnras/stad1579},
archivePrefix = {arXiv},
       eprint = {2302.10142},
 primaryClass = {astro-ph.GA},
       adsurl = {https://ui.adsabs.harvard.edu/abs/2023MNRAS.523.3516C}
}

@ARTICLE{Hayden-Pawson:2022,
       author = {{Hayden-Pawson}, Connor and {Curti}, Mirko and {Maiolino}, Roberto and {Cirasuolo}, Michele and {Belfiore}, Francesco and {Cappellari}, Michele and {Concas}, Alice and {Cresci}, Giovanni and {Cullen}, Fergus and {Kobayashi}, Chiaki and {Mannucci}, Filippo and {Marconi}, Alessandro and {Meneghetti}, Massimo and {Mercurio}, Amata and {Peng}, Yingjie and {Swinbank}, Mark and {Vincenzo}, Fiorenzo},
        title = "{The KLEVER survey: nitrogen abundances at z   2 and probing the existence of a fundamental nitrogen relation}",
      journal = {\mnras},
         year = 2022,
        month = may,
       volume = {512},
       number = {2},
        pages = {2867-2889},
          doi = {10.1093/mnras/stac584},
archivePrefix = {arXiv},
       eprint = {2110.00033},
 primaryClass = {astro-ph.GA},
       adsurl = {https://ui.adsabs.harvard.edu/abs/2022MNRAS.512.2867H}
}

@ARTICLE{Shen:2024,
       author = {{Shen}, Lu and {Papovich}, Casey and {Matharu}, Jasleen and {Pirzkal}, Nor and {Hu}, Weida and {Backhaus}, Bren E. and {Bagley}, Micaela B. and {Cheng}, Yingjie and {Cleri}, Nikko J. and {Finkelstein}, Steven L. and {Huertas-Company}, Marc and {Giavalisco}, Mauro and {Grogin}, Norman A. and {Jung}, Intae and {Kartaltepe}, Jeyhan S. and {Koekemoer}, Anton M. and {Lotz}, Jennifer M. and {Maseda}, Michael V. and {P{\'e}rez-Gonz{\'a}lez}, Pablo G. and {Rothberg}, Barry and {Simons}, Raymond C. and {Tacchella}, Sandro and {Williams}, Christina C. and {Yung}, L.~Y. Aaron},
        title = "{NGDEEP Epoch 1: Spatially Resolved H{\ensuremath{\alpha}} Observations of Disk and Bulge Growth in Star-forming Galaxies at z {\ensuremath{\sim}} 0.6{\textendash}2.2 from JWST NIRISS Slitless Spectroscopy}",
      journal = {\apjl},
         year = 2024,
        month = mar,
       volume = {963},
       number = {2},
          eid = {L49},
        pages = {L49},
          doi = {10.3847/2041-8213/ad28bd},
archivePrefix = {arXiv},
       eprint = {2310.13745},
 primaryClass = {astro-ph.GA},
       adsurl = {https://ui.adsabs.harvard.edu/abs/2024ApJ...963L..49S}
}

@TECHREPORT{2025jwst.rept.9225S,
       author = {{Sunnquist}, Ben and {Boyer}, Martha and {Brooks}, Brian and {Canipe}, Alicia and {Hilbert}, Bryan and {Rest}, Armin},
        title = "{Version 4 of the NIRCam Wisp Templates: Wisp Characterization, Stability, and Validation Testing}",
  institution = {STScI},
         year = 2025,
       number = {Technical Report JWST-STScI-009225},
 howpublished = {Technical Report JWST-STScI-009225, 12 pages},
       adsurl = {https://ui.adsabs.harvard.edu/abs/2025jwst.rept.9225S}
}

@ARTICLE{KormendyKennicutt2004,
       author = {{Kormendy}, John and {Kennicutt}, Jr., Robert C.},
        title = "{Secular Evolution and the Formation of Pseudobulges in Disk Galaxies}",
      journal = {\araa},
         year = 2004,
        month = sep,
       volume = {42},
       number = {1},
        pages = {603-683},
          doi = {10.1146/annurev.astro.42.053102.134024},
archivePrefix = {arXiv},
       eprint = {astro-ph/0407343},
 primaryClass = {astro-ph},
       adsurl = {https://ui.adsabs.harvard.edu/abs/2004ARA&A..42..603K}
}

@ARTICLE{Valentino:2023,
       author = {{Valentino}, Francesco and {Brammer}, Gabriel and {Gould}, Katriona M.~L. and {Kokorev}, Vasily and {Fujimoto}, Seiji and {Jespersen}, Christian Kragh and {Vijayan}, Aswin P. and {Weaver}, John R. and {Ito}, Kei and {Tanaka}, Masayuki and {Ilbert}, Olivier and {Magdis}, Georgios E. and {Whitaker}, Katherine E. and {Faisst}, Andreas L. and {Gallazzi}, Anna and {Gillman}, Steven and {Gim{\'e}nez-Arteaga}, Clara and {G{\'o}mez-Guijarro}, Carlos and {Kubo}, Mariko and {Heintz}, Kasper E. and {Hirschmann}, Michaela and {Oesch}, Pascal and {Onodera}, Masato and {Rizzo}, Francesca and {Lee}, Minju and {Strait}, Victoria and {Toft}, Sune},
        title = "{An Atlas of Color-selected Quiescent Galaxies at z > 3 in Public JWST Fields}",
      journal = {\apj},
         year = 2023,
        month = apr,
       volume = {947},
       number = {1},
          eid = {20},
        pages = {20},
          doi = {10.3847/1538-4357/acbefa},
archivePrefix = {arXiv},
       eprint = {2302.10936},
 primaryClass = {astro-ph.GA},
       adsurl = {https://ui.adsabs.harvard.edu/abs/2023ApJ...947...20V}
}

@ARTICLE{Li:2024,
       author = {{Li}, Juno and {Da Cunha}, Elisabete and {Gonz{\'a}lez-L{\'o}pez}, Jorge and {Aravena}, Manuel and {De Looze}, Ilse and {F{\"o}rster Schreiber}, N.~M. and {Herrera-Camus}, Rodrigo and {Spilker}, Justin and {Tadaki}, Ken-ichi and {Barcos-Munoz}, Loreto and {Battisti}, Andrew J. and {Birkin}, Jack E. and {Bowler}, Rebecca A.~A. and {Davies}, Rebecca and {D{\'\i}az-Santos}, Tanio and {Ferrara}, Andrea and {Fisher}, Deanne B. and {Hodge}, Jacqueline and {Ikeda}, Ryota and {Killi}, Meghana and {Lee}, Lilian and {Liu}, Daizhong and {Lutz}, Dieter and {Mitsuhashi}, Ikki and {Naab}, Thorsten and {Posses}, Ana and {Rela{\~n}o}, Monica and {Solimano}, Manuel and {{\"U}bler}, Hannah and {van der Giessen}, Stefan Anthony and {Villanueva}, Vicente},
        title = "{The ALMA-CRISTAL Survey: Spatially Resolved Star Formation Activity and Dust Content in 4 < z < 6 Star-forming Galaxies}",
      journal = {\apj},
         year = 2024,
        month = nov,
       volume = {976},
       number = {1},
          eid = {70},
        pages = {70},
          doi = {10.3847/1538-4357/ad7fee},
archivePrefix = {arXiv},
       eprint = {2409.10961},
 primaryClass = {astro-ph.GA},
       adsurl = {https://ui.adsabs.harvard.edu/abs/2024ApJ...976...70L}
}

@ARTICLE{Rogers:2026,
       author = {{Rogers}, Noah S.~J. and {Strom}, Allison L. and {Rudie}, Gwen C. and {Trainor}, Ryan F. and {von Raesfeld}, Caroline and {Raptis}, Menelaos and {Korhonen Cuestas}, Nathalie A. and {Miller}, Tim B. and {Steidel}, Charles C. and {Maseda}, Michael V. and {Chen}, Yuguang and {Law}, David R.},
        title = "{CECILIA: Gas-phase Physical Conditions and Multielement Chemistry at Cosmic Noon}",
      journal = {\apjl},
         year = 2026,
        month = feb,
       volume = {997},
       number = {2},
          eid = {L44},
        pages = {L44},
          doi = {10.3847/2041-8213/ae31f3},
archivePrefix = {arXiv},
       eprint = {2509.18257},
 primaryClass = {astro-ph.GA},
       adsurl = {https://ui.adsabs.harvard.edu/abs/2026ApJ...997L..44R}
}

@ARTICLE{Curti:2020klever,
       author = {{Curti}, Mirko and {Maiolino}, Roberto and {Cirasuolo}, Michele and {Mannucci}, Filippo and {Williams}, Rebecca J. and {Auger}, Matt and {Mercurio}, Amata and {Hayden-Pawson}, Connor and {Cresci}, Giovanni and {Marconi}, Alessandro and {Belfiore}, Francesco and {Cappellari}, Michele and {Cicone}, Claudia and {Cullen}, Fergus and {Meneghetti}, Massimo and {Ota}, Kazuaki and {Peng}, Yingjie and {Pettini}, Max and {Swinbank}, Mark and {Troncoso}, Paulina},
        title = "{The KLEVER Survey: spatially resolved metallicity maps and gradients in a sample of 1.2 < z < 2.5 lensed galaxies}",
      journal = {\mnras},
         year = 2020,
        month = feb,
       volume = {492},
       number = {1},
        pages = {821-842},
          doi = {10.1093/mnras/stz3379},
archivePrefix = {arXiv},
       eprint = {1910.13451},
 primaryClass = {astro-ph.GA},
       adsurl = {https://ui.adsabs.harvard.edu/abs/2020MNRAS.492..821C}
}

@ARTICLE{Lines:2025,
       author = {{Lines}, N.~E.~P. and {Bowler}, R.~A.~A. and {Adams}, N.~J. and {Fisher}, R. and {Varadaraj}, R.~G. and {Nakazato}, Y. and {Aravena}, M. and {Assef}, R.~J. and {Birkin}, J.~E. and {Ceverino}, D. and {da Cunha}, E. and {Cullen}, F. and {De Looze}, I. and {Donnan}, C.~T. and {Dunlop}, J.~S. and {Ferrara}, A. and {Grogin}, N.~A. and {Herrera-Camus}, R. and {Ikeda}, R. and {Koekemoer}, A.~M. and {Killi}, M. and {Li}, J. and {McLeod}, D.~J. and {McLure}, R.~J. and {Mitsuhashi}, I. and {P{\'e}rez-Gonz{\'a}lez}, P.~G. and {Relano}, M. and {Solimano}, M. and {Spilker}, J.~S. and {Villanueva}, V. and {Yoshida}, N.},
        title = "{JWST PRIMER: a lack of outshining in four normal z = 4 ‑ 6 galaxies from the ALMA-CRISTAL Survey}",
      journal = {\mnras},
         year = 2025,
        month = may,
       volume = {539},
       number = {3},
        pages = {2685-2706},
          doi = {10.1093/mnras/staf627},
archivePrefix = {arXiv},
       eprint = {2409.10963},
 primaryClass = {astro-ph.GA},
       adsurl = {https://ui.adsabs.harvard.edu/abs/2025MNRAS.539.2685L}
}

@ARTICLE{Sorba:2018,
       author = {{Sorba}, Robert and {Sawicki}, Marcin},
        title = "{Spatially unresolved SED fitting can underestimate galaxy masses: a solution to the missing mass problem}",
      journal = {\mnras},
         year = 2018,
        month = may,
       volume = {476},
       number = {2},
        pages = {1532-1547},
          doi = {10.1093/mnras/sty186},
archivePrefix = {arXiv},
       eprint = {1801.07368},
 primaryClass = {astro-ph.GA},
       adsurl = {https://ui.adsabs.harvard.edu/abs/2018MNRAS.476.1532S}
}

@ARTICLE{GimenezArteaga:2023,
       author = {{Gim{\'e}nez-Arteaga}, Clara and {Oesch}, Pascal A. and {Brammer}, Gabriel B. and {Valentino}, Francesco and {Mason}, Charlotte A. and {Weibel}, Andrea and {Barrufet}, Laia and {Fujimoto}, Seiji and {Heintz}, Kasper E. and {Nelson}, Erica J. and {Strait}, Victoria B. and {Suess}, Katherine A. and {Gibson}, Justus},
        title = "{Spatially Resolved Properties of Galaxies at 5 < z < 9 in the SMACS 0723 JWST ERO Field}",
      journal = {\apj},
         year = 2023,
        month = may,
       volume = {948},
       number = {2},
          eid = {126},
        pages = {126},
          doi = {10.3847/1538-4357/acc5ea},
archivePrefix = {arXiv},
       eprint = {2212.08670},
 primaryClass = {astro-ph.GA},
       adsurl = {https://ui.adsabs.harvard.edu/abs/2023ApJ...948..126G}
}

@ARTICLE{GimenezArteaga:2024,
       author = {{Gim{\'e}nez-Arteaga}, C. and {Fujimoto}, S. and {Valentino}, F. and {Brammer}, G.~B. and {Mason}, C.~A. and {Rizzo}, F. and {Rusakov}, V. and {Colina}, L. and {Prieto-Lyon}, G. and {Oesch}, P.~A. and {Espada}, D. and {Heintz}, K.~E. and {Knudsen}, K.~K. and {Dessauges-Zavadsky}, M. and {Laporte}, N. and {Lee}, M. and {Magdis}, G.~E. and {Ono}, Y. and {Ao}, Y. and {Ouchi}, M. and {Kohno}, K. and {Koekemoer}, A.~M.},
        title = "{Outshining in the spatially resolved analysis of a strongly lensed galaxy at z = 6.072 with JWST NIRCam}",
      journal = {\aap},
         year = 2024,
        month = jun,
       volume = {686},
          eid = {A63},
        pages = {A63},
          doi = {10.1051/0004-6361/202349135},
archivePrefix = {arXiv},
       eprint = {2402.17875},
 primaryClass = {astro-ph.GA},
       adsurl = {https://ui.adsabs.harvard.edu/abs/2024A&A...686A..63G}
}

@ARTICLE{Bunker:2023,
       author = {{Bunker}, Andrew J. and {Saxena}, Aayush and {Cameron}, Alex J. and {Willott}, Chris J. and {Curtis-Lake}, Emma and {Jakobsen}, Peter and {Carniani}, Stefano and {Smit}, Renske and {Maiolino}, Roberto and {Witstok}, Joris and {Curti}, Mirko and {D'Eugenio}, Francesco and {Jones}, Gareth C. and {Ferruit}, Pierre and {Arribas}, Santiago and {Charlot}, Stephane and {Chevallard}, Jacopo and {Giardino}, Giovanna and {de Graaff}, Anna and {Looser}, Tobias J. and {L{\"u}tzgendorf}, Nora and {Maseda}, Michael V. and {Rawle}, Tim and {Rix}, Hans-Walter and {Del Pino}, Bruno Rodr{\'\i}guez and {Alberts}, Stacey and {Egami}, Eiichi and {Eisenstein}, Daniel J. and {Endsley}, Ryan and {Hainline}, Kevin and {Hausen}, Ryan and {Johnson}, Benjamin D. and {Rieke}, George and {Rieke}, Marcia and {Robertson}, Brant E. and {Shivaei}, Irene and {Stark}, Daniel P. and {Sun}, Fengwu and {Tacchella}, Sandro and {Tang}, Mengtao and {Williams}, Christina C. and {Willmer}, Christopher N.~A. and {Baker}, William M. and {Baum}, Stefi and {Bhatawdekar}, Rachana and {Bowler}, Rebecca and {Boyett}, Kristan and {Chen}, Zuyi and {Circosta}, Chiara and {Helton}, Jakob M. and {Ji}, Zhiyuan and {Kumari}, Nimisha and {Lyu}, Jianwei and {Nelson}, Erica and {Parlanti}, Eleonora and {Perna}, Michele and {Sandles}, Lester and {Scholtz}, Jan and {Suess}, Katherine A. and {Topping}, Michael W. and {{\"U}bler}, Hannah and {Wallace}, Imaan E.~B. and {Whitler}, Lily},
        title = "{JADES NIRSpec Spectroscopy of GN-z11: Lyman-{\ensuremath{\alpha}} emission and possible enhanced nitrogen abundance in a z = 10.60 luminous galaxy}",
      journal = {\aap},
         year = 2023,
        month = sep,
       volume = {677},
          eid = {A88},
        pages = {A88},
          doi = {10.1051/0004-6361/202346159},
archivePrefix = {arXiv},
       eprint = {2302.07256},
 primaryClass = {astro-ph.GA},
       adsurl = {https://ui.adsabs.harvard.edu/abs/2023A&A...677A..88B}
}

@ARTICLE{Stiavelli:2025,
       author = {{Stiavelli}, Massimo and {Morishita}, Takahiro and {Chiaberge}, Marco and {Leethochawalit}, Nicha and {Norman}, Colin and {Ricotti}, Massimo and {Roberts-Borsani}, Guido and {Treu}, Tommaso and {Vanzella}, Eros and {Wyse}, Rosemary F.~G. and {Zhang}, Yechi and {Boyett}, Kit},
        title = "{What Can We Learn from the Nitrogen Abundance of High-z Galaxies?}",
      journal = {\apj},
         year = 2025,
        month = mar,
       volume = {981},
       number = {2},
          eid = {136},
        pages = {136},
          doi = {10.3847/1538-4357/adb5f3},
archivePrefix = {arXiv},
       eprint = {2412.06517},
 primaryClass = {astro-ph.GA},
       adsurl = {https://ui.adsabs.harvard.edu/abs/2025ApJ...981..136S}
}

@ARTICLE{Sanders:2023,
       author = {{Sanders}, Ryan L. and {Shapley}, Alice E. and {Clarke}, Leonardo and {Topping}, Michael W. and {Reddy}, Naveen A. and {Kriek}, Mariska and {Jones}, Tucker and {Stark}, Daniel P. and {Tang}, Mengtao},
        title = "{A Preview of JWST Metallicity Studies at Cosmic Noon: The First Detection of Auroral [O II] Emission at High Redshift}",
      journal = {\apj},
         year = 2023,
        month = feb,
       volume = {943},
       number = {2},
          eid = {75},
        pages = {75},
          doi = {10.3847/1538-4357/aca9cc},
archivePrefix = {arXiv},
       eprint = {2207.12430},
 primaryClass = {astro-ph.GA},
       adsurl = {https://ui.adsabs.harvard.edu/abs/2023ApJ...943...75S}
}

@ARTICLE{galfit,
       author = {{Peng}, Chien Y. and {Ho}, Luis C. and {Impey}, Chris D. and {Rix}, Hans-Walter},
        title = "{Detailed Structural Decomposition of Galaxy Images}",
      journal = {\aj},
         year = 2002,
        month = jul,
       volume = {124},
       number = {1},
        pages = {266-293},
          doi = {10.1086/340952},
archivePrefix = {arXiv},
       eprint = {astro-ph/0204182},
 primaryClass = {astro-ph},
       adsurl = {https://ui.adsabs.harvard.edu/abs/2002AJ....124..266P}
}

@ARTICLE{SersicProfile,
       author = {{Caon}, N. and {Capaccioli}, M. and {D'Onofrio}, M.},
        title = "{On the shape of the light profiles of early-type galaxies.}",
      journal = {\mnras},
         year = 1993,
        month = dec,
       volume = {265},
        pages = {1013-1021},
          doi = {10.1093/mnras/265.4.1013},
archivePrefix = {arXiv},
       eprint = {astro-ph/9309013},
 primaryClass = {astro-ph},
       adsurl = {https://ui.adsabs.harvard.edu/abs/1993MNRAS.265.1013C}
}

@ARTICLE{Tacchella:2015sins,
       author = {{Tacchella}, S. and {Lang}, P. and {Carollo}, C.~M. and {F{\"o}rster Schreiber}, N.~M. and {Renzini}, A. and {Shapley}, A.~E. and {Wuyts}, S. and {Cresci}, G. and {Genzel}, R. and {Lilly}, S.~J. and {Mancini}, C. and {Newman}, S.~F. and {Tacconi}, L.~J. and {Zamorani}, G. and {Davies}, R.~I. and {Kurk}, J. and {Pozzetti}, L.},
        title = "{SINS/zC-SINF Survey of z {\ensuremath{\sim}} 2 Galaxy Kinematics: Rest-frame Morphology, Structure, and Colors from Near-infrared Hubble Space Telescope Imaging}",
      journal = {\apj},
         year = 2015,
        month = apr,
       volume = {802},
       number = {2},
          eid = {101},
        pages = {101},
          doi = {10.1088/0004-637X/802/2/101},
archivePrefix = {arXiv},
       eprint = {1411.7034},
 primaryClass = {astro-ph.GA},
       adsurl = {https://ui.adsabs.harvard.edu/abs/2015ApJ...802..101T}
}

@ARTICLE{Tacchella:2015b,
       author = {{Tacchella}, S. and {Carollo}, C.~M. and {Renzini}, A. and {F{\"o}rster Schreiber}, N.~M. and {Lang}, P. and {Wuyts}, S. and {Cresci}, G. and {Dekel}, A. and {Genzel}, R. and {Lilly}, S.~J. and {Mancini}, C. and {Newman}, S. and {Onodera}, M. and {Shapley}, A. and {Tacconi}, L. and {Woo}, J. and {Zamorani}, G.},
        title = "{Evidence for mature bulges and an inside-out quenching phase 3 billion years after the Big Bang}",
      journal = {Science},
         year = 2015,
        month = apr,
       volume = {348},
       number = {6232},
        pages = {314-317},
          doi = {10.1126/science.1261094},
archivePrefix = {arXiv},
       eprint = {1504.04021},
 primaryClass = {astro-ph.GA},
       adsurl = {https://ui.adsabs.harvard.edu/abs/2015Sci...348..314T}
}

@ARTICLE{Rogers:2024,
       author = {{Rogers}, Noah S.~J. and {Strom}, Allison L. and {Rudie}, Gwen C. and {Trainor}, Ryan F. and {Raptis}, Menelaos and {von Raesfeld}, Caroline},
        title = "{CECILIA: Direct O, N, S, and Ar Abundances in Q2343-D40, a Galaxy at z {\ensuremath{\sim}} 3}",
      journal = {\apjl},
         year = 2024,
        month = mar,
       volume = {964},
       number = {1},
          eid = {L12},
        pages = {L12},
          doi = {10.3847/2041-8213/ad2f37},
archivePrefix = {arXiv},
       eprint = {2312.08427},
 primaryClass = {astro-ph.GA},
       adsurl = {https://ui.adsabs.harvard.edu/abs/2024ApJ...964L..12R}
}

@ARTICLE{Sanders:2020,
       author = {{Sanders}, Ryan L. and {Jones}, Tucker and {Shapley}, Alice E. and {Reddy}, Naveen A. and {Kriek}, Mariska and {Coil}, Alison L. and {Siana}, Brian and {Mobasher}, Bahram and {Shivaei}, Irene and {Price}, Sedona H. and {Freeman}, William R. and {Azadi}, Mojegan and {Leung}, Gene C.~K. and {Fetherolf}, Tara and {Zick}, Tom O. and {de Groot}, Laura and {Barro}, Guillermo and {Fornasini}, Francesca M.},
        title = "{The MOSDEF Survey: [S III] as a New Probe of Evolving Interstellar Medium Conditions}",
      journal = {\apjl},
         year = 2020,
        month = jan,
       volume = {888},
       number = {1},
          eid = {L11},
        pages = {L11},
          doi = {10.3847/2041-8213/ab5d40},
       adsurl = {https://ui.adsabs.harvard.edu/abs/2020ApJ...888L..11S}
}

@ARTICLE{Asplund:2009,
       author = {{Asplund}, Martin and {Grevesse}, Nicolas and {Sauval}, A. Jacques and {Scott}, Pat},
        title = "{The Chemical Composition of the Sun}",
      journal = {\araa},
         year = 2009,
        month = sep,
       volume = {47},
       number = {1},
        pages = {481-522},
          doi = {10.1146/annurev.astro.46.060407.145222},
archivePrefix = {arXiv},
       eprint = {0909.0948},
 primaryClass = {astro-ph.SR},
       adsurl = {https://ui.adsabs.harvard.edu/abs/2009ARA&A..47..481A}
}

@ARTICLE{Diaz2007,
       author = {{D{\'\i}az}, {\'A}ngeles I. and {Terlevich}, Elena and {Castellanos}, Marcelo and {H{\"a}gele}, Guillermo F.},
        title = "{The metal abundance of circumnuclear star-forming regions in early-type spirals. Spectrophotometric observations}",
      journal = {\mnras},
         year = 2007,
        month = nov,
       volume = {382},
       number = {1},
        pages = {251-269},
          doi = {10.1111/j.1365-2966.2007.12351.x},
archivePrefix = {arXiv},
       eprint = {0709.1236},
 primaryClass = {astro-ph},
       adsurl = {https://ui.adsabs.harvard.edu/abs/2007MNRAS.382..251D}
}

@ARTICLE{Diaz:2022,
       author = {{D{\'\i}az}, {\'A}ngeles I. and {Zamora}, S.},
        title = "{On the use of sulphur as a tracer for abundances in galaxies}",
      journal = {\mnras},
         year = 2022,
        month = apr,
       volume = {511},
       number = {3},
        pages = {4377-4392},
          doi = {10.1093/mnras/stac387},
archivePrefix = {arXiv},
       eprint = {2202.10302},
 primaryClass = {astro-ph.GA},
       adsurl = {https://ui.adsabs.harvard.edu/abs/2022MNRAS.511.4377D}
}

@ARTICLE{Nelson:2024,
       author = {{Nelson}, Erica and {Brammer}, Gabriel and {Gim{\'e}nez-Arteaga}, Clara and {Oesch}, Pascal A. and {Naidu}, Rohan P. and {{\"U}bler}, Hannah and {Matharu}, Jasleen and {Shapley}, Alice E. and {Whitaker}, Katherine E. and {Wisnioski}, Emily and {F{\"o}rster Schreiber}, Natascha M. and {Smit}, Renske and {van Dokkum}, Pieter and {Chisholm}, John and {Endsley}, Ryan and {Hartley}, Abigail I. and {Gibson}, Justus and {Giovinazzo}, Emma and {Illingworth}, Garth and {Labbe}, Ivo and {Maseda}, Michael V. and {Matthee}, Jorryt and {Covelo Paz}, Alba and {Price}, Sedona H. and {Reddy}, Naveen A. and {Shivaei}, Irene and {Weibel}, Andrea and {Wuyts}, Stijn and {Xiao}, Mengyuan and {Alberts}, Stacey and {Baker}, William M. and {Bunker}, Andrew J. and {Cameron}, Alex J. and {Charlot}, Stephane and {Eisenstein}, Daniel J. and {de Graaff}, Anna and {Ji}, Zhiyuan and {Johnson}, Benjamin D. and {Jones}, Gareth C. and {Maiolino}, Roberto and {Robertson}, Brant and {Sandles}, Lester and {Suess}, Katherine A. and {Tacchella}, Sandro and {Williams}, Christina C. and {Witstok}, Joris},
        title = "{Ionized Gas Kinematics with FRESCO: An Extended, Massive, Rapidly Rotating Galaxy at z = 5.4}",
      journal = {\apjl},
         year = 2024,
        month = dec,
       volume = {976},
       number = {2},
          eid = {L27},
        pages = {L27},
          doi = {10.3847/2041-8213/ad7b17},
archivePrefix = {arXiv},
       eprint = {2310.06887},
 primaryClass = {astro-ph.GA},
       adsurl = {https://ui.adsabs.harvard.edu/abs/2024ApJ...976L..27N}
}

@ARTICLE{Neufeld:2024,
       author = {{Neufeld}, Chloe and {van Dokkum}, Pieter and {Asali}, Yasmeen and {Covelo-Paz}, Alba and {Leja}, Joel and {Lin}, Jamie and {Matthee}, Jorryt and {Oesch}, Pascal A. and {Reddy}, Naveen A. and {Shivaei}, Irene and {Whitaker}, Katherine E. and {Wuyts}, Stijn and {Brammer}, Gabriel and {Marchesini}, Danilo and {Maseda}, Michael V. and {Naidu}, Rohan P. and {Nelson}, Erica J. and {Velichko}, Anna and {Weibel}, Andrea and {Xiao}, Mengyuan},
        title = "{FRESCO: The Paschen-{\ensuremath{\alpha}} Star-forming Sequence at Cosmic Noon}",
      journal = {\apj},
         year = 2024,
        month = sep,
       volume = {972},
       number = {2},
          eid = {156},
        pages = {156},
          doi = {10.3847/1538-4357/ad6158},
archivePrefix = {arXiv},
       eprint = {2404.10816},
 primaryClass = {astro-ph.GA},
       adsurl = {https://ui.adsabs.harvard.edu/abs/2024ApJ...972..156N}
}

@software{grizli,
       author = {{Brammer}, Gabriel},
        title = "{grizli}",
         year = 2023,
        month = sep,
          eid = {10.5281/zenodo.8370018},
          doi = {10.5281/zenodo.8370018},
      version = {1.9.11},
    publisher = {Zenodo},
       adsurl = {https://ui.adsabs.harvard.edu/abs/2023zndo...8370018B}
}

@ARTICLE{Oesch:2023,
       author = {{Oesch}, P.~A. and {Brammer}, G. and {Naidu}, R.~P. and {Bouwens}, R.~J. and {Chisholm}, J. and {Illingworth}, G.~D. and {Matthee}, J. and {Nelson}, E. and {Qin}, Y. and {Reddy}, N. and {Shapley}, A. and {Shivaei}, I. and {van Dokkum}, P. and {Weibel}, A. and {Whitaker}, K. and {Wuyts}, S. and {Covelo-Paz}, A. and {Endsley}, R. and {Fudamoto}, Y. and {Giovinazzo}, E. and {Herard-Demanche}, T. and {Kerutt}, J. and {Kramarenko}, I. and {Labbe}, I. and {Leonova}, E. and {Lin}, J. and {Magee}, D. and {Marchesini}, D. and {Maseda}, M. and {Mason}, C. and {Matharu}, J. and {Meyer}, R.~A. and {Neufeld}, C. and {Prieto Lyon}, G. and {Schaerer}, D. and {Sharma}, R. and {Shuntov}, M. and {Smit}, R. and {Stefanon}, M. and {Wyithe}, J.~S.~B. and {Xiao}, M.},
        title = "{The JWST FRESCO survey: legacy NIRCam/grism spectroscopy and imaging in the two GOODS fields}",
      journal = {\mnras},
         year = 2023,
        month = oct,
       volume = {525},
       number = {2},
        pages = {2864-2874},
          doi = {10.1093/mnras/stad2411},
archivePrefix = {arXiv},
       eprint = {2304.02026},
 primaryClass = {astro-ph.GA},
       adsurl = {https://ui.adsabs.harvard.edu/abs/2023MNRAS.525.2864O}
}

@ARTICLE{Ferruit:2022,
       author = {{Ferruit}, P. and {Jakobsen}, P. and {Giardino}, G. and {Rawle}, T. and {Alves de Oliveira}, C. and {Arribas}, S. and {Beck}, T.~L. and {Birkmann}, S. and {B{\"o}ker}, T. and {Bunker}, A.~J. and {Charlot}, S. and {de Marchi}, G. and {Franx}, M. and {Henry}, A. and {Karakla}, D. and {Kassin}, S.~A. and {Kumari}, N. and {L{\'o}pez-Caniego}, M. and {L{\"u}tzgendorf}, N. and {Maiolino}, R. and {Manjavacas}, E. and {Marston}, A. and {Moseley}, S.~H. and {Muzerolle}, J. and {Pirzkal}, N. and {Rauscher}, B. and {Rix}, H. -W. and {Sabbi}, E. and {Sirianni}, M. and {te Plate}, M. and {Valenti}, J. and {Willott}, C.~J. and {Zeidler}, P.},
        title = "{The Near-Infrared Spectrograph (NIRSpec) on the James Webb Space Telescope. II. Multi-object spectroscopy (MOS)}",
      journal = {\aap},
         year = 2022,
        month = may,
       volume = {661},
          eid = {A81},
        pages = {A81},
          doi = {10.1051/0004-6361/202142673},
archivePrefix = {arXiv},
       eprint = {2202.03306},
 primaryClass = {astro-ph.IM},
       adsurl = {https://ui.adsabs.harvard.edu/abs/2022A&A...661A..81F}
}

@INPROCEEDINGS{Eisenhauer2003,
       author = {{Eisenhauer}, Frank and {Abuter}, Roberto and {Bickert}, Klaus and {Biancat-Marchet}, Fabio and {Bonnet}, Henri and {Brynnel}, Joar and {Conzelmann}, Ralf D. and {Delabre}, Bernard and {Donaldson}, Robert and {Farinato}, Jacopo and {Fedrigo}, Enrico and {Genzel}, Reinhard and {Hubin}, Norbert N. and {Iserlohe}, Christof and {Kasper}, Markus E. and {Kissler-Patig}, Markus and {Monnet}, Guy J. and {Roehrle}, Claudia and {Schreiber}, Juergen and {Stroebele}, Stefan and {Tecza}, Matthias and {Thatte}, Niranjan A. and {Weisz}, Harald},
        title = "{SINFONI - Integral field spectroscopy at 50 milli-arcsecond resolution with the ESO VLT}",
    booktitle = {Instrument Design and Performance for Optical/Infrared Ground-based Telescopes},
         year = 2003,
       editor = {{Iye}, Masanori and {Moorwood}, Alan F.~M.},
       series = {Society of Photo-Optical Instrumentation Engineers (SPIE) Conference Series},
       volume = {4841},
        month = mar,
        pages = {1548-1561},
          doi = {10.1117/12.459468},
archivePrefix = {arXiv},
       eprint = {astro-ph/0306191},
 primaryClass = {astro-ph},
       adsurl = {https://ui.adsabs.harvard.edu/abs/2003SPIE.4841.1548E}
}

@ARTICLE{Eisenstein:2026,
       author = {{Eisenstein}, Daniel J. and {Willott}, Chris and {Alberts}, Stacey and {Arribas}, Santiago and {Bonaventura}, Nina and {Bunker}, Andrew J. and {Cameron}, Alex J. and {Carniani}, Stefano and {Charlot}, Stephane and {Curtis-Lake}, Emma and {D'Eugenio}, Francesco and {Ferruit}, Pierre and {Giardino}, Giovanna and {Hainline}, Kevin and {Hausen}, Ryan and {Jakobsen}, Peter and {Johnson}, Benjamin D. and {Maiolino}, Roberto and {Rauscher}, Bernard J. and {Rieke}, Marcia and {Rieke}, George and {Rix}, Hans-Walter and {Robertson}, Brant and {Stark}, Daniel P. and {Tacchella}, Sandro and {Williams}, Christina C. and {Willmer}, Christopher N.~A. and {Baker}, William M. and {Baum}, Stefi and {Bhatawdekar}, Rachana and {Boyett}, Kristan and {Chen}, Zuyi and {Chevallard}, Jacopo and {Circosta}, Chiara and {Curti}, Mirko and {Danhaive}, A. Lola and {DeCoursey}, Christa and {Endsley}, Ryan and {de Graaff}, Anna and {Dressler}, Alan and {Egami}, Eiichi and {Helton}, Jakob M. and {Hviding}, Raphael E. and {Ji}, Zhiyuan and {Jones}, Gareth C. and {Kumari}, Nimisha and {L{\"u}tzgendorf}, Nora and {Laseter}, Isaac and {Looser}, Tobias J. and {Lyu}, Jianwei and {Maseda}, Michael V. and {Nelson}, Erica and {Parlanti}, Eleonora and {Perna}, Michele and {Pusk{\'a}s}, D{\'a}vid and {Rawle}, Tim and {Rodr{\'\i}guez Del Pino}, Bruno and {Rujopakarn}, Wiphu and {Sandles}, Lester and {Saxena}, Aayush and {Scholtz}, Jan and {Sharpe}, Katherine and {Shivaei}, Irene and {Silcock}, Maddie S. and {Simmonds}, Charlotte and {Skarbinski}, Maya and {Smit}, Renske and {Stone}, Meredith and {Suess}, Katherine A. and {Sun}, Fengwu and {Tang}, Mengtao and {Topping}, Michael W. and {{\"U}bler}, Hannah and {Villanueva}, Natalia C. and {Wallace}, Imaan E.~B. and {Whitler}, Lily and {Witstok}, Joris and {Woodrum}, Charity},
        title = "{Overview of the JWST Advanced Deep Extragalactic Survey (JADES)}",
      journal = {\apjs},
         year = 2026,
        month = mar,
       volume = {283},
       number = {1},
          eid = {6},
        pages = {6},
          doi = {10.3847/1538-4365/ae3163},
archivePrefix = {arXiv},
       eprint = {2306.02465},
 primaryClass = {astro-ph.GA},
       adsurl = {https://ui.adsabs.harvard.edu/abs/2026ApJS..283....6E}
}

@ARTICLE{Espejo:2025a,
       author = {{Espejo Salcedo}, Juan M. and {Glazebrook}, Karl and {Fisher}, Deanne B. and {Sweet}, Sarah M. and {Obreschkow}, Danail and {Schreiber}, N.~M. F{\"o}rster},
        title = "{A shallow slope for the stellar mass-angular momentum relation of star-forming galaxies at 1.5 < z < 2.5}",
      journal = {\mnras},
         year = 2025,
        month = jan,
       volume = {536},
       number = {2},
        pages = {1188-1216},
          doi = {10.1093/mnras/stae2647},
archivePrefix = {arXiv},
       eprint = {2411.17312},
 primaryClass = {astro-ph.GA},
       adsurl = {https://ui.adsabs.harvard.edu/abs/2025MNRAS.536.1188E}
}

@ARTICLE{Espejo:2025,
       author = {{Espejo Salcedo}, J.~M. and {Pastras}, S. and {V{\'a}cha}, J. and {Pulsoni}, C. and {Genzel}, R. and {F{\"o}rster Schreiber}, N.~M. and {Jolly}, J. -B. and {Barfety}, C. and {Chen}, J. and {Tozzi}, G. and {Liu}, D. and {Lee}, L.~L. and {Wuyts}, S. and {Tacconi}, L.~J. and {Davies}, R. and {{\"U}bler}, H. and {Lutz}, D. and {Wisnioski}, E. and {Shangguan}, J. and {Lee}, M. and {Price}, S.~H. and {Eisenhauer}, F. and {Renzini}, A. and {Nestor Shachar}, A. and {Herrera-Camus}, R.},
        title = "{Galaxy morphologies at cosmic noon with JWST: A foundation for exploring gas transport with bars and spiral arms}",
      journal = {\aap},
         year = 2025,
        month = aug,
       volume = {700},
          eid = {A42},
        pages = {A42},
          doi = {10.1051/0004-6361/202554725},
archivePrefix = {arXiv},
       eprint = {2503.21738},
 primaryClass = {astro-ph.GA},
       adsurl = {https://ui.adsabs.harvard.edu/abs/2025A&A...700A..42E}
}

@ARTICLE{Bunker:2024,
       author = {{Bunker}, Andrew J. and {Cameron}, Alex J. and {Curtis-Lake}, Emma and {Jakobsen}, Peter and {Carniani}, Stefano and {Curti}, Mirko and {Witstok}, Joris and {Maiolino}, Roberto and {D'Eugenio}, Francesco and {Looser}, Tobias J. and {Willott}, Chris and {Bonaventura}, Nina and {Hainline}, Kevin and {{\"U}bler}, Hannah and {Willmer}, Christopher N.~A. and {Saxena}, Aayush and {Smit}, Renske and {Alberts}, Stacey and {Arribas}, Santiago and {Baker}, William M. and {Baum}, Stefi and {Bhatawdekar}, Rachana and {Bowler}, Rebecca A.~A. and {Boyett}, Kristan and {Charlot}, Stephane and {Chen}, Zuyi and {Chevallard}, Jacopo and {Circosta}, Chiara and {DeCoursey}, Christa and {de Graaff}, Anna and {Egami}, Eiichi and {Eisenstein}, Daniel J. and {Endsley}, Ryan and {Ferruit}, Pierre and {Giardino}, Giovanna and {Hausen}, Ryan and {Helton}, Jakob M. and {Hviding}, Raphael E. and {Ji}, Zhiyuan and {Johnson}, Benjamin D. and {Jones}, Gareth C. and {Kumari}, Nimisha and {Laseter}, Isaac and {L{\"u}tzgendorf}, Nora and {Maseda}, Michael V. and {Nelson}, Erica and {Parlanti}, Eleonora and {Perna}, Michele and {Rauscher}, Bernard J. and {Rawle}, Tim and {Rix}, Hans-Walter and {Rieke}, Marcia and {Robertson}, Brant and {Rodr{\'\i}guez Del Pino}, Bruno and {Sandles}, Lester and {Scholtz}, Jan and {Sharpe}, Katherine and {Skarbinski}, Maya and {Stark}, Daniel P. and {Sun}, Fengwu and {Tacchella}, Sandro and {Topping}, Michael W. and {Villanueva}, Natalia C. and {Wallace}, Imaan E.~B. and {Williams}, Christina C. and {Woodrum}, Charity},
        title = "{JADES NIRSpec initial data release for the Hubble Ultra Deep Field: Redshifts and line fluxes of distant galaxies from the deepest JWST Cycle 1 NIRSpec multi-object spectroscopy}",
      journal = {\aap},
         year = 2024,
        month = oct,
       volume = {690},
          eid = {A288},
        pages = {A288},
          doi = {10.1051/0004-6361/202347094},
archivePrefix = {arXiv},
       eprint = {2306.02467},
 primaryClass = {astro-ph.GA},
       adsurl = {https://ui.adsabs.harvard.edu/abs/2024A&A...690A.288B}
}

@ARTICLE{Genzel:2023,
       author = {{Genzel}, R. and {Jolly}, J. -B. and {Liu}, D. and {Price}, S.~H. and {Lee}, L.~L. and {F{\"o}rster Schreiber}, N.~M. and {Tacconi}, L.~J. and {Herrera-Camus}, R. and {Barfety}, C. and {Burkert}, A. and {Cao}, Y. and {Davies}, R.~I. and {Dekel}, A. and {Lee}, M.~M. and {Lutz}, D. and {Naab}, T. and {Neri}, R. and {Nestor Shachar}, A. and {Pastras}, S. and {Pulsoni}, C. and {Renzini}, A. and {Schuster}, K. and {Shimizu}, T.~T. and {Stanley}, F. and {Sternberg}, A. and {{\"U}bler}, H.},
        title = "{Evidence for Large-scale, Rapid Gas Inflows in z   2 Star-forming Disks}",
      journal = {\apj},
         year = 2023,
        month = nov,
       volume = {957},
       number = {1},
          eid = {48},
        pages = {48},
          doi = {10.3847/1538-4357/acef1a},
archivePrefix = {arXiv},
       eprint = {2305.02959},
 primaryClass = {astro-ph.GA},
       adsurl = {https://ui.adsabs.harvard.edu/abs/2023ApJ...957...48G}
}

@ARTICLE{Boquien:2019,
       author = {{Boquien}, M. and {Burgarella}, D. and {Roehlly}, Y. and {Buat}, V. and {Ciesla}, L. and {Corre}, D. and {Inoue}, A.~K. and {Salas}, H.},
        title = "{CIGALE: a python Code Investigating GALaxy Emission}",
      journal = {\aap},
         year = 2019,
        month = feb,
       volume = {622},
          eid = {A103},
        pages = {A103},
          doi = {10.1051/0004-6361/201834156},
archivePrefix = {arXiv},
       eprint = {1811.03094},
 primaryClass = {astro-ph.GA},
       adsurl = {https://ui.adsabs.harvard.edu/abs/2019A&A...622A.103B}
}

@ARTICLE{Chabrier:2003,
       author = {{Chabrier}, Gilles},
        title = "{Galactic Stellar and Substellar Initial Mass Function}",
      journal = {\pasp},
         year = 2003,
        month = jul,
       volume = {115},
       number = {809},
        pages = {763-795},
          doi = {10.1086/376392},
archivePrefix = {arXiv},
       eprint = {astro-ph/0304382},
 primaryClass = {astro-ph},
       adsurl = {https://ui.adsabs.harvard.edu/abs/2003PASP..115..763C}
}

@ARTICLE{Bruzual:2003,
       author = {{Bruzual}, G. and {Charlot}, S.},
        title = "{Stellar population synthesis at the resolution of 2003}",
      journal = {\mnras},
         year = 2003,
        month = oct,
       volume = {344},
       number = {4},
        pages = {1000-1028},
          doi = {10.1046/j.1365-8711.2003.06897.x},
archivePrefix = {arXiv},
       eprint = {astro-ph/0309134},
 primaryClass = {astro-ph},
       adsurl = {https://ui.adsabs.harvard.edu/abs/2003MNRAS.344.1000B}
}

@ARTICLE{Vazdekis:2010,
       author = {{Vazdekis}, A. and {S{\'a}nchez-Bl{\'a}zquez}, P. and {Falc{\'o}n-Barroso}, J. and {Cenarro}, A.~J. and {Beasley}, M.~A. and {Cardiel}, N. and {Gorgas}, J. and {Peletier}, R.~F.},
        title = "{Evolutionary stellar population synthesis with MILES - I. The base models and a new line index system}",
      journal = {\mnras},
         year = 2010,
        month = jun,
       volume = {404},
       number = {4},
        pages = {1639-1671},
          doi = {10.1111/j.1365-2966.2010.16407.x},
archivePrefix = {arXiv},
       eprint = {1004.4439},
 primaryClass = {astro-ph.CO},
       adsurl = {https://ui.adsabs.harvard.edu/abs/2010MNRAS.404.1639V}
}

@BOOK{Osterbrock:2006,
       author = {{Osterbrock}, Donald E. and {Ferland}, Gary J.},
        title = "{Astrophysics of gaseous nebulae and active galactic nuclei}",
         year = 2006,
       adsurl = {https://ui.adsabs.harvard.edu/abs/2006agna.book.....O}
}

@ARTICLE{Koekemoer:2011,
       author = {{Koekemoer}, Anton M. and {Faber}, S.~M. and {Ferguson}, Henry C. and {Grogin}, Norman A. and {Kocevski}, Dale D. and {Koo}, David C. and {Lai}, Kamson and {Lotz}, Jennifer M. and {Lucas}, Ray A. and {McGrath}, Elizabeth J. and {Ogaz}, Sara and {Rajan}, Abhijith and {Riess}, Adam G. and {Rodney}, Steve A. and {Strolger}, Louis and {Casertano}, Stefano and {Castellano}, Marco and {Dahlen}, Tomas and {Dickinson}, Mark and {Dolch}, Timothy and {Fontana}, Adriano and {Giavalisco}, Mauro and {Grazian}, Andrea and {Guo}, Yicheng and {Hathi}, Nimish P. and {Huang}, Kuang-Han and {van der Wel}, Arjen and {Yan}, Hao-Jing and {Acquaviva}, Viviana and {Alexander}, David M. and {Almaini}, Omar and {Ashby}, Matthew L.~N. and {Barden}, Marco and {Bell}, Eric F. and {Bournaud}, Fr{\'e}d{\'e}ric and {Brown}, Thomas M. and {Caputi}, Karina I. and {Cassata}, Paolo and {Challis}, Peter J. and {Chary}, Ranga-Ram and {Cheung}, Edmond and {Cirasuolo}, Michele and {Conselice}, Christopher J. and {Roshan Cooray}, Asantha and {Croton}, Darren J. and {Daddi}, Emanuele and {Dav{\'e}}, Romeel and {de Mello}, Duilia F. and {de Ravel}, Loic and {Dekel}, Avishai and {Donley}, Jennifer L. and {Dunlop}, James S. and {Dutton}, Aaron A. and {Elbaz}, David and {Fazio}, Giovanni G. and {Filippenko}, Alexei V. and {Finkelstein}, Steven L. and {Frazer}, Chris and {Gardner}, Jonathan P. and {Garnavich}, Peter M. and {Gawiser}, Eric and {Gruetzbauch}, Ruth and {Hartley}, Will G. and {H{\"a}ussler}, Boris and {Herrington}, Jessica and {Hopkins}, Philip F. and {Huang}, Jia-Sheng and {Jha}, Saurabh W. and {Johnson}, Andrew and {Kartaltepe}, Jeyhan S. and {Khostovan}, Ali A. and {Kirshner}, Robert P. and {Lani}, Caterina and {Lee}, Kyoung-Soo and {Li}, Weidong and {Madau}, Piero and {McCarthy}, Patrick J. and {McIntosh}, Daniel H. and {McLure}, Ross J. and {McPartland}, Conor and {Mobasher}, Bahram and {Moreira}, Heidi and {Mortlock}, Alice and {Moustakas}, Leonidas A. and {Mozena}, Mark and {Nandra}, Kirpal and {Newman}, Jeffrey A. and {Nielsen}, Jennifer L. and {Niemi}, Sami and {Noeske}, Kai G. and {Papovich}, Casey J. and {Pentericci}, Laura and {Pope}, Alexandra and {Primack}, Joel R. and {Ravindranath}, Swara and {Reddy}, Naveen A. and {Renzini}, Alvio and {Rix}, Hans-Walter and {Robaina}, Aday R. and {Rosario}, David J. and {Rosati}, Piero and {Salimbeni}, Sara and {Scarlata}, Claudia and {Siana}, Brian and {Simard}, Luc and {Smidt}, Joseph and {Snyder}, Diana and {Somerville}, Rachel S. and {Spinrad}, Hyron and {Straughn}, Amber N. and {Telford}, Olivia and {Teplitz}, Harry I. and {Trump}, Jonathan R. and {Vargas}, Carlos and {Villforth}, Carolin and {Wagner}, Cory R. and {Wandro}, Pat and {Wechsler}, Risa H. and {Weiner}, Benjamin J. and {Wiklind}, Tommy and {Wild}, Vivienne and {Wilson}, Grant and {Wuyts}, Stijn and {Yun}, Min S.},
        title = "{CANDELS: The Cosmic Assembly Near-infrared Deep Extragalactic Legacy Survey{\textemdash}The Hubble Space Telescope Observations, Imaging Data Products, and Mosaics}",
      journal = {\apjs},
         year = 2011,
        month = dec,
       volume = {197},
       number = {2},
          eid = {36},
        pages = {36},
          doi = {10.1088/0067-0049/197/2/36},
archivePrefix = {arXiv},
       eprint = {1105.3754},
 primaryClass = {astro-ph.CO},
       adsurl = {https://ui.adsabs.harvard.edu/abs/2011ApJS..197...36K}
}

@ARTICLE{Lang:2014,
       author = {{Lang}, Philipp and {Wuyts}, Stijn and {Somerville}, Rachel S. and {F{\"o}rster Schreiber}, Natascha M. and {Genzel}, Reinhard and {Bell}, Eric F. and {Brammer}, Gabe and {Dekel}, Avishai and {Faber}, Sandra M. and {Ferguson}, Henry C. and {Grogin}, Norman A. and {Kocevski}, Dale D. and {Koekemoer}, Anton M. and {Lutz}, Dieter and {McGrath}, Elizabeth J. and {Momcheva}, Ivelina and {Nelson}, Erica J. and {Primack}, Joel R. and {Rosario}, David J. and {Skelton}, Rosalind E. and {Tacconi}, Linda J. and {van Dokkum}, Pieter G. and {Whitaker}, Katherine E.},
        title = "{Bulge Growth and Quenching since z = 2.5 in CANDELS/3D-HST}",
      journal = {\apj},
         year = 2014,
        month = jun,
       volume = {788},
       number = {1},
          eid = {11},
        pages = {11},
          doi = {10.1088/0004-637X/788/1/11},
archivePrefix = {arXiv},
       eprint = {1402.0866},
 primaryClass = {astro-ph.GA},
       adsurl = {https://ui.adsabs.harvard.edu/abs/2014ApJ...788...11L}
}

@ARTICLE{Asplund:2021,
       author = {{Asplund}, M. and {Amarsi}, A.~M. and {Grevesse}, N.},
        title = "{The chemical make-up of the Sun: A 2020 vision}",
      journal = {\aap},
         year = 2021,
        month = sep,
       volume = {653},
          eid = {A141},
        pages = {A141},
          doi = {10.1051/0004-6361/202140445},
archivePrefix = {arXiv},
       eprint = {2105.01661},
 primaryClass = {astro-ph.SR},
       adsurl = {https://ui.adsabs.harvard.edu/abs/2021A&A...653A.141A}
}

@ARTICLE{Lutz:2011,
       author = {{Lutz}, D. and {Poglitsch}, A. and {Altieri}, B. and {Andreani}, P. and {Aussel}, H. and {Berta}, S. and {Bongiovanni}, A. and {Brisbin}, D. and {Cava}, A. and {Cepa}, J. and {Cimatti}, A. and {Daddi}, E. and {Dominguez-Sanchez}, H. and {Elbaz}, D. and {F{\"o}rster Schreiber}, N.~M. and {Genzel}, R. and {Grazian}, A. and {Gruppioni}, C. and {Harwit}, M. and {Le Floc'h}, E. and {Magdis}, G. and {Magnelli}, B. and {Maiolino}, R. and {Nordon}, R. and {P{\'e}rez Garc{\'\i}a}, A.~M. and {Popesso}, P. and {Pozzi}, F. and {Riguccini}, L. and {Rodighiero}, G. and {Saintonge}, A. and {Sanchez Portal}, M. and {Santini}, P. and {Shao}, L. and {Sturm}, E. and {Tacconi}, L.~J. and {Valtchanov}, I. and {Wetzstein}, M. and {Wieprecht}, E.},
        title = "{PACS Evolutionary Probe (PEP) - A Herschel key program}",
      journal = {\aap},
         year = 2011,
        month = aug,
       volume = {532},
          eid = {A90},
        pages = {A90},
          doi = {10.1051/0004-6361/201117107},
archivePrefix = {arXiv},
       eprint = {1106.3285},
 primaryClass = {astro-ph.CO},
       adsurl = {https://ui.adsabs.harvard.edu/abs/2011A&A...532A..90L}
}

@ARTICLE{Forster:2019,
       author = {{F{\"o}rster Schreiber}, N.~M. and {{\"U}bler}, H. and {Davies}, R.~L. and {Genzel}, R. and {Wisnioski}, E. and {Belli}, S. and {Shimizu}, T. and {Lutz}, D. and {Fossati}, M. and {Herrera-Camus}, R. and {Mendel}, J.~T. and {Tacconi}, L.~J. and {Wilman}, D. and {Beifiori}, A. and {Brammer}, G.~B. and {Burkert}, A. and {Carollo}, C.~M. and {Davies}, R.~I. and {Eisenhauer}, F. and {Fabricius}, M. and {Lilly}, S.~J. and {Momcheva}, I. and {Naab}, T. and {Nelson}, E.~J. and {Price}, S.~H. and {Renzini}, A. and {Saglia}, R. and {Sternberg}, A. and {van Dokkum}, P. and {Wuyts}, S.},
        title = "{The KMOS$^{3D}$ Survey: Demographics and Properties of Galactic Outflows at z = 0.6-2.7}",
      journal = {\apj},
         year = 2019,
        month = apr,
       volume = {875},
       number = {1},
          eid = {21},
        pages = {21},
          doi = {10.3847/1538-4357/ab0ca2},
archivePrefix = {arXiv},
       eprint = {1807.04738},
 primaryClass = {astro-ph.GA},
       adsurl = {https://ui.adsabs.harvard.edu/abs/2019ApJ...875...21F}
}

@ARTICLE{Maraston:2010,
       author = {{Maraston}, Claudia and {Pforr}, Janine and {Renzini}, Alvio and {Daddi}, Emanuele and {Dickinson}, Mark and {Cimatti}, Andrea and {Tonini}, Chiara},
        title = "{Star formation rates and masses of z \raisebox{-0.5ex}\textasciitilde 2 galaxies from multicolour photometry}",
      journal = {\mnras},
         year = 2010,
        month = sep,
       volume = {407},
       number = {2},
        pages = {830-845},
          doi = {10.1111/j.1365-2966.2010.16973.x},
archivePrefix = {arXiv},
       eprint = {1004.4546},
 primaryClass = {astro-ph.CO},
       adsurl = {https://ui.adsabs.harvard.edu/abs/2010MNRAS.407..830M}
}

@ARTICLE{Brammer:2012,
       author = {{Brammer}, Gabriel B. and {van Dokkum}, Pieter G. and {Franx}, Marijn and {Fumagalli}, Mattia and {Patel}, Shannon and {Rix}, Hans-Walter and {Skelton}, Rosalind E. and {Kriek}, Mariska and {Nelson}, Erica and {Schmidt}, Kasper B. and {Bezanson}, Rachel and {da Cunha}, Elisabete and {Erb}, Dawn K. and {Fan}, Xiaohui and {F{\"o}rster Schreiber}, Natascha and {Illingworth}, Garth D. and {Labb{\'e}}, Ivo and {Leja}, Joel and {Lundgren}, Britt and {Magee}, Dan and {Marchesini}, Danilo and {McCarthy}, Patrick and {Momcheva}, Ivelina and {Muzzin}, Adam and {Quadri}, Ryan and {Steidel}, Charles C. and {Tal}, Tomer and {Wake}, David and {Whitaker}, Katherine E. and {Williams}, Anna},
        title = "{3D-HST: A Wide-field Grism Spectroscopic Survey with the Hubble Space Telescope}",
      journal = {\apjs},
         year = 2012,
        month = jun,
       volume = {200},
       number = {2},
          eid = {13},
        pages = {13},
          doi = {10.1088/0067-0049/200/2/13},
archivePrefix = {arXiv},
       eprint = {1204.2829},
 primaryClass = {astro-ph.CO},
       adsurl = {https://ui.adsabs.harvard.edu/abs/2012ApJS..200...13B}
}

@ARTICLE{Toomre:1964,
       author = {{Toomre}, A.},
        title = "{On the gravitational stability of a disk of stars.}",
      journal = {\apj},
         year = 1964,
        month = may,
       volume = {139},
        pages = {1217-1238},
          doi = {10.1086/147861},
       adsurl = {https://ui.adsabs.harvard.edu/abs/1964ApJ...139.1217T}
}

@ARTICLE{Arriagada:2025,
       author = {{Arriagada-Neira}, Sebasti{\'a}n and {Herrera-Camus}, Rodrigo and {Villanueva}, Vicente and {F{\"o}rster Schreiber}, Natascha M. and {Lee}, Minju and {Bolatto}, Alberto and {Chen}, Jianhang and {Genzel}, Reinhard and {Liu}, Daizhong and {Renzini}, Alvio and {Tacconi}, Linda J. and {Tozzi}, Giulia and {{\"U}bler}, Hannah},
        title = "{Deep kiloparsec view of the molecular gas in a massive star-forming galaxy at cosmic noon}",
      journal = {\aap},
         year = 2025,
        month = apr,
       volume = {696},
          eid = {A83},
        pages = {A83},
          doi = {10.1051/0004-6361/202452652},
archivePrefix = {arXiv},
       eprint = {2410.14781},
 primaryClass = {astro-ph.GA},
       adsurl = {https://ui.adsabs.harvard.edu/abs/2025A&A...696A..83A}
}

@ARTICLE{Mannucci:2010,
       author = {{Mannucci}, F. and {Cresci}, G. and {Maiolino}, R. and {Marconi}, A. and {Gnerucci}, A.},
        title = "{A fundamental relation between mass, star formation rate and metallicity in local and high-redshift galaxies}",
      journal = {\mnras},
         year = 2010,
        month = nov,
       volume = {408},
       number = {4},
        pages = {2115-2127},
          doi = {10.1111/j.1365-2966.2010.17291.x},
archivePrefix = {arXiv},
       eprint = {1005.0006},
 primaryClass = {astro-ph.CO},
       adsurl = {https://ui.adsabs.harvard.edu/abs/2010MNRAS.408.2115M}
}

@ARTICLE{hilyBlant:2022,
       author = {{Hily-Blant}, P. and {Pineau des For{\^e}ts}, G. and {Faure}, A. and {Lique}, F.},
        title = "{Sulfur gas-phase abundance in dense cores}",
      journal = {\aap},
         year = 2022,
        month = feb,
       volume = {658},
          eid = {A168},
        pages = {A168},
          doi = {10.1051/0004-6361/201936498},
archivePrefix = {arXiv},
       eprint = {2112.01076},
 primaryClass = {astro-ph.GA},
       adsurl = {https://ui.adsabs.harvard.edu/abs/2022A&A...658A.168H}
}

@ARTICLE{Skelton:2014,
       author = {{Skelton}, Rosalind E. and {Whitaker}, Katherine E. and {Momcheva}, Ivelina G. and {Brammer}, Gabriel B. and {van Dokkum}, Pieter G. and {Labb{\'e}}, Ivo and {Franx}, Marijn and {van der Wel}, Arjen and {Bezanson}, Rachel and {Da Cunha}, Elisabete and {Fumagalli}, Mattia and {F{\"o}rster Schreiber}, Natascha and {Kriek}, Mariska and {Leja}, Joel and {Lundgren}, Britt F. and {Magee}, Daniel and {Marchesini}, Danilo and {Maseda}, Michael V. and {Nelson}, Erica J. and {Oesch}, Pascal and {Pacifici}, Camilla and {Patel}, Shannon G. and {Price}, Sedona and {Rix}, Hans-Walter and {Tal}, Tomer and {Wake}, David A. and {Wuyts}, Stijn},
        title = "{3D-HST WFC3-selected Photometric Catalogs in the Five CANDELS/3D-HST Fields: Photometry, Photometric Redshifts, and Stellar Masses}",
      journal = {\apjs},
         year = 2014,
        month = oct,
       volume = {214},
       number = {2},
          eid = {24},
        pages = {24},
          doi = {10.1088/0067-0049/214/2/24},
archivePrefix = {arXiv},
       eprint = {1403.3689},
 primaryClass = {astro-ph.GA},
       adsurl = {https://ui.adsabs.harvard.edu/abs/2014ApJS..214...24S}
}

@ARTICLE{Shapley:2015,
       author = {{Shapley}, Alice E. and {Reddy}, Naveen A. and {Kriek}, Mariska and {Freeman}, William R. and {Sanders}, Ryan L. and {Siana}, Brian and {Coil}, Alison L. and {Mobasher}, Bahram and {Shivaei}, Irene and {Price}, Sedona H. and {de Groot}, Laura},
        title = "{The MOSDEF Survey: Excitation Properties of z {\ensuremath{\sim}} 2.3 Star-forming Galaxies}",
      journal = {\apj},
         year = 2015,
        month = mar,
       volume = {801},
       number = {2},
          eid = {88},
        pages = {88},
          doi = {10.1088/0004-637X/801/2/88},
archivePrefix = {arXiv},
       eprint = {1409.7071},
 primaryClass = {astro-ph.GA},
       adsurl = {https://ui.adsabs.harvard.edu/abs/2015ApJ...801...88S}
}

@ARTICLE{Markov:2025,
       author = {{Markov}, Vladan and {Gallerani}, Simona and {Ferrara}, Andrea and {Pallottini}, Andrea and {Parlanti}, Eleonora and {Mascia}, Fabio Di and {Sommovigo}, Laura and {Kohandel}, Mahsa},
        title = "{The evolution of dust attenuation in z {\ensuremath{\approx}} 2-12 galaxies observed by JWST}",
      journal = {Nature Astronomy},
         year = 2025,
        month = mar,
       volume = {9},
        pages = {458-468},
          doi = {10.1038/s41550-024-02426-1},
archivePrefix = {arXiv},
       eprint = {2402.05996},
 primaryClass = {astro-ph.GA},
       adsurl = {https://ui.adsabs.harvard.edu/abs/2025NatAs...9..458M}
}

@ARTICLE{Reddy:2026,
       author = {{Reddy}, Naveen A. and {Shapley}, Alice E. and {Sanders}, Ryan L. and {Topping}, Michael W. and {Ellis}, Richard S. and {Pettini}, Max and {Brammer}, Gabriel and {Cullen}, Fergus and {F{\"o}rster Schreiber}, Natascha M. and {Khostovan}, Ali A. and {McLeod}, Derek J. and {McLure}, Ross J. and {Narayanan}, Desika and {Oesch}, Pascal A. and {Pahl}, Anthony J. and {Steidel}, Charles C. and {Berg}, Danielle A.},
        title = "{The AURORA Survey: Multiple Balmer and Paschen Emission Lines for Individual Star-forming Galaxies at z = 1.5─4.4. I. A Diversity of Nebular Attenuation Curves and Evidence for Non-unity Dust Covering Fractions}",
      journal = {\apj},
         year = 2026,
        month = mar,
       volume = {999},
       number = {1},
          eid = {15},
        pages = {15},
          doi = {10.3847/1538-4357/ae38da},
archivePrefix = {arXiv},
       eprint = {2506.17396},
 primaryClass = {astro-ph.GA},
       adsurl = {https://ui.adsabs.harvard.edu/abs/2026ApJ...999...15R}
}

@ARTICLE{Sanders:2025,
       author = {{Sanders}, Ryan L. and {Shapley}, Alice E. and {Topping}, Michael W. and {Reddy}, Naveen A. and {Berg}, Danielle A. and {Khostovan}, Ali Ahmad and {Bouwens}, Rychard J. and {Brammer}, Gabriel and {Carnall}, Adam C. and {Cullen}, Fergus and {Dav{\'e}}, Romeel and {Dunlop}, James S. and {Ellis}, Richard S. and {F{\"o}rster Schreiber}, N.~M. and {Furlanetto}, Steven R. and {Glazebrook}, Karl and {Illingworth}, Garth D. and {Jones}, Tucker and {Kriek}, Mariska and {McLeod}, Derek J. and {McLure}, Ross J. and {Narayanan}, Desika and {Oesch}, Pascal A. and {Pahl}, Anthony J. and {Pettini}, Max and {Schaerer}, Daniel and {Stark}, Daniel P. and {Steidel}, Charles C. and {Tang}, Mengtao and {Clarke}, Leonardo and {Donnan}, Callum T. and {Kehoe}, Emily},
        title = "{The AURORA Survey: High-Redshift Empirical Metallicity Calibrations from Electron Temperature Measurements at z=2-10}",
      journal = {arXiv e-prints},
         year = 2025,
        month = aug,
          eid = {arXiv:2508.10099},
        pages = {arXiv:2508.10099},
          doi = {10.48550/arXiv.2508.10099},
archivePrefix = {arXiv},
       eprint = {2508.10099},
 primaryClass = {astro-ph.GA},
       adsurl = {https://ui.adsabs.harvard.edu/abs/2025arXiv250810099S}
}

@ARTICLE{Park:2024,
       author = {{Park}, Minjung and {Belli}, Sirio and {Conroy}, Charlie and {Johnson}, Benjamin D. and {Davies}, Rebecca L. and {Leja}, Joel and {Tacchella}, Sandro and {Mendel}, J. Trevor and {Benton}, Chlo{\"e} and {Bugiani}, Letizia and {Emami}, Razieh and {Khoram}, Amir H. and {Li}, Yijia and {Maheson}, Gabriel and {Mathews}, Elijah P. and {Naidu}, Rohan P. and {Nelson}, Erica J. and {Terrazas}, Bryan A. and {Weinberger}, Rainer},
        title = "{Widespread Rapid Quenching at Cosmic Noon Revealed by JWST Deep Spectroscopy}",
      journal = {\apj},
         year = 2024,
        month = nov,
       volume = {976},
       number = {1},
          eid = {72},
        pages = {72},
          doi = {10.3847/1538-4357/ad7e15},
archivePrefix = {arXiv},
       eprint = {2404.17945},
 primaryClass = {astro-ph.GA},
       adsurl = {https://ui.adsabs.harvard.edu/abs/2024ApJ...976...72P}
}

@ARTICLE{Nestor:2025,
       author = {{Nestor Shachar}, A. and {Sternberg}, A. and {Genzel}, R. and {Liu}, D. and {Price}, S.~H. and {Pulsoni}, C. and {Tacconi}, L.~J. and {Herrera-Camus}, R. and {F{\"o}rster Schreiber}, N.~M. and {Burkert}, A. and {Jolly}, J.~B. and {Lutz}, D. and {Wuyts}, S. and {Barfety}, C. and {Cao}, Y. and {Chen}, J. and {Davies}, R. and {Eisenhauer}, F. and {Espejo Salcedo}, J.~M. and {Lee}, L.~L. and {Lee}, M. and {Naab}, T. and {Pastras}, S. and {Shimizu}, T.~T. and {Sturm}, E. and {Tozzi}, G. and {{\"U}bler}, H.},
        title = "{A Large-scale Ring Galaxy at z = 2.2 Revealed by JWST/NIRCam: Kinematic Observations and Analytical Modelling}",
      journal = {\apj},
         year = 2025,
        month = aug,
       volume = {988},
       number = {2},
          eid = {182},
        pages = {182},
          doi = {10.3847/1538-4357/ade2d0},
archivePrefix = {arXiv},
       eprint = {2503.00839},
 primaryClass = {astro-ph.GA},
       adsurl = {https://ui.adsabs.harvard.edu/abs/2025ApJ...988..182N}
}

@ARTICLE{Walter:2020,
       author = {{Walter}, Fabian and {Carilli}, Chris and {Neeleman}, Marcel and {Decarli}, Roberto and {Popping}, Gerg{\"o} and {Somerville}, Rachel S. and {Aravena}, Manuel and {Bertoldi}, Frank and {Boogaard}, Leindert and {Cox}, Pierre and {da Cunha}, Elisabete and {Magnelli}, Benjamin and {Obreschkow}, Danail and {Riechers}, Dominik and {Rix}, Hans-Walter and {Smail}, Ian and {Weiss}, Axel and {Assef}, Roberto J. and {Bauer}, Franz and {Bouwens}, Rychard and {Contini}, Thierry and {Cortes}, Paulo C. and {Daddi}, Emanuele and {Diaz-Santos}, Tanio and {Gonz{\'a}lez-L{\'o}pez}, Jorge and {Hennawi}, Joseph and {Hodge}, Jacqueline A. and {Inami}, Hanae and {Ivison}, Rob and {Oesch}, Pascal and {Sargent}, Mark and {van der Werf}, Paul and {Wagg}, Jeff and {Yung}, L.~Y. Aaron},
        title = "{The Evolution of the Baryons Associated with Galaxies Averaged over Cosmic Time and Space}",
      journal = {\apj},
         year = 2020,
        month = oct,
       volume = {902},
       number = {2},
          eid = {111},
        pages = {111},
          doi = {10.3847/1538-4357/abb82e},
archivePrefix = {arXiv},
       eprint = {2009.11126},
 primaryClass = {astro-ph.GA},
       adsurl = {https://ui.adsabs.harvard.edu/abs/2020ApJ...902..111W}
}

@ARTICLE{Whitaker:2014,
       author = {{Whitaker}, Katherine E. and {Franx}, Marijn and {Leja}, Joel and {van Dokkum}, Pieter G. and {Henry}, Alaina and {Skelton}, Rosalind E. and {Fumagalli}, Mattia and {Momcheva}, Ivelina G. and {Brammer}, Gabriel B. and {Labb{\'e}}, Ivo and {Nelson}, Erica J. and {Rigby}, Jane R.},
        title = "{Constraining the Low-mass Slope of the Star Formation Sequence at 0.5 < z < 2.5}",
      journal = {\apj},
         year = 2014,
        month = nov,
       volume = {795},
       number = {2},
          eid = {104},
        pages = {104},
          doi = {10.1088/0004-637X/795/2/104},
archivePrefix = {arXiv},
       eprint = {1407.1843},
 primaryClass = {astro-ph.GA},
       adsurl = {https://ui.adsabs.harvard.edu/abs/2014ApJ...795..104W}
}

@ARTICLE{Rieke:2023,
       author = {{Rieke}, Marcia J. and {Kelly}, Douglas M. and {Misselt}, Karl and {Stansberry}, John and {Boyer}, Martha and {Beatty}, Thomas and {Egami}, Eiichi and {Florian}, Michael and {Greene}, Thomas P. and {Hainline}, Kevin and {Leisenring}, Jarron and {Roellig}, Thomas and {Schlawin}, Everett and {Sun}, Fengwu and {Tinnin}, Lee and {Williams}, Christina C. and {Willmer}, Christopher N.~A. and {Wilson}, Debra and {Clark}, Charles R. and {Rohrbach}, Scott and {Brooks}, Brian and {Canipe}, Alicia and {Correnti}, Matteo and {DiFelice}, Audrey and {Gennaro}, Mario and {Girard}, Julien H. and {Hartig}, George and {Hilbert}, Bryan and {Koekemoer}, Anton M. and {Nikolov}, Nikolay K. and {Pirzkal}, Norbert and {Rest}, Armin and {Robberto}, Massimo and {Sunnquist}, Ben and {Telfer}, Randal and {Wu}, Chi Rai and {Ferry}, Malcolm and {Lewis}, Dan and {Baum}, Stefi and {Beichman}, Charles and {Doyon}, Ren{\'e} and {Dressler}, Alan and {Eisenstein}, Daniel J. and {Ferrarese}, Laura and {Hodapp}, Klaus and {Horner}, Scott and {Jaffe}, Daniel T. and {Johnstone}, Doug and {Krist}, John and {Martin}, Peter and {McCarthy}, Donald W. and {Meyer}, Michael and {Rieke}, George H. and {Trauger}, John and {Young}, Erick T.},
        title = "{Performance of NIRCam on JWST in Flight}",
      journal = {\pasp},
         year = 2023,
        month = feb,
       volume = {135},
       number = {1044},
          eid = {028001},
        pages = {028001},
          doi = {10.1088/1538-3873/acac53},
archivePrefix = {arXiv},
       eprint = {2212.12069},
 primaryClass = {astro-ph.IM},
       adsurl = {https://ui.adsabs.harvard.edu/abs/2023PASP..135b8001R}
}

@ARTICLE{Lorenz:2025,
       author = {{Lorenz}, Brian and {Suess}, Katherine A. and {Kriek}, Mariska and {Price}, Sedona H. and {Leja}, Joel and {Nelson}, Erica and {Atek}, Hakim and {Bezanson}, Rachel and {Brammer}, Gabriel and {Cutler}, Sam E. and {Dayal}, Pratika and {de Graaff}, Anna and {Greene}, Jenny E. and {Furtak}, Lukas J. and {Labb{\'e}}, Ivo and {Marchesini}, Danilo and {Maseda}, Michael V. and {Miller}, Tim B. and {Mintz}, Abby and {Mitsuhashi}, Ikki and {Pan}, Richard and {Porraz Barrera}, Natalia and {Wang}, Bingjie and {Weaver}, John R. and {Williams}, Christina C. and {Whitaker}, Katherine E.},
        title = "{Measuring Emission Lines with JWST MegaScience Medium Bands: A New Window into Dust and Star Formation at Cosmic Noon}",
      journal = {\apjl},
         year = 2025,
        month = jul,
       volume = {988},
       number = {1},
          eid = {L20},
        pages = {L20},
          doi = {10.3847/2041-8213/ade887},
archivePrefix = {arXiv},
       eprint = {2505.10632},
 primaryClass = {astro-ph.GA},
       adsurl = {https://ui.adsabs.harvard.edu/abs/2025ApJ...988L..20L}
}

@ARTICLE{Slob:2025,
       author = {{Slob}, Martje and {Kriek}, Mariska and {de Graaff}, Anna and {Cheng}, Chloe M. and {Beverage}, Aliza G. and {Bezanson}, Rachel and {F{\"o}rster Schreiber}, Natascha M. and {Lorenz}, Brian and {Mancera Pi{\~n}a}, Pavel E. and {Marchesini}, Danilo and {Muzzin}, Adam and {Newman}, Andrew B. and {Price}, Sedona H. and {Suess}, Katherine A. and {van de Sande}, Jesse and {van Dokkum}, Pieter and {Weisz}, Daniel R.},
        title = "{Fast rotators at cosmic noon: Stellar kinematics for 15 quiescent galaxies from JWST-SUSPENSE}",
      journal = {\aap},
         year = 2025,
        month = oct,
       volume = {702},
          eid = {A110},
        pages = {A110},
          doi = {10.1051/0004-6361/202555812},
archivePrefix = {arXiv},
       eprint = {2506.04310},
 primaryClass = {astro-ph.GA},
       adsurl = {https://ui.adsabs.harvard.edu/abs/2025A&A...702A.110S}
}

@ARTICLE{Ju:2025,
       author = {{Ju}, Mengting and {Wang}, Xin and {Jones}, Tucker and {Bari{\v{s}}i{\'c}}, Ivana and {Nanayakkara}, Themiya and {Bundy}, Kevin and {Faucher-Gigu{\`e}re}, Claude-Andr{\'e} and {Feng}, Shuai and {Glazebrook}, Karl and {Henry}, Alaina and {Malkan}, Matthew A. and {Obreschkow}, Danail and {Roy}, Namrata and {Sanders}, Ryan L. and {Sun}, Xunda and {Treu}, Tommaso and {Zhou}, Qianqiao},
        title = "{MSA-3D: Metallicity Gradients in Galaxies at z {\ensuremath{\sim}} 1 with JWST/NIRSpec Slit-stepping Spectroscopy}",
      journal = {\apjl},
         year = 2025,
        month = jan,
       volume = {978},
       number = {2},
          eid = {L39},
        pages = {L39},
          doi = {10.3847/2041-8213/ada150},
archivePrefix = {arXiv},
       eprint = {2409.01616},
 primaryClass = {astro-ph.GA},
       adsurl = {https://ui.adsabs.harvard.edu/abs/2025ApJ...978L..39J}
}

@ARTICLE{Naidu:2026,
       author = {{Naidu}, Rohan P. and {Oesch}, Pascal A. and {Brammer}, Gabriel and {Weibel}, Andrea and {Li}, Yijia and {Matthee}, Jorryt and {Chisolm}, John and {Pollock}, Clara L. and {Heintz}, Kasper E. and {Johnson}, Benjamin D. and {Shen}, Xuejian and {Hviding}, Raphael E. and {Leja}, Joel and {Tacchella}, Sandro and {Ganguly}, Arpita and {Witten}, Callum and {Atek}, Hakim and {Belli}, Siro and {Bose}, Sownak and {Bouwens}, Rychard and {Dayal}, Pratika and {Decarli}, Roberto and {de Graaff}, Anna and {Fudamoto}, Yoshinobu and {Giovinazzo}, Emma and {Greene}, Jenny E. and {Illingworth}, Garth and {Inoue}, Akio K. and {Kane}, Sarah G. and {Labbe}, Ivo and {Leonova}, Ecaterina and {Marques-Chaves}, Rui and {Meyer}, Roman A. and {Nelson}, Erica J. and {Roberts-Borsani}, Guido and {Schaerer}, Daniel and {Simcoe}, Robert A. and {Stefanon}, Mauro and {Sugahara}, Yuma and {Toft}, Sune and {van der Wel}, Arjen and {van Dokkum}, Pieter and {Walter}, Fabian and {Watson}, Darrach and {Weaver}, John R. and {Whitaker}, Katherine E.},
        title = "{A Cosmic Miracle: A Remarkably Luminous Galaxy at zspec = 14.44 Confirmed with JWST}",
      journal = {The Open Journal of Astrophysics},
         year = 2026,
        month = jan,
       volume = {9},
        pages = {56033},
          doi = {10.33232/001c.156033},
archivePrefix = {arXiv},
       eprint = {2505.11263},
 primaryClass = {astro-ph.GA},
       adsurl = {https://ui.adsabs.harvard.edu/abs/2026OJAp....956033N}
}

@ARTICLE{Carniani:2024,
       author = {{Carniani}, Stefano and {Hainline}, Kevin and {D'Eugenio}, Francesco and {Eisenstein}, Daniel J. and {Jakobsen}, Peter and {Witstok}, Joris and {Johnson}, Benjamin D. and {Chevallard}, Jacopo and {Maiolino}, Roberto and {Helton}, Jakob M. and {Willott}, Chris and {Robertson}, Brant and {Alberts}, Stacey and {Arribas}, Santiago and {Baker}, William M. and {Bhatawdekar}, Rachana and {Boyett}, Kristan and {Bunker}, Andrew J. and {Cameron}, Alex J. and {Cargile}, Phillip A. and {Charlot}, St{\'e}phane and {Curti}, Mirko and {Curtis-Lake}, Emma and {Egami}, Eiichi and {Giardino}, Giovanna and {Isaak}, Kate and {Ji}, Zhiyuan and {Jones}, Gareth C. and {Kumari}, Nimisha and {Maseda}, Michael V. and {Parlanti}, Eleonora and {P{\'e}rez-Gonz{\'a}lez}, Pablo G. and {Rawle}, Tim and {Rieke}, George and {Rieke}, Marcia and {Del Pino}, Bruno Rodr{\'\i}guez and {Saxena}, Aayush and {Scholtz}, Jan and {Smit}, Renske and {Sun}, Fengwu and {Tacchella}, Sandro and {{\"U}bler}, Hannah and {Venturi}, Giacomo and {Williams}, Christina C. and {Willmer}, Christopher N.~A.},
        title = "{Spectroscopic confirmation of two luminous galaxies at a redshift of 14}",
      journal = {\nat},
         year = 2024,
        month = sep,
       volume = {633},
       number = {8029},
        pages = {318-322},
          doi = {10.1038/s41586-024-07860-9},
archivePrefix = {arXiv},
       eprint = {2405.18485},
 primaryClass = {astro-ph.GA},
       adsurl = {https://ui.adsabs.harvard.edu/abs/2024Natur.633..318C}
}

@ARTICLE{CurtisLake:2023,
       author = {{Curtis-Lake}, Emma and {Carniani}, Stefano and {Cameron}, Alex and {Charlot}, Stephane and {Jakobsen}, Peter and {Maiolino}, Roberto and {Bunker}, Andrew and {Witstok}, Joris and {Smit}, Renske and {Chevallard}, Jacopo and {Willott}, Chris and {Ferruit}, Pierre and {Arribas}, Santiago and {Bonaventura}, Nina and {Curti}, Mirko and {D'Eugenio}, Francesco and {Franx}, Marijn and {Giardino}, Giovanna and {Looser}, Tobias J. and {L{\"u}tzgendorf}, Nora and {Maseda}, Michael V. and {Rawle}, Tim and {Rix}, Hans-Walter and {Rodr{\'\i}guez del Pino}, Bruno and {{\"U}bler}, Hannah and {Sirianni}, Marco and {Dressler}, Alan and {Egami}, Eiichi and {Eisenstein}, Daniel J. and {Endsley}, Ryan and {Hainline}, Kevin and {Hausen}, Ryan and {Johnson}, Benjamin D. and {Rieke}, Marcia and {Robertson}, Brant and {Shivaei}, Irene and {Stark}, Daniel P. and {Tacchella}, Sandro and {Williams}, Christina C. and {Willmer}, Christopher N.~A. and {Bhatawdekar}, Rachana and {Bowler}, Rebecca and {Boyett}, Kristan and {Chen}, Zuyi and {de Graaff}, Anna and {Helton}, Jakob M. and {Hviding}, Raphael E. and {Jones}, Gareth C. and {Kumari}, Nimisha and {Lyu}, Jianwei and {Nelson}, Erica and {Perna}, Michele and {Sandles}, Lester and {Saxena}, Aayush and {Suess}, Katherine A. and {Sun}, Fengwu and {Topping}, Michael W. and {Wallace}, Imaan E.~B. and {Whitler}, Lily},
        title = "{Spectroscopic confirmation of four metal-poor galaxies at z = 10.3-13.2}",
      journal = {Nature Astronomy},
         year = 2023,
        month = may,
       volume = {7},
        pages = {622-632},
          doi = {10.1038/s41550-023-01918-w},
archivePrefix = {arXiv},
       eprint = {2212.04568},
 primaryClass = {astro-ph.GA},
       adsurl = {https://ui.adsabs.harvard.edu/abs/2023NatAs...7..622C}
}

@ARTICLE{Finkelstein:2023,
       author = {{Finkelstein}, Steven L. and {Bagley}, Micaela B. and {Ferguson}, Henry C. and {Wilkins}, Stephen M. and {Kartaltepe}, Jeyhan S. and {Papovich}, Casey and {Yung}, L.~Y. Aaron and {Arrabal Haro}, Pablo and {Behroozi}, Peter and {Dickinson}, Mark and {Kocevski}, Dale D. and {Koekemoer}, Anton M. and {Larson}, Rebecca L. and {Le Bail}, Aur{\'e}lien and {Morales}, Alexa M. and {P{\'e}rez-Gonz{\'a}lez}, Pablo G. and {Burgarella}, Denis and {Dav{\'e}}, Romeel and {Hirschmann}, Michaela and {Somerville}, Rachel S. and {Wuyts}, Stijn and {Bromm}, Volker and {Casey}, Caitlin M. and {Fontana}, Adriano and {Fujimoto}, Seiji and {Gardner}, Jonathan P. and {Giavalisco}, Mauro and {Grazian}, Andrea and {Grogin}, Norman A. and {Hathi}, Nimish P. and {Hutchison}, Taylor A. and {Jha}, Saurabh W. and {Jogee}, Shardha and {Kewley}, Lisa J. and {Kirkpatrick}, Allison and {Long}, Arianna S. and {Lotz}, Jennifer M. and {Pentericci}, Laura and {Pierel}, Justin D.~R. and {Pirzkal}, Nor and {Ravindranath}, Swara and {Ryan}, Russell E. and {Trump}, Jonathan R. and {Yang}, Guang and {Bhatawdekar}, Rachana and {Bisigello}, Laura and {Buat}, V{\'e}ronique and {Calabr{\`o}}, Antonello and {Castellano}, Marco and {Cleri}, Nikko J. and {Cooper}, M.~C. and {Croton}, Darren and {Daddi}, Emanuele and {Dekel}, Avishai and {Elbaz}, David and {Franco}, Maximilien and {Gawiser}, Eric and {Holwerda}, Benne W. and {Huertas-Company}, Marc and {Jaskot}, Anne E. and {Leung}, Gene C.~K. and {Lucas}, Ray A. and {Mobasher}, Bahram and {Pandya}, Viraj and {Tacchella}, Sandro and {Weiner}, Benjamin J. and {Zavala}, Jorge A.},
        title = "{CEERS Key Paper. I. An Early Look into the First 500 Myr of Galaxy Formation with JWST}",
      journal = {\apjl},
         year = 2023,
        month = mar,
       volume = {946},
       number = {1},
          eid = {L13},
        pages = {L13},
          doi = {10.3847/2041-8213/acade4},
archivePrefix = {arXiv},
       eprint = {2211.05792},
 primaryClass = {astro-ph.GA},
       adsurl = {https://ui.adsabs.harvard.edu/abs/2023ApJ...946L..13F}
}

@ARTICLE{Tacchella:2018,
       author = {{Tacchella}, S. and {Carollo}, C.~M. and {F{\"o}rster Schreiber}, N.~M. and {Renzini}, A. and {Dekel}, A. and {Genzel}, R. and {Lang}, P. and {Lilly}, S.~J. and {Mancini}, C. and {Onodera}, M. and {Tacconi}, L.~J. and {Wuyts}, S. and {Zamorani}, G.},
        title = "{Dust Attenuation, Bulge Formation, and Inside-out Quenching of Star Formation in Star-forming Main Sequence Galaxies at z {\ensuremath{\sim}} 2}",
      journal = {\apj},
         year = 2018,
        month = may,
       volume = {859},
       number = {1},
          eid = {56},
        pages = {56},
          doi = {10.3847/1538-4357/aabf8b},
archivePrefix = {arXiv},
       eprint = {1704.00733},
 primaryClass = {astro-ph.GA},
       adsurl = {https://ui.adsabs.harvard.edu/abs/2018ApJ...859...56T}
}

@ARTICLE{Kalita:2025,
       author = {{Kalita}, Boris S. and {Suzuki}, Tomoko L. and {Kashino}, Daichi and {Silverman}, John D. and {Daddi}, Emanuele and {Ho}, Luis C. and {Ding}, Xuheng and {Mercier}, Wilfried and {Faisst}, Andreas L. and {Sheth}, Kartik and {Valentino}, Francesco and {Puglisi}, Annagrazia and {Saito}, Toshiki and {Kakkad}, Darshan and {Ilbert}, Olivier and {Khostovan}, Ali Ahmad and {Liu}, Zhaoxuan and {Tanaka}, Takumi and {Magdis}, Georgios and {Zavala}, Jorge A. and {Tan}, Qinghua and {Kartaltepe}, Jeyhan S. and {Yang}, Lilan and {Koekemoer}, Anton M. and {McKinney}, Jed and {Robertson}, Brant E. and {Jin}, Shuowen and {Hayward}, Christopher C. and {Hirschmann}, Michaela and {Franco}, Maximilien and {Shuntov}, Marko and {Gozaliasl}, Ghassem and {Kaminsky}, Aidan and {Rich}, R. Michael},
        title = "{Clumps as multiscale structures in cosmic noon galaxies}",
      journal = {\mnras},
         year = 2025,
        month = jan,
       volume = {536},
       number = {3},
        pages = {3090-3111},
          doi = {10.1093/mnras/stae2781},
archivePrefix = {arXiv},
       eprint = {2501.03328},
 primaryClass = {astro-ph.GA},
       adsurl = {https://ui.adsabs.harvard.edu/abs/2025MNRAS.536.3090K}
}

@ARTICLE{Ferreira:2023,
       author = {{Ferreira}, Leonardo and {Conselice}, Christopher J. and {Sazonova}, Elizaveta and {Ferrari}, Fabricio and {Caruana}, Joseph and {Tohill}, Cl{\'a}r-Br{\'\i}d and {Lucatelli}, Geferson and {Adams}, Nathan and {Irodotou}, Dimitrios and {Marshall}, Madeline A. and {Roper}, Will J. and {Lovell}, Christopher C. and {Verma}, Aprajita and {Austin}, Duncan and {Trussler}, James and {Wilkins}, Stephen M.},
        title = "{The JWST Hubble Sequence: The Rest-frame Optical Evolution of Galaxy Structure at 1.5 < z < 6.5}",
      journal = {\apj},
         year = 2023,
        month = oct,
       volume = {955},
       number = {2},
          eid = {94},
        pages = {94},
          doi = {10.3847/1538-4357/acec76},
archivePrefix = {arXiv},
       eprint = {2210.01110},
 primaryClass = {astro-ph.GA},
       adsurl = {https://ui.adsabs.harvard.edu/abs/2023ApJ...955...94F}
}

@ARTICLE{Pilyugin:2012,
       author = {{Pilyugin}, L.~S. and {V{\'\i}lchez}, J.~M. and {Mattsson}, L. and {Thuan}, T.~X.},
        title = "{Abundance determination from global emission-line SDSS spectra: exploring objects with high N/O ratios}",
      journal = {\mnras},
         year = 2012,
        month = apr,
       volume = {421},
       number = {2},
        pages = {1624-1634},
          doi = {10.1111/j.1365-2966.2012.20420.x},
archivePrefix = {arXiv},
       eprint = {1201.1554},
 primaryClass = {astro-ph.CO},
       adsurl = {https://ui.adsabs.harvard.edu/abs/2012MNRAS.421.1624P}
}

@ARTICLE{Levesque:2013,
       author = {{Levesque}, Emily M. and {Leitherer}, Claus},
        title = "{Modeling Tracers of Young Stellar Population Age in Star-forming Galaxies}",
      journal = {\apj},
         year = 2013,
        month = dec,
       volume = {779},
       number = {2},
          eid = {170},
        pages = {170},
          doi = {10.1088/0004-637X/779/2/170},
archivePrefix = {arXiv},
       eprint = {1311.1202},
 primaryClass = {astro-ph.GA},
       adsurl = {https://ui.adsabs.harvard.edu/abs/2013ApJ...779..170L}
}

@ARTICLE{Matharu:2022,
       author = {{Matharu}, Jasleen and {Papovich}, Casey and {Simons}, Raymond C. and {Momcheva}, Ivelina and {Brammer}, Gabriel and {Ji}, Zhiyuan and {Backhaus}, Bren E. and {Cleri}, Nikko J. and {Estrada-Carpenter}, Vicente and {Finkelstein}, Steven L. and {Finlator}, Kristian and {Giavalisco}, Mauro and {Jung}, Intae and {Muzzin}, Adam and {Nelson}, Erica J. and {Pillepich}, Annalisa and {Trump}, Jonathan R. and {Weiner}, Benjamin},
        title = "{CLEAR: The Evolution of Spatially Resolved Star Formation in Galaxies between 0.5 {\ensuremath{\lesssim}} z {\ensuremath{\lesssim}} 1.7 Using H{\ensuremath{\alpha}} Emission Line Maps}",
      journal = {\apj},
         year = 2022,
        month = sep,
       volume = {937},
       number = {1},
          eid = {16},
        pages = {16},
          doi = {10.3847/1538-4357/ac8471},
archivePrefix = {arXiv},
       eprint = {2205.08543},
 primaryClass = {astro-ph.GA},
       adsurl = {https://ui.adsabs.harvard.edu/abs/2022ApJ...937...16M}
}

@ARTICLE{Price:2014,
       author = {{Price}, Sedona H. and {Kriek}, Mariska and {Brammer}, Gabriel B. and {Conroy}, Charlie and {F{\"o}rster Schreiber}, Natascha M. and {Franx}, Marijn and {Fumagalli}, Mattia and {Lundgren}, Britt and {Momcheva}, Ivelina and {Nelson}, Erica J. and {Skelton}, Rosalind E. and {van Dokkum}, Pieter G. and {Whitaker}, Katherine E. and {Wuyts}, Stijn},
        title = "{Direct Measurements of Dust Attenuation in z \raisebox{-0.5ex}\textasciitilde 1.5 Star-forming Galaxies from 3D-HST: Implications for Dust Geometry and Star Formation Rates}",
      journal = {\apj},
         year = 2014,
        month = jun,
       volume = {788},
       number = {1},
          eid = {86},
        pages = {86},
          doi = {10.1088/0004-637X/788/1/86},
archivePrefix = {arXiv},
       eprint = {1310.4177},
 primaryClass = {astro-ph.CO},
       adsurl = {https://ui.adsabs.harvard.edu/abs/2014ApJ...788...86P}
}

@ARTICLE{Nelson:2016,
       author = {{Nelson}, Erica June and {van Dokkum}, Pieter G. and {F{\"o}rster Schreiber}, Natascha M. and {Franx}, Marijn and {Brammer}, Gabriel B. and {Momcheva}, Ivelina G. and {Wuyts}, Stijn and {Whitaker}, Katherine E. and {Skelton}, Rosalind E. and {Fumagalli}, Mattia and {Hayward}, Christopher C. and {Kriek}, Mariska and {Labb{\'e}}, Ivo and {Leja}, Joel and {Rix}, Hans-Walter and {Tacconi}, Linda J. and {van der Wel}, Arjen and {van den Bosch}, Frank C. and {Oesch}, Pascal A. and {Dickey}, Claire and {Ulf Lange}, Johannes},
        title = "{Where Stars Form: Inside-out Growth and Coherent Star Formation from HST H{\ensuremath{\alpha}} Maps of 3200 Galaxies across the Main Sequence at 0.7 < z < 1.5}",
      journal = {\apj},
         year = 2016,
        month = sep,
       volume = {828},
       number = {1},
          eid = {27},
        pages = {27},
          doi = {10.3847/0004-637X/828/1/27},
archivePrefix = {arXiv},
       eprint = {1507.03999},
 primaryClass = {astro-ph.GA},
       adsurl = {https://ui.adsabs.harvard.edu/abs/2016ApJ...828...27N}
}

@ARTICLE{Reddy:2018,
       author = {{Reddy}, Naveen A. and {Shapley}, Alice E. and {Sanders}, Ryan L. and {Kriek}, Mariska and {Coil}, Alison L. and {Shivaei}, Irene and {Freeman}, William R. and {Mobasher}, Bahram and {Siana}, Brian and {Azadi}, Mojegan and {Fetherolf}, Tara and {Fornasini}, Francesca M. and {Leung}, Gene and {Price}, Sedona H. and {Zick}, Tom and {Barro}, Guillermo},
        title = "{The MOSDEF Survey: Significant Evolution in the Rest-frame Optical Emission Line Equivalent Widths of Star-forming Galaxies at z = 1.4-3.8}",
      journal = {\apj},
         year = 2018,
        month = dec,
       volume = {869},
       number = {2},
          eid = {92},
        pages = {92},
          doi = {10.3847/1538-4357/aaed1e},
archivePrefix = {arXiv},
       eprint = {1811.11767},
 primaryClass = {astro-ph.GA},
       adsurl = {https://ui.adsabs.harvard.edu/abs/2018ApJ...869...92R}
}

@ARTICLE{PerezDiaz:2024,
       author = {{P{\'e}rez-D{\'\i}az}, Borja and {P{\'e}rez-Montero}, Enrique and {Fern{\'a}ndez-Ontiveros}, Juan A. and {V{\'\i}lchez}, Jos{\'e} M. and {Hern{\'a}n-Caballero}, Antonio and {Amor{\'\i}n}, Ricardo},
        title = "{Chemical abundances and deviations from the solar S/O ratio in the gas-phase interstellar medium of galaxies based on infrared emission lines}",
      journal = {\aap},
         year = 2024,
        month = may,
       volume = {685},
          eid = {A168},
        pages = {A168},
          doi = {10.1051/0004-6361/202348318},
archivePrefix = {arXiv},
       eprint = {2403.02903},
 primaryClass = {astro-ph.GA},
       adsurl = {https://ui.adsabs.harvard.edu/abs/2024A&A...685A.168P}
}

@ARTICLE{Izotov:2006,
       author = {{Izotov}, Y.~I. and {Stasi{\'n}ska}, G. and {Meynet}, G. and {Guseva}, N.~G. and {Thuan}, T.~X.},
        title = "{The chemical composition of metal-poor emission-line galaxies in the Data Release 3 of the Sloan Digital Sky Survey}",
      journal = {\aap},
         year = 2006,
        month = mar,
       volume = {448},
       number = {3},
        pages = {955-970},
          doi = {10.1051/0004-6361:20053763},
archivePrefix = {arXiv},
       eprint = {astro-ph/0511644},
 primaryClass = {astro-ph},
       adsurl = {https://ui.adsabs.harvard.edu/abs/2006A&A...448..955I}
}

@ARTICLE{Topping:2025,
       author = {{Topping}, Michael W. and {Sanders}, Ryan L. and {Shapley}, Alice E. and {Pahl}, Anthony J. and {Reddy}, Naveen A. and {Stark}, Daniel P. and {Berg}, Danielle A. and {Clarke}, Leonardo and {Cullen}, Fergus and {Dunlop}, James S. and {Ellis}, Richard S. and {Schreiber}, N.~M. F{\"o}rster and {Illingworth}, Garth D. and {Jones}, Tucker and {Narayanan}, Desika and {Pettini}, Max and {Schaerer}, Daniel},
        title = "{The AURORA survey: the evolution of multiphase electron densities at high redshift}",
      journal = {\mnras},
         year = 2025,
        month = aug,
       volume = {541},
       number = {2},
        pages = {1707-1721},
          doi = {10.1093/mnras/staf903},
archivePrefix = {arXiv},
       eprint = {2502.08712},
 primaryClass = {astro-ph.GA},
       adsurl = {https://ui.adsabs.harvard.edu/abs/2025MNRAS.541.1707T}
}

@ARTICLE{Davies:2021,
       author = {{Davies}, Rebecca L. and {F{\"o}rster Schreiber}, N.~M. and {Genzel}, R. and {Shimizu}, T.~T. and {Davies}, R.~I. and {Schruba}, A. and {Tacconi}, L.~J. and {{\"U}bler}, H. and {Wisnioski}, E. and {Wuyts}, S. and {Fossati}, M. and {Herrera-Camus}, R. and {Lutz}, D. and {Mendel}, J.~T. and {Naab}, T. and {Price}, S.~H. and {Renzini}, A. and {Wilman}, D. and {Beifiori}, A. and {Belli}, S. and {Burkert}, A. and {Chan}, J. and {Contursi}, A. and {Fabricius}, M. and {Lee}, M.~M. and {Saglia}, R.~P. and {Sternberg}, A.},
        title = "{The KMOS$^{3D}$ Survey: Investigating the Origin of the Elevated Electron Densities in Star-forming Galaxies at 1 {\ensuremath{\lesssim}} z {\ensuremath{\lesssim}} 3}",
      journal = {\apj},
         year = 2021,
        month = mar,
       volume = {909},
       number = {1},
          eid = {78},
        pages = {78},
          doi = {10.3847/1538-4357/abd551},
archivePrefix = {arXiv},
       eprint = {2012.10445},
 primaryClass = {astro-ph.GA},
       adsurl = {https://ui.adsabs.harvard.edu/abs/2021ApJ...909...78D}
}

@ARTICLE{Jakobsen:2022,
       author = {{Jakobsen}, P. and {Ferruit}, P. and {Alves de Oliveira}, C. and {Arribas}, S. and {Bagnasco}, G. and {Barho}, R. and {Beck}, T.~L. and {Birkmann}, S. and {B{\"o}ker}, T. and {Bunker}, A.~J. and {Charlot}, S. and {de Jong}, P. and {de Marchi}, G. and {Ehrenwinkler}, R. and {Falcolini}, M. and {Fels}, R. and {Franx}, M. and {Franz}, D. and {Funke}, M. and {Giardino}, G. and {Gnata}, X. and {Holota}, W. and {Honnen}, K. and {Jensen}, P.~L. and {Jentsch}, M. and {Johnson}, T. and {Jollet}, D. and {Karl}, H. and {Kling}, G. and {K{\"o}hler}, J. and {Kolm}, M. -G. and {Kumari}, N. and {Lander}, M.~E. and {Lemke}, R. and {L{\'o}pez-Caniego}, M. and {L{\"u}tzgendorf}, N. and {Maiolino}, R. and {Manjavacas}, E. and {Marston}, A. and {Maschmann}, M. and {Maurer}, R. and {Messerschmidt}, B. and {Moseley}, S.~H. and {Mosner}, P. and {Mott}, D.~B. and {Muzerolle}, J. and {Pirzkal}, N. and {Pittet}, J. -F. and {Plitzke}, A. and {Posselt}, W. and {Rapp}, B. and {Rauscher}, B.~J. and {Rawle}, T. and {Rix}, H. -W. and {R{\"o}del}, A. and {Rumler}, P. and {Sabbi}, E. and {Salvignol}, J. -C. and {Schmid}, T. and {Sirianni}, M. and {Smith}, C. and {Strada}, P. and {te Plate}, M. and {Valenti}, J. and {Wettemann}, T. and {Wiehe}, T. and {Wiesmayer}, M. and {Willott}, C.~J. and {Wright}, R. and {Zeidler}, P. and {Zincke}, C.},
        title = "{The Near-Infrared Spectrograph (NIRSpec) on the James Webb Space Telescope. I. Overview of the instrument and its capabilities}",
      journal = {\aap},
         year = 2022,
        month = may,
       volume = {661},
          eid = {A80},
        pages = {A80},
          doi = {10.1051/0004-6361/202142663},
archivePrefix = {arXiv},
       eprint = {2202.03305},
 primaryClass = {astro-ph.IM},
       adsurl = {https://ui.adsabs.harvard.edu/abs/2022A&A...661A..80J}
}

@article{Planck:2015,
	adsurl = {https://ui.adsabs.harvard.edu/abs/2016A&A...594A..13P},
	archiveprefix = {arXiv},
	author = {{Planck Collaboration} and {Ade}, P.~A.~R. and {Aghanim}, N. and {Arnaud}, M. and {Ashdown}, M. and {Aumont}, J. and {Baccigalupi}, C. and {Banday}, A.~J. and {Barreiro}, R.~B. and {Bartlett}, J.~G. and {Bartolo}, N. and {Battaner}, E. and {Battye}, R. and {Benabed}, K. and {Beno{\^\i}t}, A. and {Benoit-L{\'e}vy}, A. and {Bernard}, J. -P. and {Bersanelli}, M. and {Bielewicz}, P. and {Bock}, J.~J. and {Bonaldi}, A. and {Bonavera}, L. and {Bond}, J.~R. and {Borrill}, J. and {Bouchet}, F.~R. and {Boulanger}, F. and {Bucher}, M. and {Burigana}, C. and {Butler}, R.~C. and {Calabrese}, E. and {Cardoso}, J. -F. and {Catalano}, A. and {Challinor}, A. and {Chamballu}, A. and {Chary}, R. -R. and {Chiang}, H.~C. and {Chluba}, J. and {Christensen}, P.~R. and {Church}, S. and {Clements}, D.~L. and {Colombi}, S. and {Colombo}, L.~P.~L. and {Combet}, C. and {Coulais}, A. and {Crill}, B.~P. and {Curto}, A. and {Cuttaia}, F. and {Danese}, L. and {Davies}, R.~D. and {Davis}, R.~J. and {de Bernardis}, P. and {de Rosa}, A. and {de Zotti}, G. and {Delabrouille}, J. and {D{\'e}sert}, F. -X. and {Di Valentino}, E. and {Dickinson}, C. and {Diego}, J.~M. and {Dolag}, K. and {Dole}, H. and {Donzelli}, S. and {Dor{\'e}}, O. and {Douspis}, M. and {Ducout}, A. and {Dunkley}, J. and {Dupac}, X. and {Efstathiou}, G. and {Elsner}, F. and {En{\ss}lin}, T.~A. and {Eriksen}, H.~K. and {Farhang}, M. and {Fergusson}, J. and {Finelli}, F. and {Forni}, O. and {Frailis}, M. and {Fraisse}, A.~A. and {Franceschi}, E. and {Frejsel}, A. and {Galeotta}, S. and {Galli}, S. and {Ganga}, K. and {Gauthier}, C. and {Gerbino}, M. and {Ghosh}, T. and {Giard}, M. and {Giraud-H{\'e}raud}, Y. and {Giusarma}, E. and {Gjerl{\o}w}, E. and {Gonz{\'a}lez-Nuevo}, J. and {G{\'o}rski}, K.~M. and {Gratton}, S. and {Gregorio}, A. and {Gruppuso}, A. and {Gudmundsson}, J.~E. and {Hamann}, J. and {Hansen}, F.~K. and {Hanson}, D. and {Harrison}, D.~L. and {Helou}, G. and {Henrot-Versill{\'e}}, S. and {Hern{\'a}ndez-Monteagudo}, C. and {Herranz}, D. and {Hildebrandt}, S.~R. and {Hivon}, E. and {Hobson}, M. and {Holmes}, W.~A. and {Hornstrup}, A. and {Hovest}, W. and {Huang}, Z. and {Huffenberger}, K.~M. and {Hurier}, G. and {Jaffe}, A.~H. and {Jaffe}, T.~R. and {Jones}, W.~C. and {Juvela}, M. and {Keih{\"a}nen}, E. and {Keskitalo}, R. and {Kisner}, T.~S. and {Kneissl}, R. and {Knoche}, J. and {Knox}, L. and {Kunz}, M. and {Kurki-Suonio}, H. and {Lagache}, G. and {L{\"a}hteenm{\"a}ki}, A. and {Lamarre}, J. -M. and {Lasenby}, A. and {Lattanzi}, M. and {Lawrence}, C.~R. and {Leahy}, J.~P. and {Leonardi}, R. and {Lesgourgues}, J. and {Levrier}, F. and {Lewis}, A. and {Liguori}, M. and {Lilje}, P.~B. and {Linden-V{\o}rnle}, M. and {L{\'o}pez-Caniego}, M. and {Lubin}, P.~M. and {Mac{\'\i}as-P{\'e}rez}, J.~F. and {Maggio}, G. and {Maino}, D. and {Mandolesi}, N. and {Mangilli}, A. and {Marchini}, A. and {Maris}, M. and {Martin}, P.~G. and {Martinelli}, M. and {Mart{\'\i}nez-Gonz{\'a}lez}, E. and {Masi}, S. and {Matarrese}, S. and {McGehee}, P. and {Meinhold}, P.~R. and {Melchiorri}, A. and {Melin}, J. -B. and {Mendes}, L. and {Mennella}, A. and {Migliaccio}, M. and {Millea}, M. and {Mitra}, S. and {Miville-Desch{\^e}nes}, M. -A. and {Moneti}, A. and {Montier}, L. and {Morgante}, G. and {Mortlock}, D. and {Moss}, A. and {Munshi}, D. and {Murphy}, J.~A. and {Naselsky}, P. and {Nati}, F. and {Natoli}, P. and {Netterfield}, C.~B. and {N{\o}rgaard-Nielsen}, H.~U. and {Noviello}, F. and {Novikov}, D. and {Novikov}, I. and {Oxborrow}, C.~A. and {Paci}, F. and {Pagano}, L. and {Pajot}, F. and {Paladini}, R. and {Paoletti}, D. and {Partridge}, B. and {Pasian}, F. and {Patanchon}, G. and {Pearson}, T.~J. and {Perdereau}, O. and {Perotto}, L. and {Perrotta}, F. and {Pettorino}, V. and {Piacentini}, F. and {Piat}, M. and {Pierpaoli}, E. and {Pietrobon}, D. and {Plaszczynski}, S. and {Pointecouteau}, E. and {Polenta}, G. and {Popa}, L. and {Pratt}, G.~W. and {Pr{\'e}zeau}, G. and {Prunet}, S. and {Puget}, J. -L. and {Rachen}, J.~P. and {Reach}, W.~T. and {Rebolo}, R. and {Reinecke}, M. and {Remazeilles}, M. and {Renault}, C. and {Renzi}, A. and {Ristorcelli}, I. and {Rocha}, G. and {Rosset}, C. and {Rossetti}, M. and {Roudier}, G. and {Rouill{\'e} d'Orfeuil}, B. and {Rowan-Robinson}, M. and {Rubi{\~n}o-Mart{\'\i}n}, J.~A. and {Rusholme}, B. and {Said}, N. and {Salvatelli}, V. and {Salvati}, L. and {Sandri}, M. and {Santos}, D. and {Savelainen}, M. and {Savini}, G. and {Scott}, D. and {Seiffert}, M.~D. and {Serra}, P. and {Shellard}, E.~P.~S. and {Spencer}, L.~D. and {Spinelli}, M. and {Stolyarov}, V. and {Stompor}, R. and {Sudiwala}, R. and {Sunyaev}, R. and {Sutton}, D. and {Suur-Uski}, A. -S. and {Sygnet}, J. -F. and {Tauber}, J.~A. and {Terenzi}, L. and {Toffolatti}, L. and {Tomasi}, M. and {Tristram}, M. and {Trombetti}, T. and {Tucci}, M. and {Tuovinen}, J. and {T{\"u}rler}, M. and {Umana}, G. and {Valenziano}, L. and {Valiviita}, J. and {Van Tent}, F. and {Vielva}, P. and {Villa}, F. and {Wade}, L.~A. and {Wandelt}, B.~D. and {Wehus}, I.~K. and {White}, M. and {White}, S.~D.~M. and {Wilkinson}, A. and {Yvon}, D. and {Zacchei}, A. and {Zonca}, A.},
	doi = {10.1051/0004-6361/201525830},
	eid = {A13},
	eprint = {1502.01589},
	journal = {\aap},
	month = sep,
	pages = {A13},
	primaryclass = {astro-ph.CO},
	title = {{Planck 2015 results. XIII. Cosmological parameters}},
	volume = {594},
	year = 2016}

@ARTICLE{Madau:2014,
       author = {{Madau}, Piero and {Dickinson}, Mark},
        title = "{Cosmic Star-Formation History}",
      journal = {\araa},
         year = 2014,
        month = aug,
       volume = {52},
        pages = {415-486},
          doi = {10.1146/annurev-astro-081811-125615},
archivePrefix = {arXiv},
       eprint = {1403.0007},
 primaryClass = {astro-ph.CO},
       adsurl = {https://ui.adsabs.harvard.edu/abs/2014ARA&A..52..415M}
}

@ARTICLE{Ashby:2013,
       author = {{Ashby}, M.~L.~N. and {Willner}, S.~P. and {Fazio}, G.~G. and {Huang}, J. -S. and {Arendt}, R. and {Barmby}, P. and {Barro}, G. and {Bell}, E.~F. and {Bouwens}, R. and {Cattaneo}, A. and {Croton}, D. and {Dav{\'e}}, R. and {Dunlop}, J.~S. and {Egami}, E. and {Faber}, S. and {Finlator}, K. and {Grogin}, N.~A. and {Guhathakurta}, P. and {Hernquist}, L. and {Hora}, J.~L. and {Illingworth}, G. and {Kashlinsky}, A. and {Koekemoer}, A.~M. and {Koo}, D.~C. and {Labb{\'e}}, I. and {Li}, Y. and {Lin}, L. and {Moseley}, H. and {Nandra}, K. and {Newman}, J. and {Noeske}, K. and {Ouchi}, M. and {Peth}, M. and {Rigopoulou}, D. and {Robertson}, B. and {Sarajedini}, V. and {Simard}, L. and {Smith}, H.~A. and {Wang}, Z. and {Wechsler}, R. and {Weiner}, B. and {Wilson}, G. and {Wuyts}, S. and {Yamada}, T. and {Yan}, H.},
        title = "{SEDS: The Spitzer Extended Deep Survey. Survey Design, Photometry, and Deep IRAC Source Counts}",
      journal = {\apj},
         year = 2013,
        month = may,
       volume = {769},
       number = {1},
          eid = {80},
        pages = {80},
          doi = {10.1088/0004-637X/769/1/80},
       adsurl = {https://ui.adsabs.harvard.edu/abs/2013ApJ...769...80A}
}

@ARTICLE{Dale:2014,
       author = {{Dale}, Daniel A. and {Helou}, George and {Magdis}, Georgios E. and {Armus}, Lee and {D{\'\i}az-Santos}, Tanio and {Shi}, Yong},
        title = "{A Two-parameter Model for the Infrared/Submillimeter/Radio Spectral Energy Distributions of Galaxies and Active Galactic Nuclei}",
      journal = {\apj},
         year = 2014,
        month = mar,
       volume = {784},
       number = {1},
          eid = {83},
        pages = {83},
          doi = {10.1088/0004-637X/784/1/83},
archivePrefix = {arXiv},
       eprint = {1402.1495},
 primaryClass = {astro-ph.GA},
       adsurl = {https://ui.adsabs.harvard.edu/abs/2014ApJ...784...83D}
}

@ARTICLE{Calzetti:2000,
       author = {{Calzetti}, Daniela and {Armus}, Lee and {Bohlin}, Ralph C. and {Kinney}, Anne L. and {Koornneef}, Jan and {Storchi-Bergmann}, Thaisa},
        title = "{The Dust Content and Opacity of Actively Star-forming Galaxies}",
      journal = {\apj},
         year = 2000,
        month = apr,
       volume = {533},
       number = {2},
        pages = {682-695},
          doi = {10.1086/308692},
archivePrefix = {arXiv},
       eprint = {astro-ph/9911459},
 primaryClass = {astro-ph},
       adsurl = {https://ui.adsabs.harvard.edu/abs/2000ApJ...533..682C}
}

@INPROCEEDINGS{Dickinson:2003,
       author = {{Dickinson}, Mark and {Giavalisco}, Mauro and {GOODS Team}},
        title = "{The Great Observatories Origins Deep Survey}",
    booktitle = {The Mass of Galaxies at Low and High Redshift},
         year = 2003,
       editor = {{Bender}, Ralf and {Renzini}, Alvio},
        month = jan,
        pages = {324},
          doi = {10.1007/10899892_78},
archivePrefix = {arXiv},
       eprint = {astro-ph/0204213},
 primaryClass = {astro-ph},
       adsurl = {https://ui.adsabs.harvard.edu/abs/2003mglh.conf..324D}
}

@ARTICLE{Reddy:2023a,
       author = {{Reddy}, Naveen A. and {Topping}, Michael W. and {Sanders}, Ryan L. and {Shapley}, Alice E. and {Brammer}, Gabriel},
        title = "{Paschen-line Constraints on Dust Attenuation and Star Formation at z   1-3 with JWST/NIRSpec}",
      journal = {\apj},
         year = 2023,
        month = may,
       volume = {948},
       number = {2},
          eid = {83},
        pages = {83},
          doi = {10.3847/1538-4357/acc869},
archivePrefix = {arXiv},
       eprint = {2301.07249},
 primaryClass = {astro-ph.GA},
       adsurl = {https://ui.adsabs.harvard.edu/abs/2023ApJ...948...83R}
}

@ARTICLE{Reddy:2023,
       author = {{Reddy}, Naveen A. and {Topping}, Michael W. and {Sanders}, Ryan L. and {Shapley}, Alice E. and {Brammer}, Gabriel},
        title = "{A JWST/NIRSpec Exploration of the Connection between Ionization Parameter, Electron Density, and Star-formation-rate Surface Density in z = 2.7-6.3 Galaxies}",
      journal = {\apj},
         year = 2023,
        month = aug,
       volume = {952},
       number = {2},
          eid = {167},
        pages = {167},
          doi = {10.3847/1538-4357/acd754},
archivePrefix = {arXiv},
       eprint = {2303.11397},
 primaryClass = {astro-ph.GA},
       adsurl = {https://ui.adsabs.harvard.edu/abs/2023ApJ...952..167R}
}

@ARTICLE{Giavalisco:2004,
       author = {{Giavalisco}, M. and {Ferguson}, H.~C. and {Koekemoer}, A.~M. and {Dickinson}, M. and {Alexander}, D.~M. and {Bauer}, F.~E. and {Bergeron}, J. and {Biagetti}, C. and {Brandt}, W.~N. and {Casertano}, S. and {Cesarsky}, C. and {Chatzichristou}, E. and {Conselice}, C. and {Cristiani}, S. and {Da Costa}, L. and {Dahlen}, T. and {de Mello}, D. and {Eisenhardt}, P. and {Erben}, T. and {Fall}, S.~M. and {Fassnacht}, C. and {Fosbury}, R. and {Fruchter}, A. and {Gardner}, J.~P. and {Grogin}, N. and {Hook}, R.~N. and {Hornschemeier}, A.~E. and {Idzi}, R. and {Jogee}, S. and {Kretchmer}, C. and {Laidler}, V. and {Lee}, K.~S. and {Livio}, M. and {Lucas}, R. and {Madau}, P. and {Mobasher}, B. and {Moustakas}, L.~A. and {Nonino}, M. and {Padovani}, P. and {Papovich}, C. and {Park}, Y. and {Ravindranath}, S. and {Renzini}, A. and {Richardson}, M. and {Riess}, A. and {Rosati}, P. and {Schirmer}, M. and {Schreier}, E. and {Somerville}, R.~S. and {Spinrad}, H. and {Stern}, D. and {Stiavelli}, M. and {Strolger}, L. and {Urry}, C.~M. and {Vandame}, B. and {Williams}, R. and {Wolf}, C.},
        title = "{The Great Observatories Origins Deep Survey: Initial Results from Optical and Near-Infrared Imaging}",
      journal = {\apjl},
         year = 2004,
        month = jan,
       volume = {600},
       number = {2},
        pages = {L93-L98},
          doi = {10.1086/379232},
archivePrefix = {arXiv},
       eprint = {astro-ph/0309105},
 primaryClass = {astro-ph},
       adsurl = {https://ui.adsabs.harvard.edu/abs/2004ApJ...600L..93G}
}

@ARTICLE{Sanders:2016,
       author = {{Sanders}, Ryan L. and {Shapley}, Alice E. and {Kriek}, Mariska and {Reddy}, Naveen A. and {Freeman}, William R. and {Coil}, Alison L. and {Siana}, Brian and {Mobasher}, Bahram and {Shivaei}, Irene and {Price}, Sedona H. and {de Groot}, Laura},
        title = "{The MOSDEF Survey: Electron Density and Ionization Parameter at z \raisebox{-0.5ex}\textasciitilde 2.3}",
      journal = {\apj},
         year = 2016,
        month = jan,
       volume = {816},
       number = {1},
          eid = {23},
        pages = {23},
          doi = {10.3847/0004-637X/816/1/23},
archivePrefix = {arXiv},
       eprint = {1509.03636},
 primaryClass = {astro-ph.GA},
       adsurl = {https://ui.adsabs.harvard.edu/abs/2016ApJ...816...23S}
}

@ARTICLE{Curti:2020,
       author = {{Curti}, Mirko and {Mannucci}, Filippo and {Cresci}, Giovanni and {Maiolino}, Roberto},
        title = "{The mass-metallicity and the fundamental metallicity relation revisited on a fully T$_{e}$-based abundance scale for galaxies}",
      journal = {\mnras},
         year = 2020,
        month = jan,
       volume = {491},
       number = {1},
        pages = {944-964},
          doi = {10.1093/mnras/stz2910},
archivePrefix = {arXiv},
       eprint = {1910.00597},
 primaryClass = {astro-ph.GA},
       adsurl = {https://ui.adsabs.harvard.edu/abs/2020MNRAS.491..944C}
}

@ARTICLE{Davies:2023,
       author = {{Davies}, R. and {Absil}, O. and {Agapito}, G. and {Agudo Berbel}, A. and {Baruffolo}, A. and {Biliotti}, V. and {Black}, M. and {Bonaglia}, M. and {Bonse}, M. and {Briguglio}, R. and {Campana}, P. and {Cao}, Y. and {Carbonaro}, L. and {Cortes}, A. and {Cresci}, G. and {Dallilar}, Y. and {Dannert}, F. and {De Rosa}, R.~J. and {Deysenroth}, M. and {Di Antonio}, I. and {Di Cianno}, A. and {Di Rico}, G. and {Doelman}, D. and {Dolci}, M. and {Dorn}, R. and {Eisenhauer}, F. and {Esposito}, S. and {Fantinel}, D. and {Ferruzzi}, D. and {Feuchtgruber}, H. and {Finger}, G. and {F{\"o}rster Schreiber}, N.~M. and {Gao}, X. and {Gemperlein}, H. and {Genzel}, R. and {Gillessen}, S. and {Ginski}, C. and {Glauser}, A.~M. and {Glindemann}, A. and {Grani}, P. and {Hartl}, M. and {Hayoz}, J. and {Heida}, M. and {Henry}, D. and {Hofmann}, R. and {Huber}, H. and {Kasper}, M. and {Keller}, C. and {Kenworthy}, M. and {Kravchenko}, K. and {Kuntschner}, H. and {Lacour}, S. and {Lightfoot}, J. and {Lunney}, D. and {Lutz}, D. and {Macintosh}, M. and {Mannucci}, F. and {Marsset}, M. and {Modigliani}, A. and {Neeser}, M. and {Orban de Xivry}, G. and {Ott}, T. and {Pallanca}, L. and {Patapis}, P. and {Pearson}, D. and {Pe{\~n}a}, E. and {Percheron}, I. and {Puglisi}, A. and {Quanz}, S.~P. and {Rabien}, S. and {Rau}, C. and {Riccardi}, A. and {Salasnich}, B. and {Schmid}, H. -M. and {Schubert}, J. and {Serra}, B. and {Shimizu}, T. and {Snik}, F. and {Sturm}, E. and {Tacconi}, L. and {Taylor}, W. and {Valentini}, A. and {Waring}, C. and {Wiezorrek}, E. and {Xompero}, M.},
        title = "{The Enhanced Resolution Imager and Spectrograph for the VLT}",
      journal = {\aap},
         year = 2023,
        month = jun,
       volume = {674},
          eid = {A207},
        pages = {A207},
          doi = {10.1051/0004-6361/202346559},
archivePrefix = {arXiv},
       eprint = {2304.02343},
 primaryClass = {astro-ph.IM},
       adsurl = {https://ui.adsabs.harvard.edu/abs/2023A&A...674A.207D}
}

@ARTICLE{Kennicutt:1998,
       author = {{Kennicutt}, Jr., Robert C.},
        title = "{Star Formation in Galaxies Along the Hubble Sequence}",
      journal = {\araa},
         year = 1998,
        month = jan,
       volume = {36},
        pages = {189-232},
          doi = {10.1146/annurev.astro.36.1.189},
archivePrefix = {arXiv},
       eprint = {astro-ph/9807187},
 primaryClass = {astro-ph},
       adsurl = {https://ui.adsabs.harvard.edu/abs/1998ARA&A..36..189K}
}

@ARTICLE{Kennicutt:2012,
       author = {{Kennicutt}, Robert C. and {Evans}, Neal J.},
        title = "{Star Formation in the Milky Way and Nearby Galaxies}",
      journal = {\araa},
         year = 2012,
        month = sep,
       volume = {50},
        pages = {531-608},
          doi = {10.1146/annurev-astro-081811-125610},
archivePrefix = {arXiv},
       eprint = {1204.3552},
 primaryClass = {astro-ph.GA},
       adsurl = {https://ui.adsabs.harvard.edu/abs/2012ARA&A..50..531K}
}

@article{Genzel:2011,
	adsurl = {https://ui.adsabs.harvard.edu/abs/2011ApJ...733..101G},
	archiveprefix = {arXiv},
	author = {{Genzel}, R. and {Newman}, S. and {Jones}, T. and {F{\"o}rster Schreiber}, N.~M. and {Shapiro}, K. and {Genel}, S. and {Lilly}, S.~J. and {Renzini}, A. and {Tacconi}, L.~J. and {Bouch{\'e}}, N. and {Burkert}, A. and {Cresci}, G. and {Buschkamp}, P. and {Carollo}, C.~M. and {Ceverino}, D. and {Davies}, R. and {Dekel}, A. and {Eisenhauer}, F. and {Hicks}, E. and {Kurk}, J. and {Lutz}, D. and {Mancini}, C. and {Naab}, T. and {Peng}, Y. and {Sternberg}, A. and {Vergani}, D. and {Zamorani}, G.},
	doi = {10.1088/0004-637X/733/2/101},
	eid = {101},
	eprint = {1011.5360},
	journal = {\apj},
	month = jun,
	number = {2},
	pages = {101},
	primaryclass = {astro-ph.CO},
	title = {{The Sins Survey of z \raisebox{-0.5ex}\textasciitilde 2 Galaxy Kinematics: Properties of the Giant Star-forming Clumps}},
	volume = {733},
	year = 2011}

@ARTICLE{Genzel:2014,
       author = {{Genzel}, R. and {F{\"o}rster Schreiber}, N.~M. and {Rosario}, D. and {Lang}, P. and {Lutz}, D. and {Wisnioski}, E. and {Wuyts}, E. and {Wuyts}, S. and {Bandara}, K. and {Bender}, R. and {Berta}, S. and {Kurk}, J. and {Mendel}, J.~T. and {Tacconi}, L.~J. and {Wilman}, D. and {Beifiori}, A. and {Brammer}, G. and {Burkert}, A. and {Buschkamp}, P. and {Chan}, J. and {Carollo}, C.~M. and {Davies}, R. and {Eisenhauer}, F. and {Fabricius}, M. and {Fossati}, M. and {Kriek}, M. and {Kulkarni}, S. and {Lilly}, S.~J. and {Mancini}, C. and {Momcheva}, I. and {Naab}, T. and {Nelson}, E.~J. and {Renzini}, A. and {Saglia}, R. and {Sharples}, R.~M. and {Sternberg}, A. and {Tacchella}, S. and {van Dokkum}, P.},
        title = "{Evidence for Wide-spread Active Galactic Nucleus-driven Outflows in the Most Massive z \raisebox{-0.5ex}\textasciitilde 1-2 Star-forming Galaxies}",
      journal = {\apj},
         year = 2014,
        month = nov,
       volume = {796},
       number = {1},
          eid = {7},
        pages = {7},
          doi = {10.1088/0004-637X/796/1/7},
archivePrefix = {arXiv},
       eprint = {1406.0183},
 primaryClass = {astro-ph.GA},
       adsurl = {https://ui.adsabs.harvard.edu/abs/2014ApJ...796....7G}
}

@article{Forster:2018,
	adsurl = {https://ui.adsabs.harvard.edu/abs/2018ApJS..238...21F},
	archiveprefix = {arXiv},
	author = {{F{\"o}rster Schreiber}, N.~M. and {Renzini}, A. and {Mancini}, C. and {Genzel}, R. and {Bouch{\'e}}, N. and {Cresci}, G. and {Hicks}, E.~K.~S. and {Lilly}, S.~J. and {Peng}, Y. and {Burkert}, A. and {Carollo}, C.~M. and {Cimatti}, A. and {Daddi}, E. and {Davies}, R.~I. and {Genel}, S. and {Kurk}, J.~D. and {Lang}, P. and {Lutz}, D. and {Mainieri}, V. and {McCracken}, H.~J. and {Mignoli}, M. and {Naab}, T. and {Oesch}, P. and {Pozzetti}, L. and {Scodeggio}, M. and {Shapiro Griffin}, K. and {Shapley}, A.~E. and {Sternberg}, A. and {Tacchella}, S. and {Tacconi}, L.~J. and {Wuyts}, S. and {Zamorani}, G.},
	doi = {10.3847/1538-4365/aadd49},
	eid = {21},
	eprint = {1802.07276},
	journal = {\apjs},
	month = oct,
	number = {2},
	pages = {21},
	primaryclass = {astro-ph.GA},
	title = {{The SINS/zC-SINF Survey of z {\ensuremath{\sim}} 2 Galaxy Kinematics: SINFONI Adaptive Optics-assisted Data and Kiloparsec-scale Emission-line Properties}},
	volume = {238},
	year = 2018}

@software{Brammer:2019,
       author = {{Brammer}, Gabe},
        title = "{Grizli: Grism redshift and line analysis software}",
 howpublished = {Astrophysics Source Code Library, record ascl:1905.001},
         year = 2019,
        month = may,
          eid = {ascl:1905.001},
       adsurl = {https://ui.adsabs.harvard.edu/abs/2019ascl.soft05001B}
}

@ARTICLE{Grogin:2011,
       author = {{Grogin}, Norman A. and {Kocevski}, Dale D. and {Faber}, S.~M. and {Ferguson}, Henry C. and {Koekemoer}, Anton M. and {Riess}, Adam G. and {Acquaviva}, Viviana and {Alexander}, David M. and {Almaini}, Omar and {Ashby}, Matthew L.~N. and {Barden}, Marco and {Bell}, Eric F. and {Bournaud}, Fr{\'e}d{\'e}ric and {Brown}, Thomas M. and {Caputi}, Karina I. and {Casertano}, Stefano and {Cassata}, Paolo and {Castellano}, Marco and {Challis}, Peter and {Chary}, Ranga-Ram and {Cheung}, Edmond and {Cirasuolo}, Michele and {Conselice}, Christopher J. and {Roshan Cooray}, Asantha and {Croton}, Darren J. and {Daddi}, Emanuele and {Dahlen}, Tomas and {Dav{\'e}}, Romeel and {de Mello}, Du{\'\i}lia F. and {Dekel}, Avishai and {Dickinson}, Mark and {Dolch}, Timothy and {Donley}, Jennifer L. and {Dunlop}, James S. and {Dutton}, Aaron A. and {Elbaz}, David and {Fazio}, Giovanni G. and {Filippenko}, Alexei V. and {Finkelstein}, Steven L. and {Fontana}, Adriano and {Gardner}, Jonathan P. and {Garnavich}, Peter M. and {Gawiser}, Eric and {Giavalisco}, Mauro and {Grazian}, Andrea and {Guo}, Yicheng and {Hathi}, Nimish P. and {H{\"a}ussler}, Boris and {Hopkins}, Philip F. and {Huang}, Jia-Sheng and {Huang}, Kuang-Han and {Jha}, Saurabh W. and {Kartaltepe}, Jeyhan S. and {Kirshner}, Robert P. and {Koo}, David C. and {Lai}, Kamson and {Lee}, Kyoung-Soo and {Li}, Weidong and {Lotz}, Jennifer M. and {Lucas}, Ray A. and {Madau}, Piero and {McCarthy}, Patrick J. and {McGrath}, Elizabeth J. and {McIntosh}, Daniel H. and {McLure}, Ross J. and {Mobasher}, Bahram and {Moustakas}, Leonidas A. and {Mozena}, Mark and {Nandra}, Kirpal and {Newman}, Jeffrey A. and {Niemi}, Sami-Matias and {Noeske}, Kai G. and {Papovich}, Casey J. and {Pentericci}, Laura and {Pope}, Alexandra and {Primack}, Joel R. and {Rajan}, Abhijith and {Ravindranath}, Swara and {Reddy}, Naveen A. and {Renzini}, Alvio and {Rix}, Hans-Walter and {Robaina}, Aday R. and {Rodney}, Steven A. and {Rosario}, David J. and {Rosati}, Piero and {Salimbeni}, Sara and {Scarlata}, Claudia and {Siana}, Brian and {Simard}, Luc and {Smidt}, Joseph and {Somerville}, Rachel S. and {Spinrad}, Hyron and {Straughn}, Amber N. and {Strolger}, Louis-Gregory and {Telford}, Olivia and {Teplitz}, Harry I. and {Trump}, Jonathan R. and {van der Wel}, Arjen and {Villforth}, Carolin and {Wechsler}, Risa H. and {Weiner}, Benjamin J. and {Wiklind}, Tommy and {Wild}, Vivienne and {Wilson}, Grant and {Wuyts}, Stijn and {Yan}, Hao-Jing and {Yun}, Min S.},
        title = "{CANDELS: The Cosmic Assembly Near-infrared Deep Extragalactic Legacy Survey}",
      journal = {\apjs},
         year = 2011,
        month = dec,
       volume = {197},
       number = {2},
          eid = {35},
        pages = {35},
          doi = {10.1088/0067-0049/197/2/35},
archivePrefix = {arXiv},
       eprint = {1105.3753},
 primaryClass = {astro-ph.CO},
       adsurl = {https://ui.adsabs.harvard.edu/abs/2011ApJS..197...35G}
}

@article{Forster:2009,
	adsurl = {https://ui.adsabs.harvard.edu/abs/2009ApJ...706.1364F},
	archiveprefix = {arXiv},
	author = {{F{\"o}rster Schreiber}, N.~M. and {Genzel}, R. and {Bouch{\'e}}, N. and {Cresci}, G. and {Davies}, R. and {Buschkamp}, P. and {Shapiro}, K. and {Tacconi}, L.~J. and {Hicks}, E.~K.~S. and {Genel}, S. and {Shapley}, A.~E. and {Erb}, D.~K. and {Steidel}, C.~C. and {Lutz}, D. and {Eisenhauer}, F. and {Gillessen}, S. and {Sternberg}, A. and {Renzini}, A. and {Cimatti}, A. and {Daddi}, E. and {Kurk}, J. and {Lilly}, S. and {Kong}, X. and {Lehnert}, M.~D. and {Nesvadba}, N. and {Verma}, A. and {McCracken}, H. and {Arimoto}, N. and {Mignoli}, M. and {Onodera}, M.},
	doi = {10.1088/0004-637X/706/2/1364},
	eprint = {0903.1872},
	journal = {\apj},
	month = dec,
	number = {2},
	pages = {1364-1428},
	primaryclass = {astro-ph.CO},
	title = {{The SINS Survey: SINFONI Integral Field Spectroscopy of z \raisebox{-0.5ex}\textasciitilde 2 Star-forming Galaxies}},
	volume = {706},
	year = 2009}

\begin{appendix}

\section{Spectral fitting of NIRSpec MSA data}\label{apx:spectral_fitting}

In this appendix,  we provide details on the spectral fitting of the NIRSpec MSA spectra of \target, at low spectral resolution and separately for the medium- and high-resolution gratings. In the PRISM/CLEAR spectra, stellar continuum is well detected and close emission lines are blended together (e.g., \hb+\,\oiii, H$\alpha$\,+\,\nii, \sii line doublet). On the contrary, in medium- and high-resolution data continuum emission is faint and the emission lines are separately detected, although spectrally resolved only with the high-resolution ($R$\,$\sim$\,2700) grating G395H/F290LP. The best-fit results obtained for the emission lines detected at $S/N$\,>\,3 are listed in Table \ref{tab:results_line_fitting}, separately for data at different spectral resolution. Lines simultaneously observed with different gratings have mostly fluxes consistent within the errors,  or they differ by less than $\sim$\,20\%, compatible with the known (undesired) differences in the flux calibration at different resolution \citep[e.g.,][]{Deugenio:2025}. For this reason, we caution on the absolute value of the inferred line fluxes, which should preferably be used for a relative comparison between different galaxy regions.

\subsection{Low-resolution spectra}

As shown in Fig. \ref{fig:Nircam_RGB_and_prism}, the PRISM/CLEAR spectra of \target exhibit a prominent continuum emission (especially Clump-N, Clump-S, and Inter-Arm), in addition to numerous (spectrally unresolved) emission lines. We fit the full (continuum plus emission lines) spectra by means of the Penalized PiXel-Fitting code (\texttt{pPXF}; \citealt{Cappellari:2023}), which models continuum emission through a linear combination of simple stellar population (SSP) models, and emission lines by means of Gaussian components. We use stellar templates from the MILES SSP library \citep{Vazdekis:2010}, and constrain all emission lines to have the same kinematics (i.e., velocity and velocity dispersion). The line ratios of \oiii$\lambda$5007/\oiii$\lambda$4959 and  \nii$\lambda$6584/\nii$\lambda$6548 are also fixed to 2.98 and 2.94, respectively, according to atomic physics prescriptions \citep[see][]{Osterbrock:2006}. The best-fit continuum emission is then used as a continuum input for the subsequent fitting of the emission lines.

After this first fitting with \texttt{pPXF}, we subtract the best-fit continuum model from the data and perform a refined modeling of the emission lines only, with a custom \texttt{python} script to properly account for the poor and variable spectral resolution of the PRISM/CLEAR data (varying from R$\sim$30 at $\sim$1.1$\mu$m to R$\sim$400 at the reddest wavelength; see \citealt{Jakobsen:2022}), and accurately estimate the flux of the emission lines and the relative error. As done in the previous fitting with \texttt{pPXF}, we constrain all lines to the same velocity and fit \oiii\ and \nii\ line doublets with a fixed line ratio. To account for the varying spectral resolution of the PRISM data, we tie the velocity dispersion only between lines that are close in wavelength (e.g., \oiii\ and \hb). Moreover, due to the poor spectral resolution of PRISM data (i.e., $R$\,=\,100), \nii\ and \ha\ emission lines appear blended together. Since they are separately resolved in the medium-resolution spectra and we aim to estimate the \ha\ flux for computing the \ha\ equivalent width (EW(\ha); see Sect. \ref{sec:stellar_age}), we fixed the \nii$\lambda$6584/\ha\ line ratio to either the value measured in the corresponding G235M/F170LP data from the same shutter (if detected) or to that measured in the G235M/F170LP data from the adjacent shutter. The \oiid\ and the \siid\ line doublets are also unresolved in the PRISM spectra. For simplicity, we fit each doublet with a single Gaussian component.

\begin{figure*}
    \centering
    \includegraphics[width=0.45\linewidth]{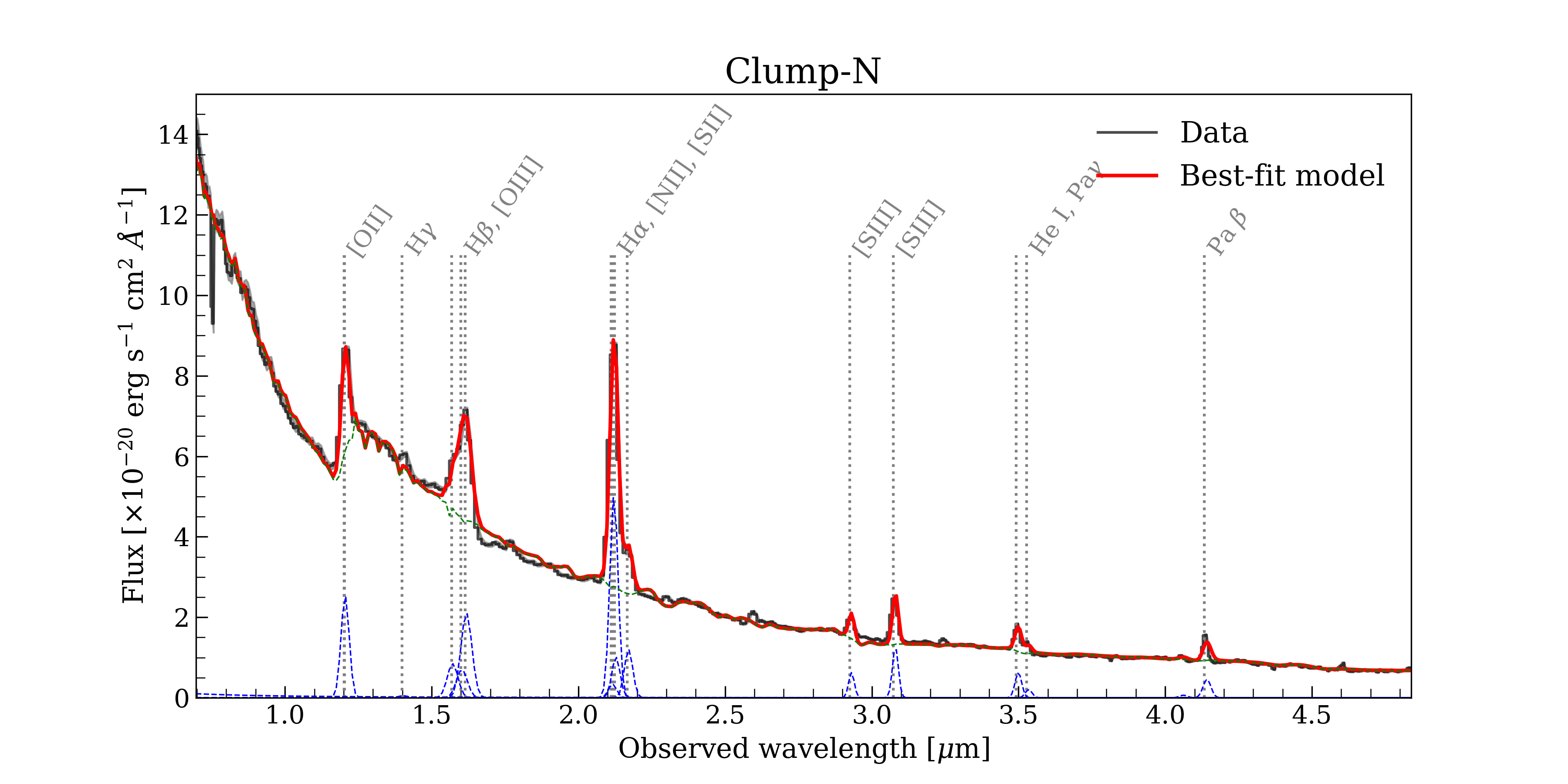}
    \includegraphics[width=0.45\linewidth]{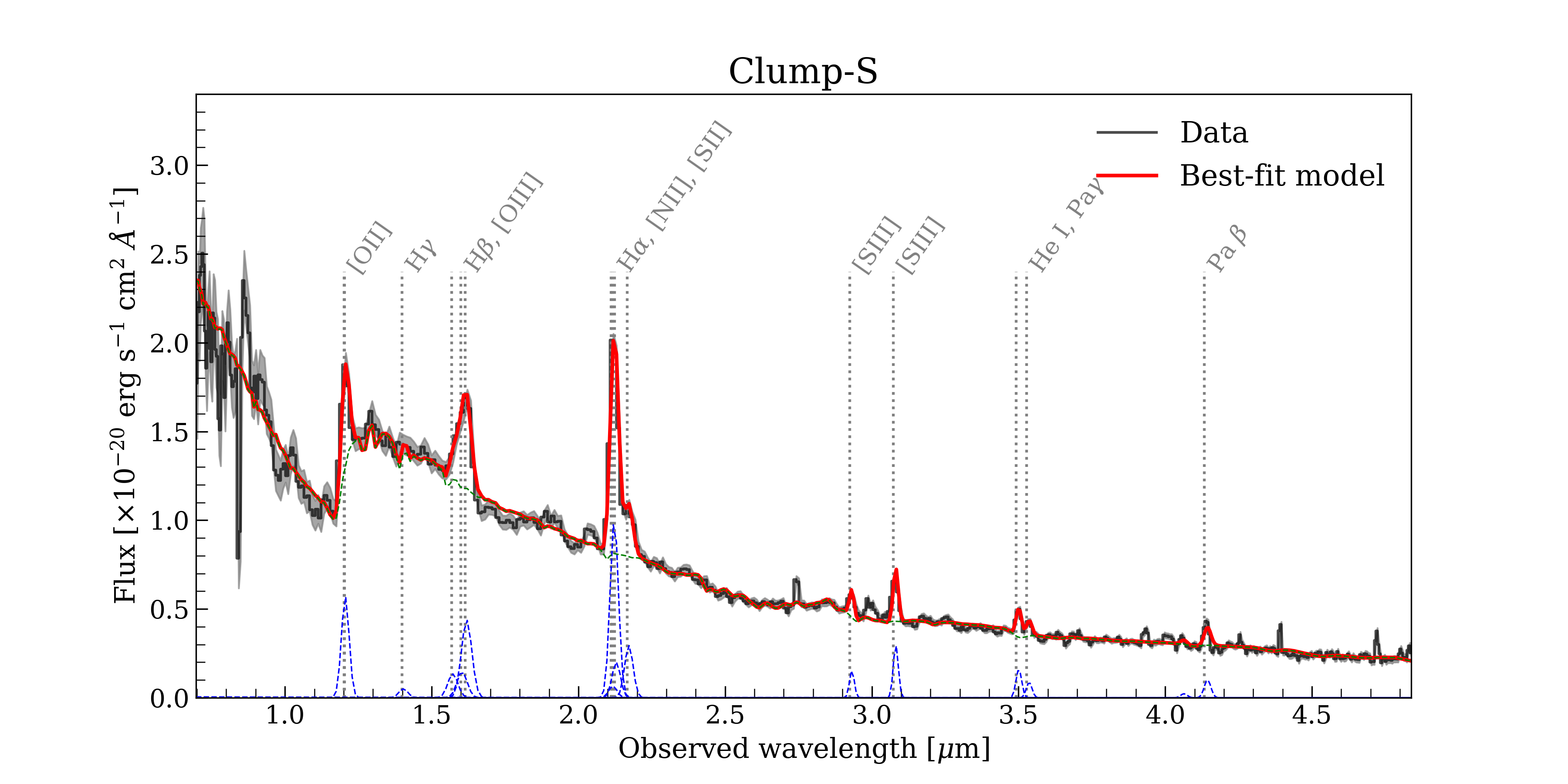}
    \includegraphics[width=0.45\linewidth]{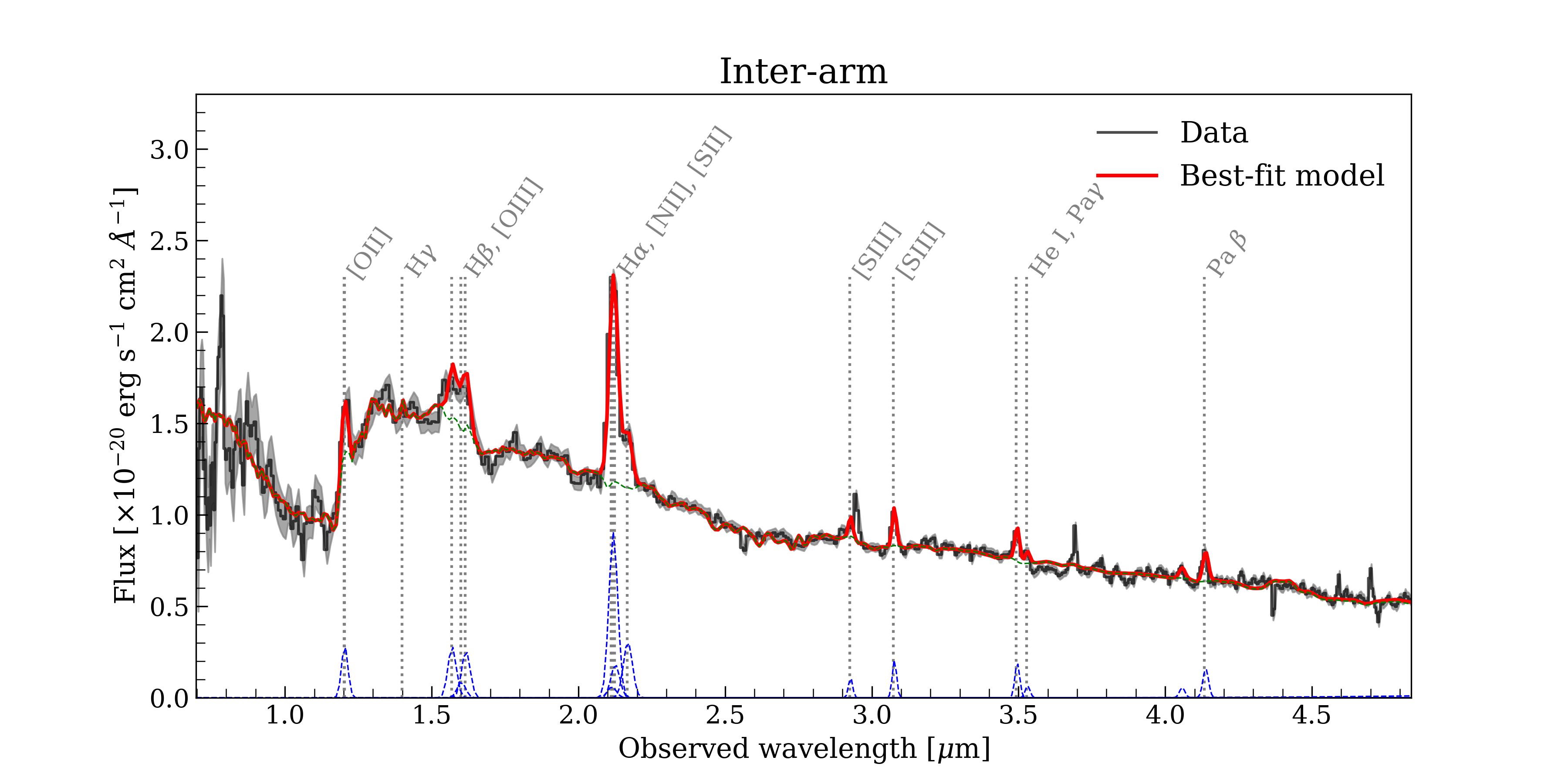}
    \includegraphics[width=0.45\linewidth]{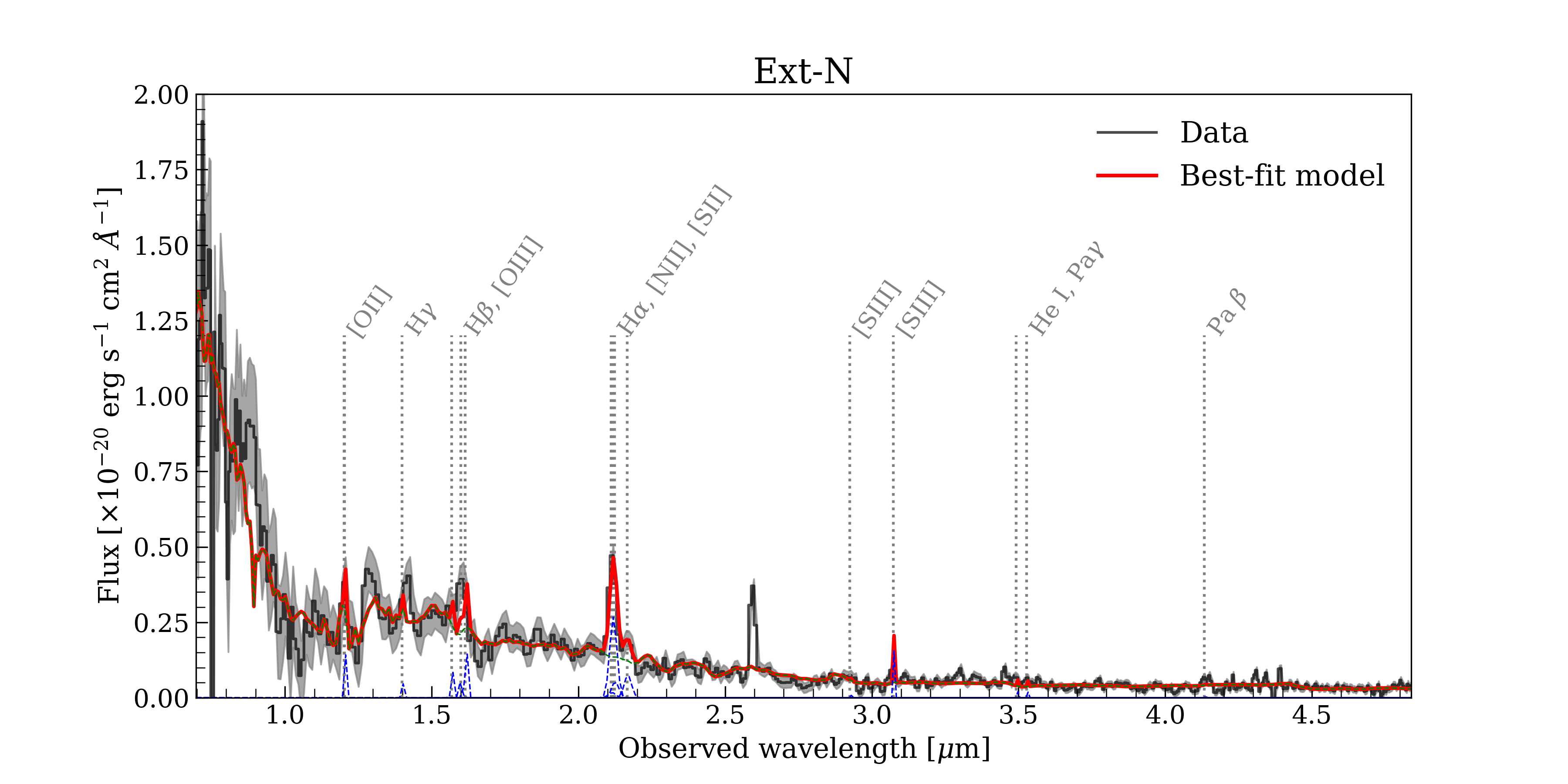}
    \includegraphics[width=0.45\linewidth]{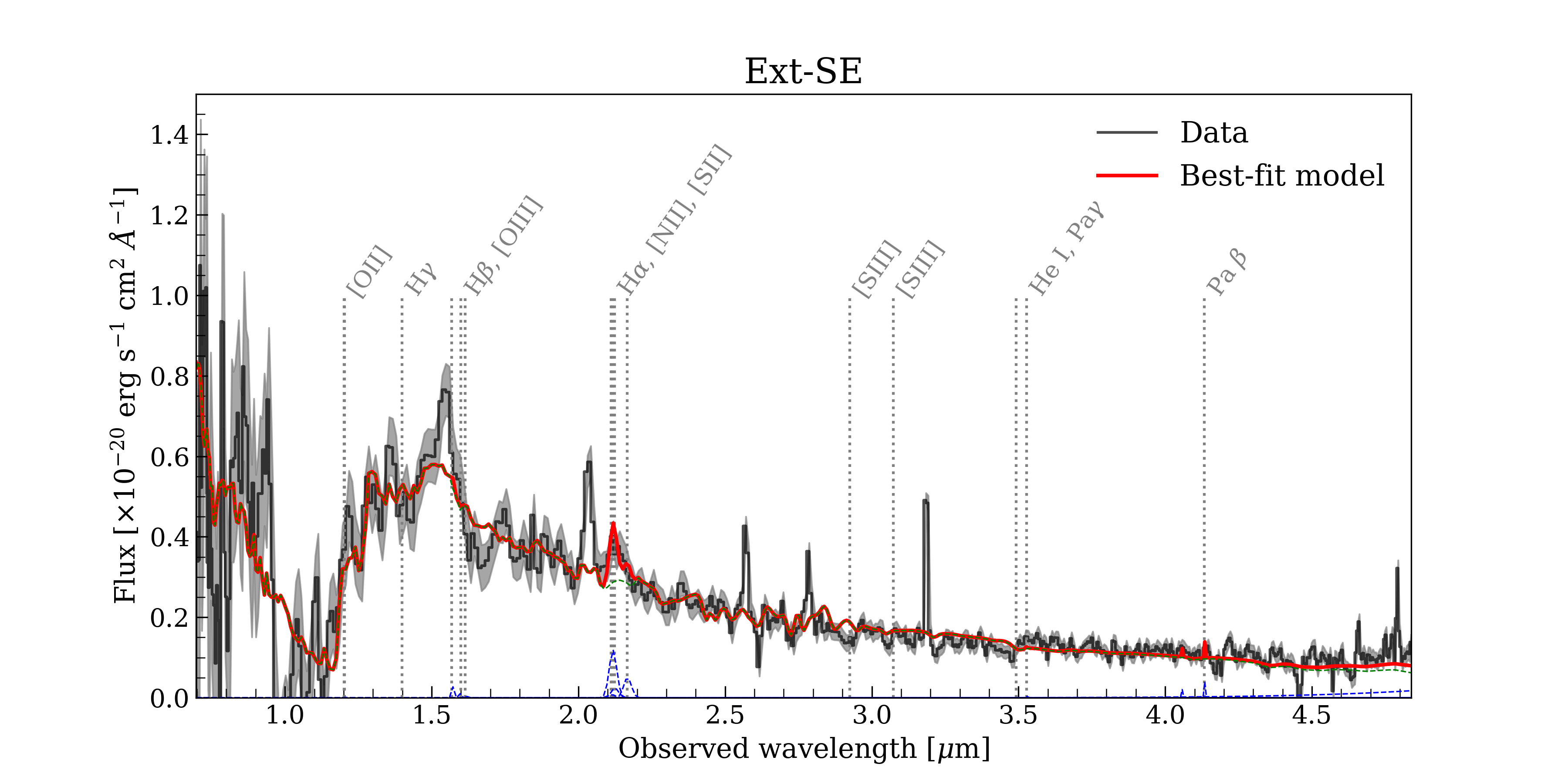}
    \includegraphics[width=0.45\linewidth]{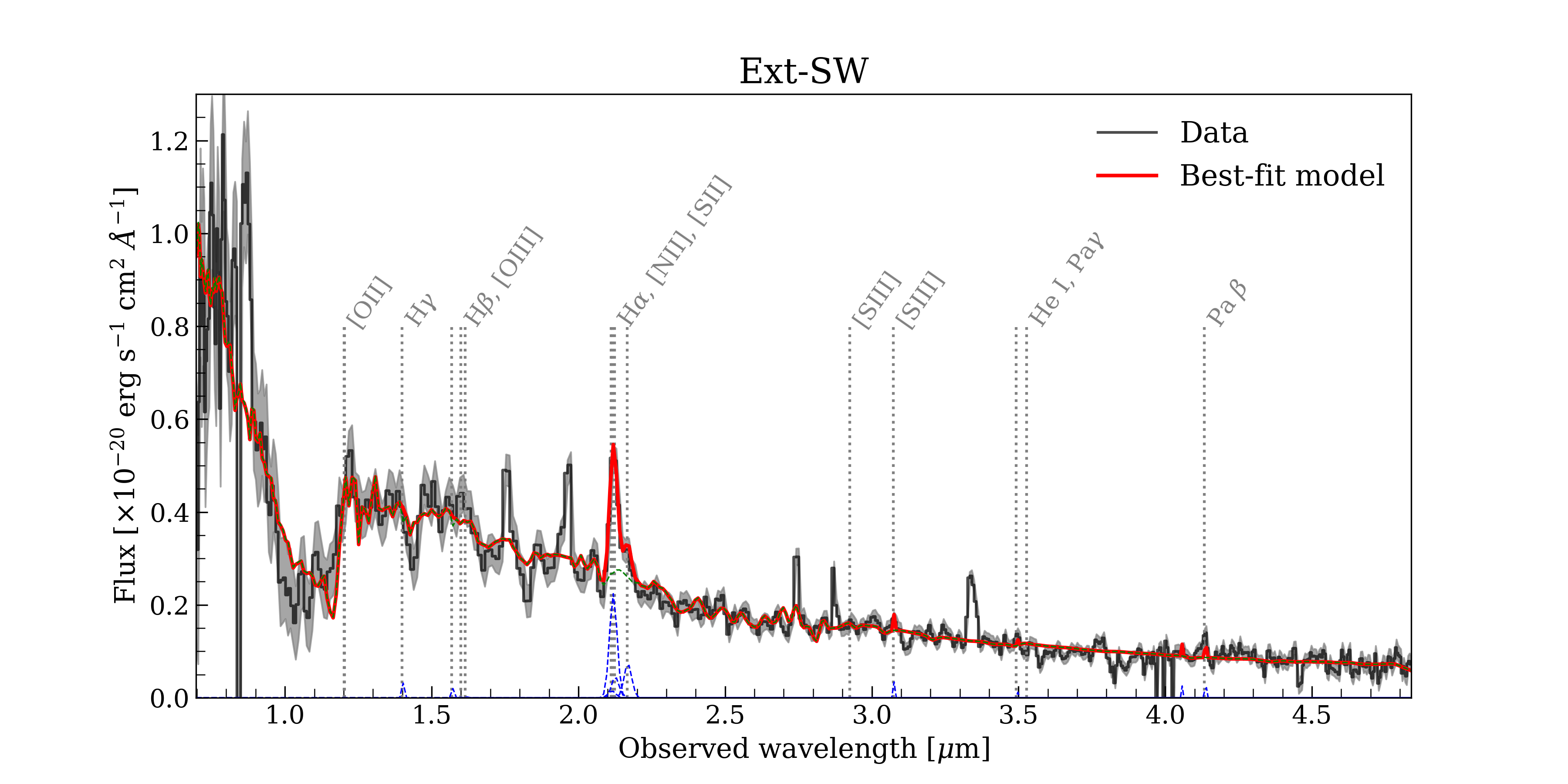}
    \includegraphics[width=0.45\linewidth]{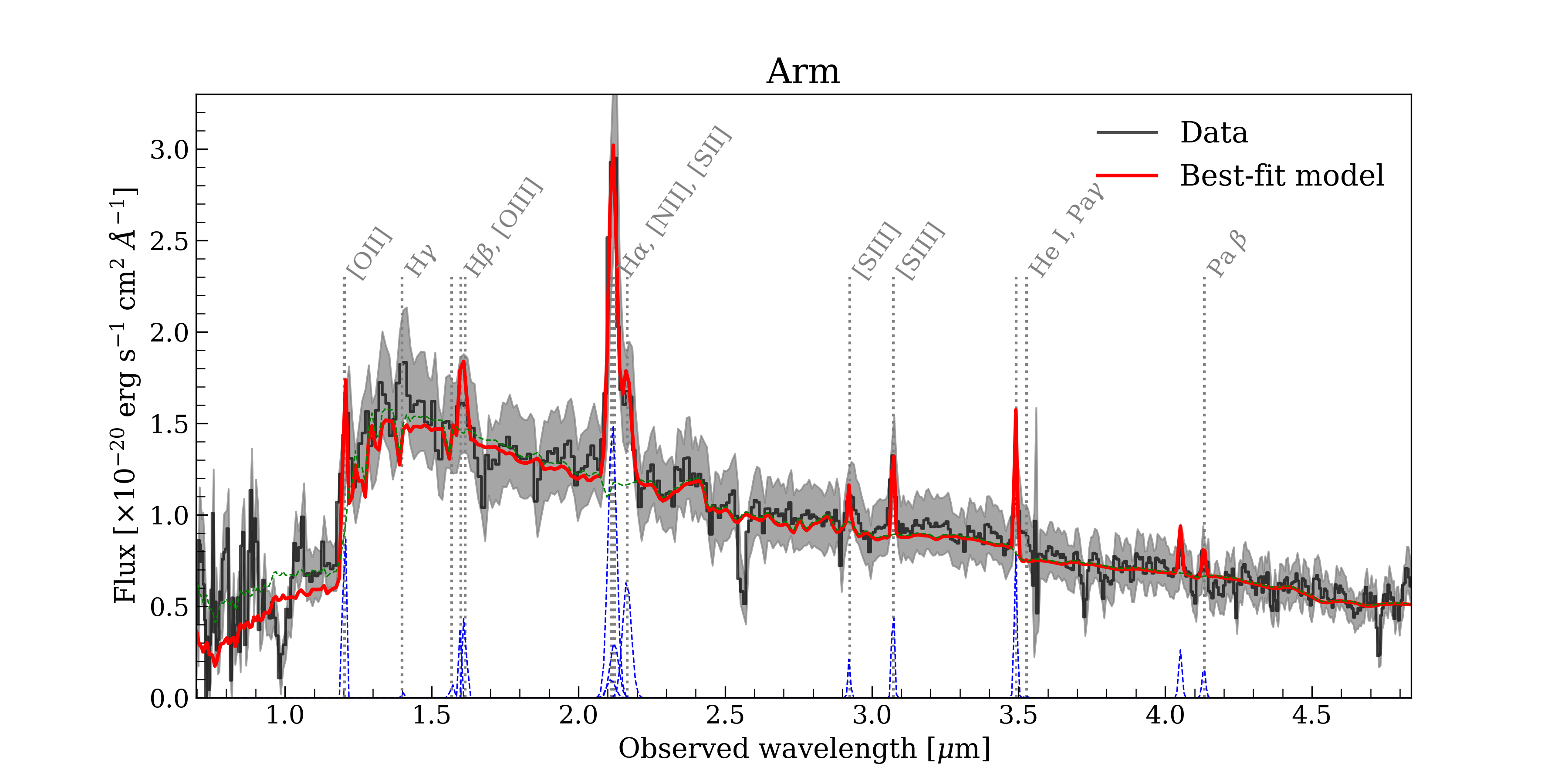}

    \caption{NIRSpec MSA PRISM/CLEAR spectra ($R$\,=\,100) of \target, from all six shutters. Data are shown in black, with the best-fit model overplotted in red. All spectra exhibit both stellar continuum and emission lines, except for the External-SE spectrum, where we only detect continuum emission.}
    \label{fig:bestfit_prism}
\end{figure*}

Following all the prescriptions above, we perform the spectral fitting using the \texttt{python} package \verb|LMFIT|, which delivers the best-fit values of the amplitude, line width, and velocity of each emission line, which minimize the $\chi^2$. In Fig. \ref{fig:bestfit_prism}, we show all PRISM spectra available along with their best-fit model, given by the sum of the continuum model obtained with \texttt{pPXF} and Gaussian emission lines from \verb|LMFIT|. Stellar continuum and emission lines are detected in all spectra but the External-SE shutter (in pink in Fig. \ref{fig:Nircam_RGB_and_prism}), where only continuum emission is detected. To accurately estimate line fluxes and errors, we follow a Monte Carlo approach. Therefore, we perturb the observed spectrum with its associated noise spectrum (from the "error" extension of the data) and repeat the emission-line fitting 200 times, with the line setting described above. In Table \ref{tab:results_line_fitting}, we report the median flux value and the 1$\sigma$ uncertainty obtained via a Monte Carlo technique for all emission lines detected with a $S/N>3$.

\subsection{Medium- and high-resolution spectra}

Following slightly different prescriptions, we fit the medium- and high-resolution spectra (each grating separately). Unlike in the PRISM data, continuum emission is only barely detected in $R$\,=\,1000 and $R$\,=\,2700 data, and does not show significant stellar features to require accurate modeling with SSPs. Therefore, there is no need to use \texttt{pPXF}, and directly perform the emission line modeling with \verb|LMFIT|, including a second-order polynomial to reproduce the faint detected continuum emission. Since the spectral resolution has a smaller variation with wavelength in the medium- and high-resolution grating \citep[see][]{Jakobsen:2022} (also because each grating covers a total smaller wavelength range), we constrain all lines within the same grating to have the same velocity and velocity dispersion. We still fix the flux ratios of the \nii and \oiii line doublets to the theoretical ratios, whereas we fit each line of the \siid doublet with two separate Gaussian components, since the two lines are here resolved. The \oiid line doublet, unfortunately, falls into the detector gap of the G140M/F070LP data; hence it is not detected. As for the PRISM data, in Table \ref{tab:results_line_fitting} we report the median flux values and the 1$\sigma$ errors for each emission line with a $S/N>3$, as resulting from a Monte Carlo simulation.

Given that the velocity dispersion measured for \target ranges from 20 to 140 \kms \citep[see][Pulsoni et al. in prep]{Forster:2018}, the lines remain spectrally unresolved in the medium-resolution data at $R$\,$\sim$\,1000 (corresponding to a FWHM of $\sim$ 300 km/s) but are resolved at $R$\,$\sim$\,2700. Table \ref{tab:results_line_fitting} lists their intrinsic velocity dispersions, derived after correcting the observed values for instrumental resolution.

\begin{figure*}
    \centering
    \includegraphics[width=0.45\linewidth]{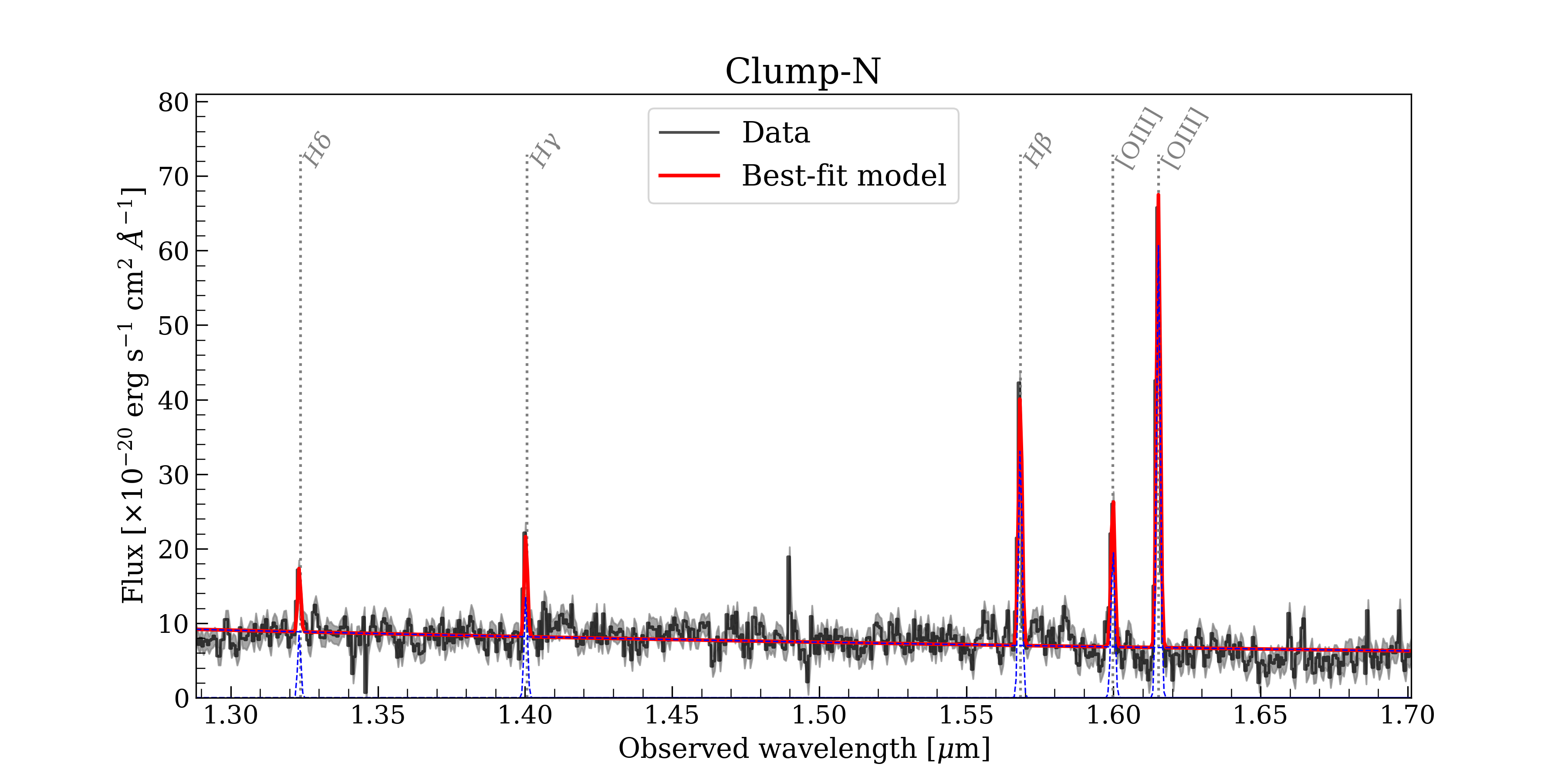}
    \includegraphics[width=0.45\linewidth]{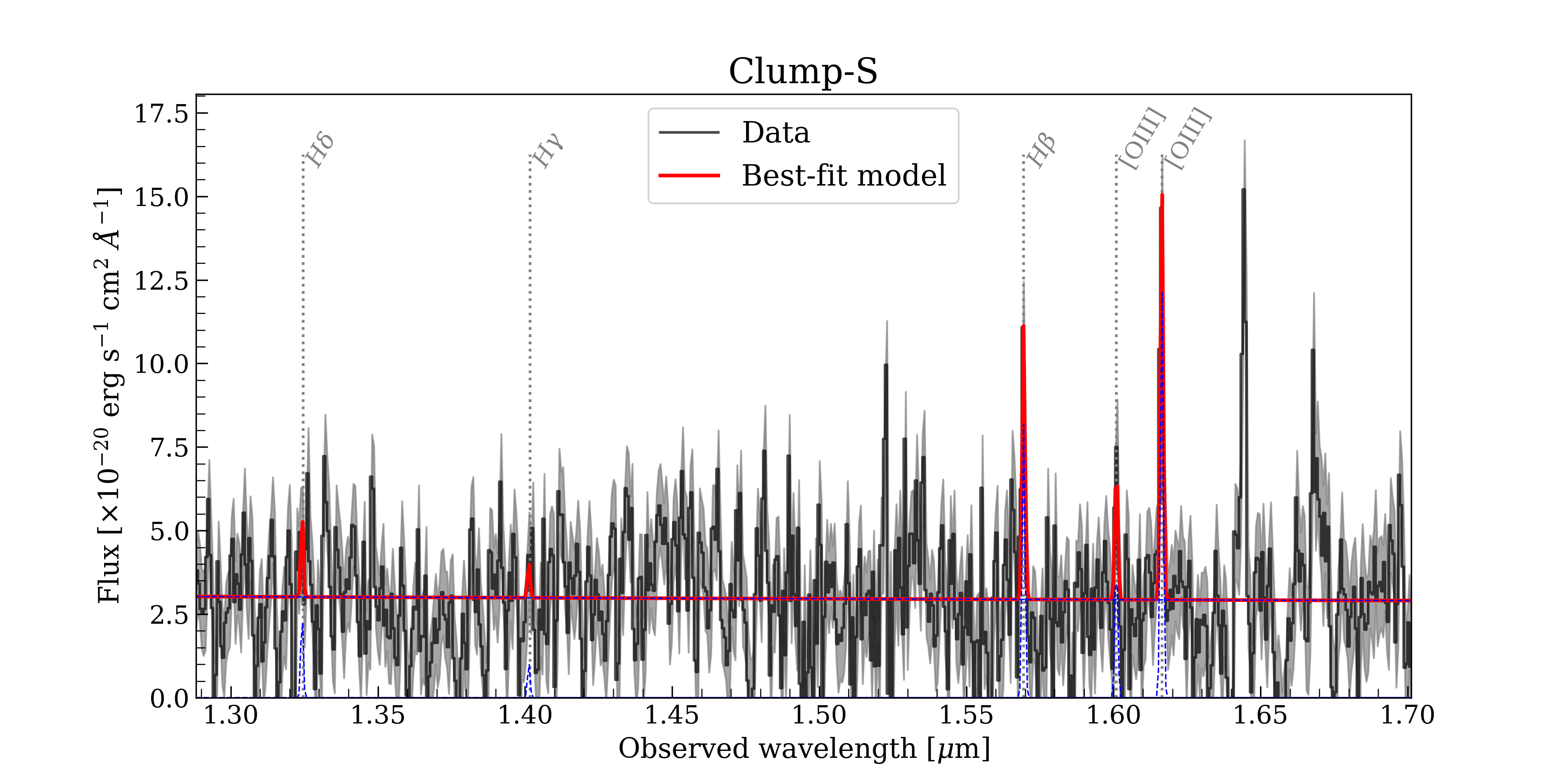}

    \caption{NIRSpec MSA medium-resolution spectra ($R$\,=\,1000) of \target, obtained with the G140M/F070LP grating, for the shutters where we detect emission lines at S/N>3. Data are shown in black, with the best-fit model overplotted in red.}
    \label{fig:bestfit_g140m}
\end{figure*}

\begin{figure*}
    \centering
    \includegraphics[width=0.45\linewidth]{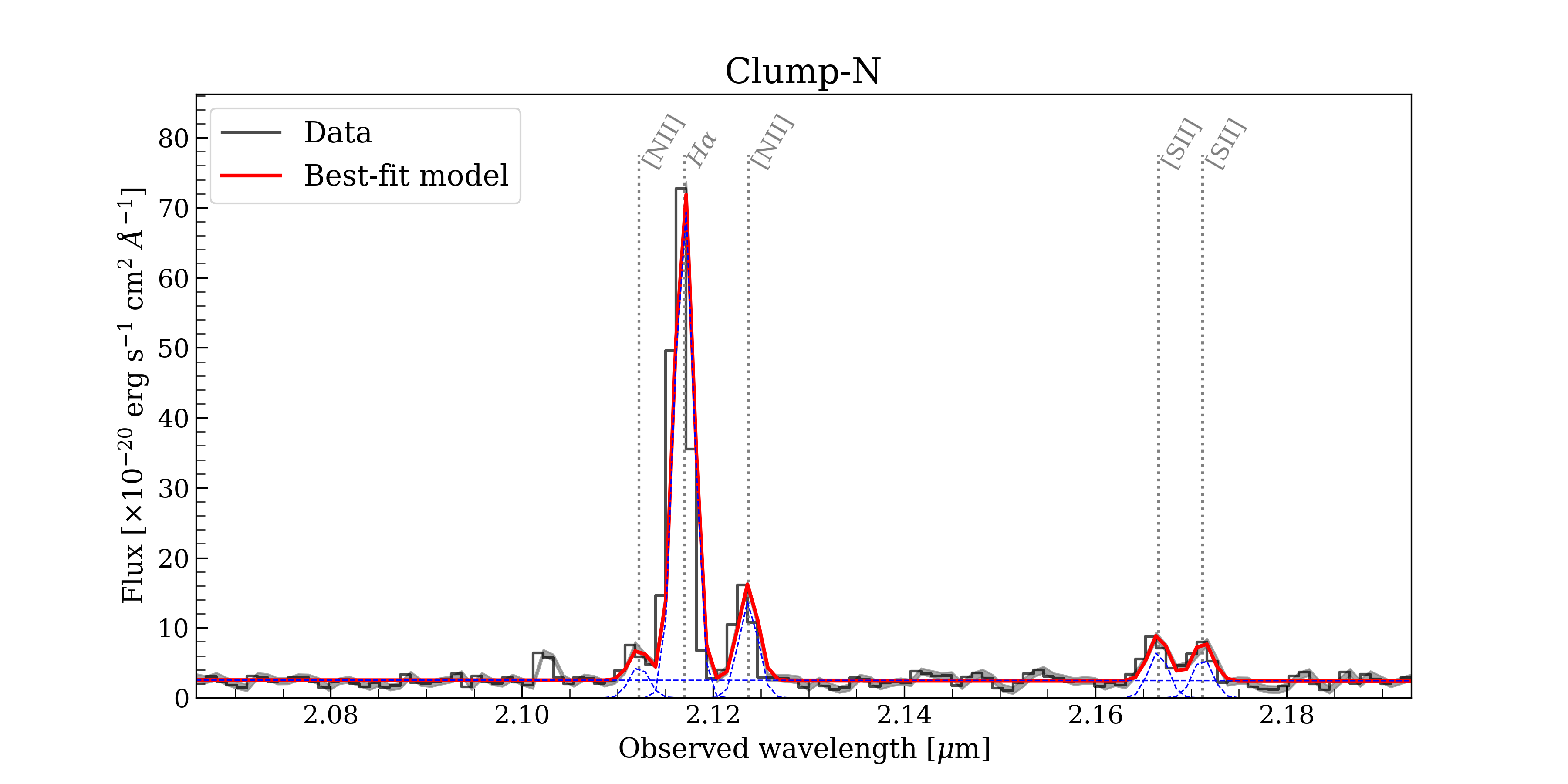}
    \includegraphics[width=0.45\linewidth]{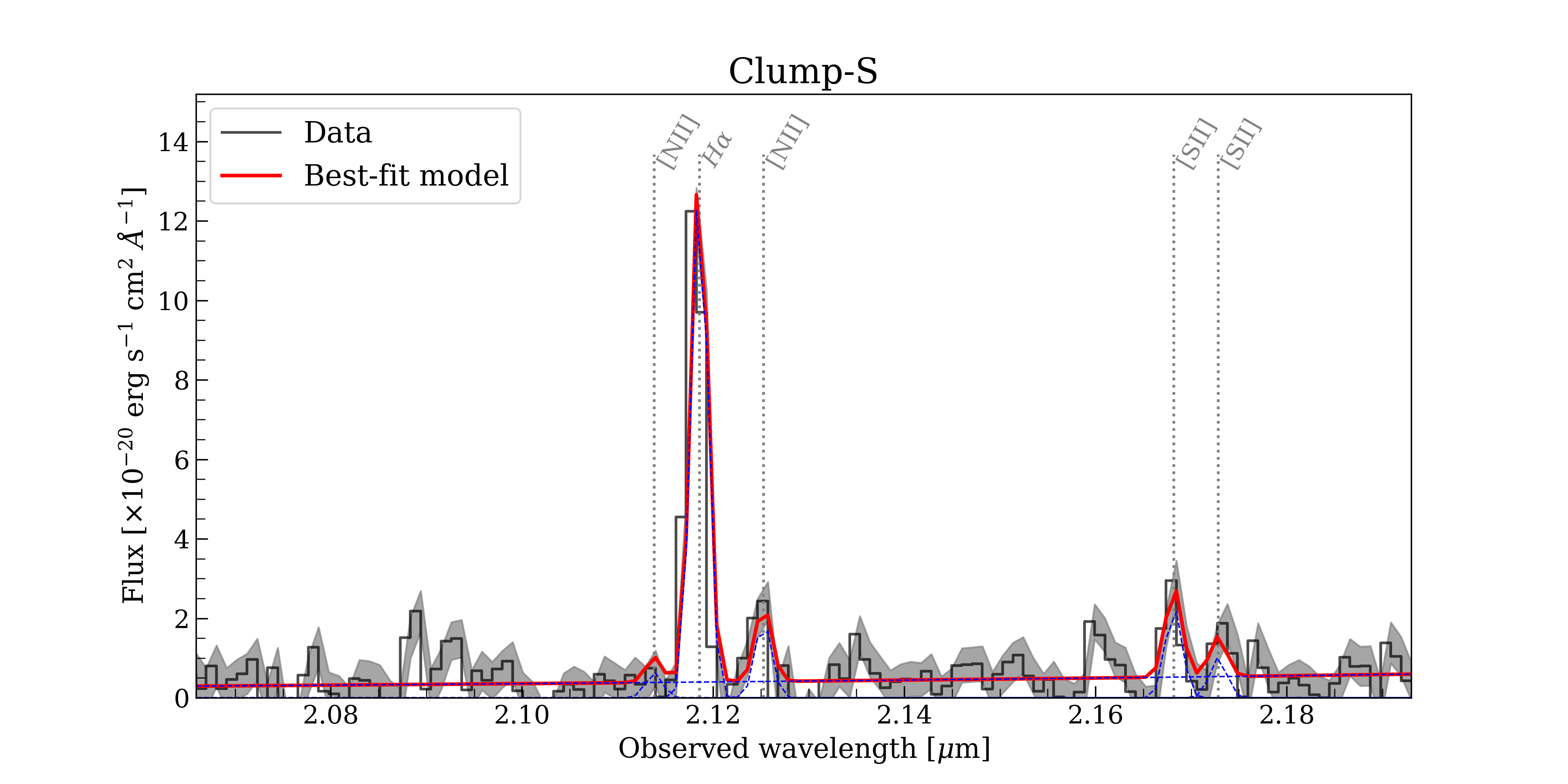}
    \includegraphics[width=0.45\linewidth]{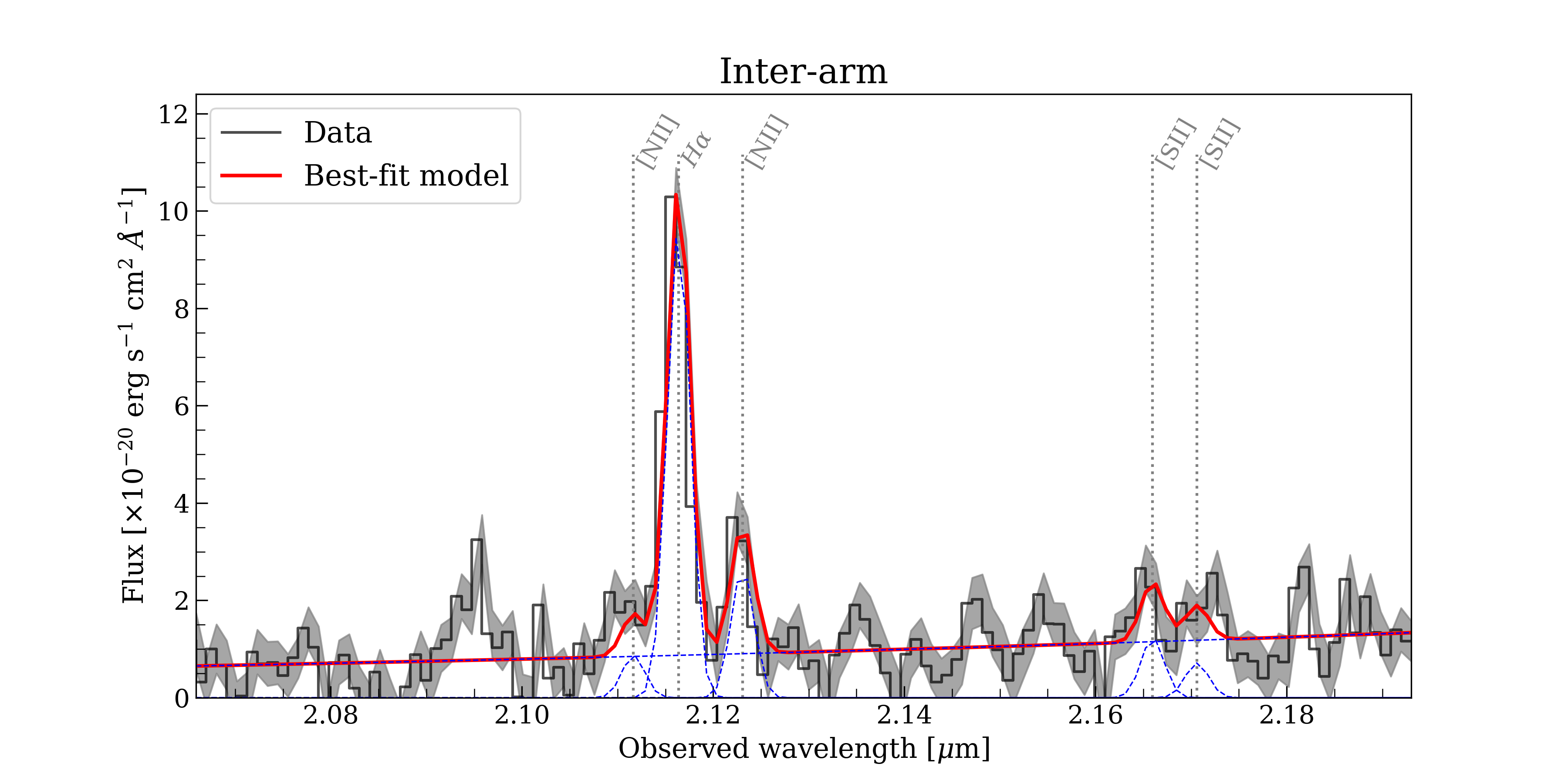}
    \includegraphics[width=0.45\linewidth]{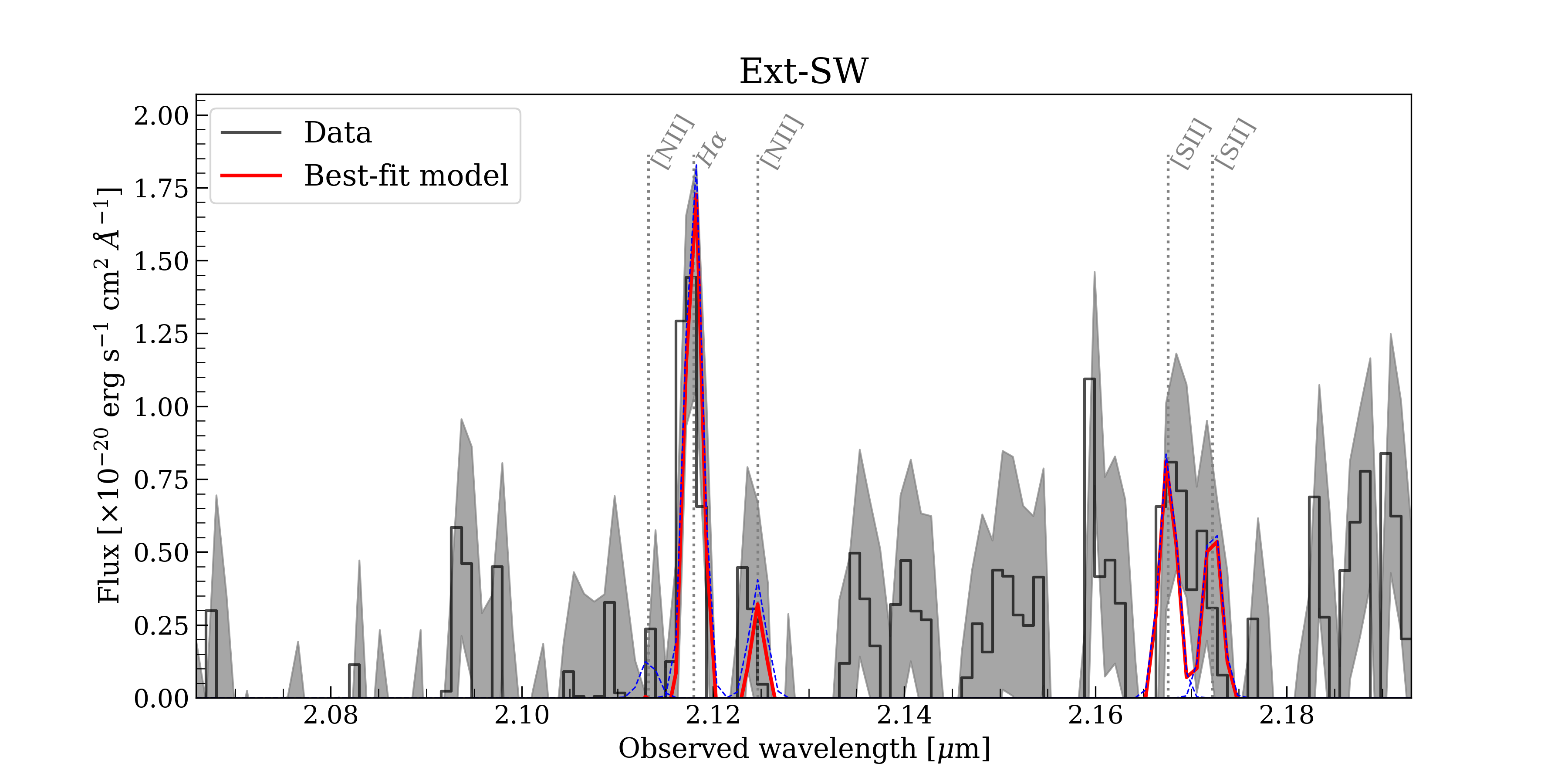}
    \includegraphics[width=0.45\linewidth]{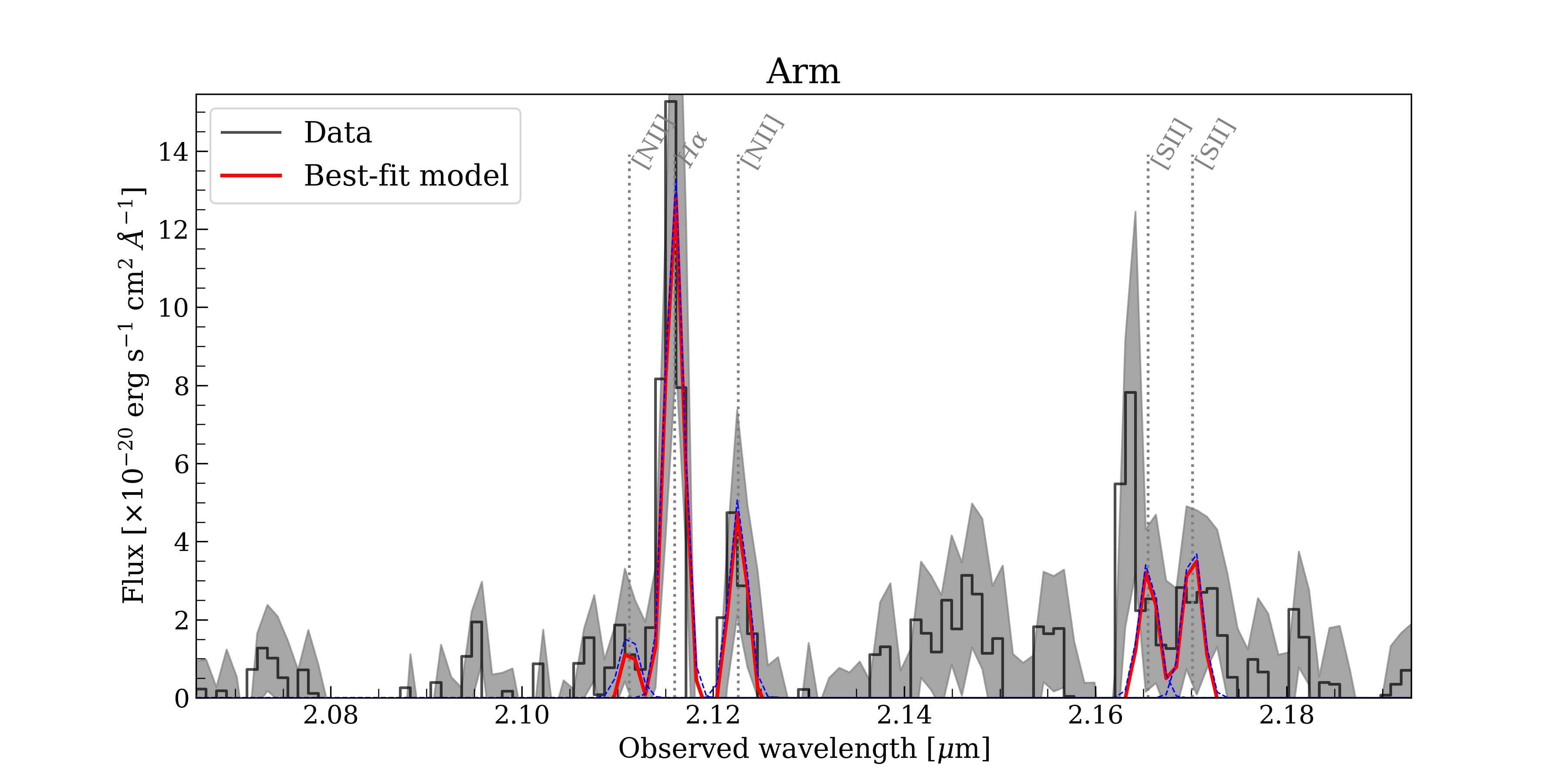}

    \caption{NIRSpec MSA medium-resolution spectra ($R$\,=\,1000) of \target, obtained with the G235M/F170LP grating around the \ha, \nii\ complex for the shutters where we detect emission lines at S/N>3. Data are shown in black, with the best-fit model overplotted in red.}
    \label{fig:bestfit_g235m}
\end{figure*}

\begin{figure*}
    \centering
    \includegraphics[width=0.45\linewidth]{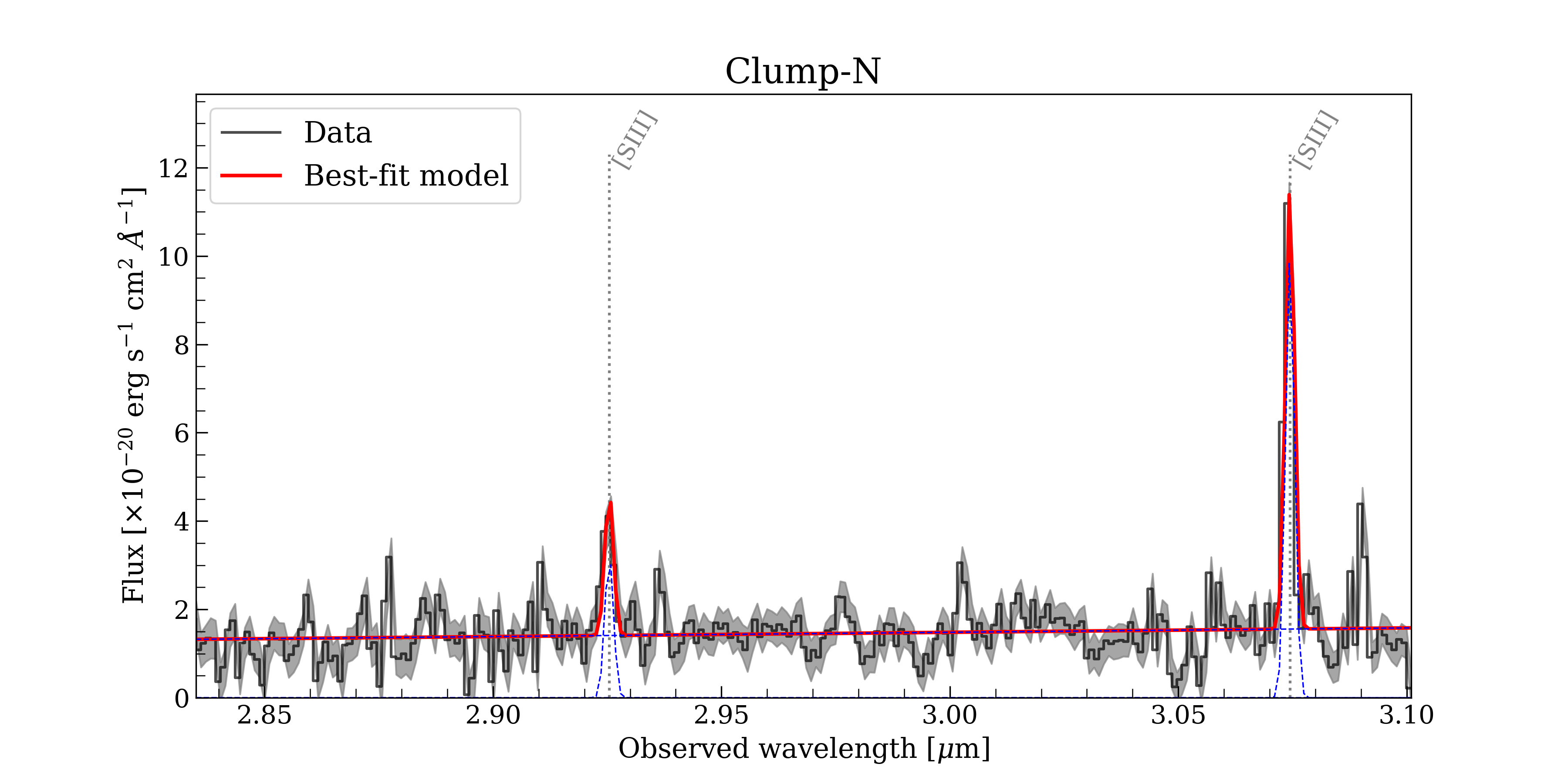}
    \includegraphics[width=0.45\linewidth]{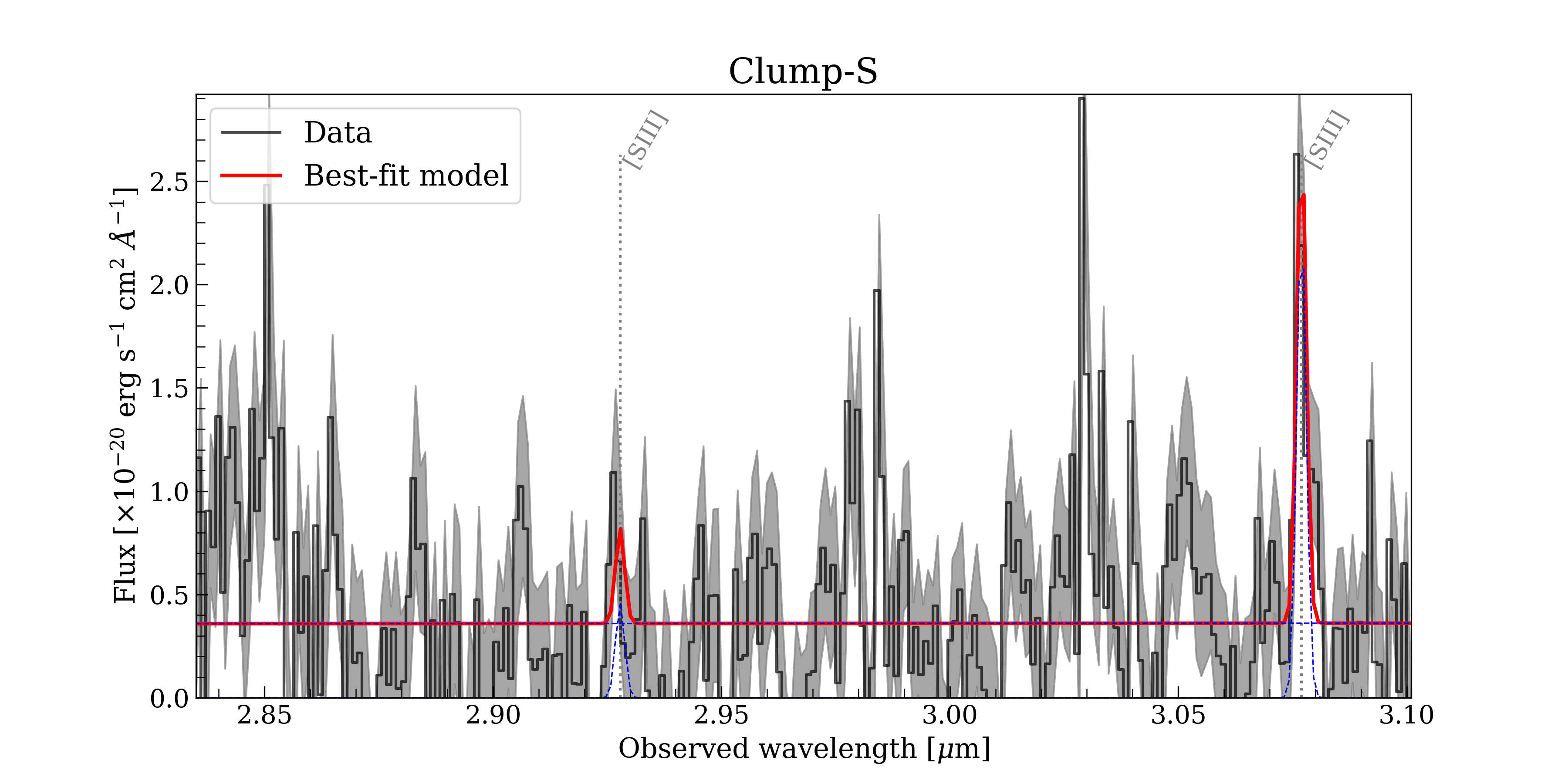}

    \caption{NIRSpec MSA medium-resolution spectra ($R$\,=\,1000) of \target, obtained with the G235M/F170LP grating around the \siii\ emission lines for the shutters where we detect emission lines at S/N>3. Data are shown in black, with the best-fit model overplotted in red.}
    \label{fig:bestfit_g235m_sii}
\end{figure*}

\begin{figure*}
    \centering
    \includegraphics[width=0.45\linewidth]{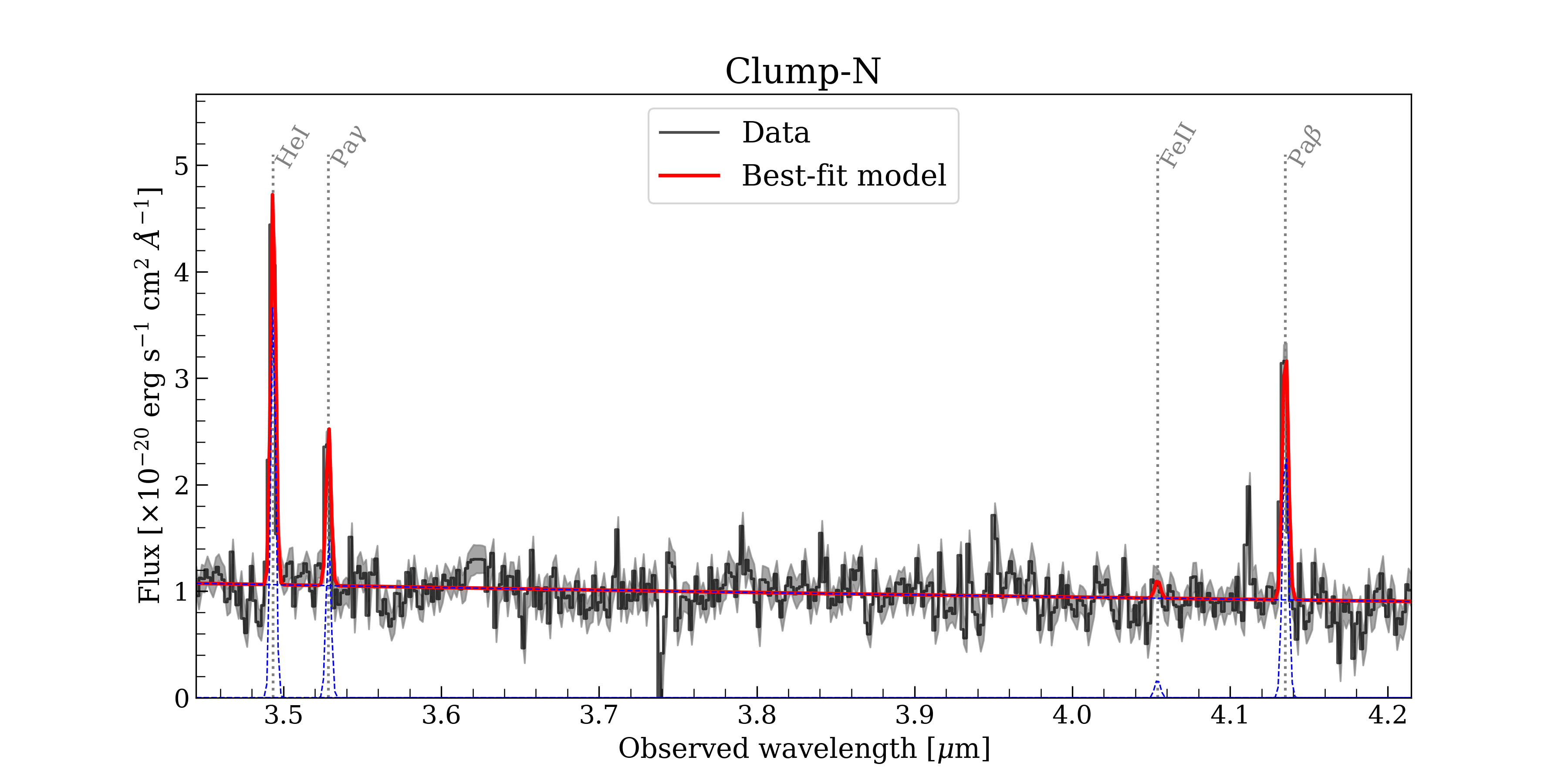}
    \includegraphics[width=0.45\linewidth]{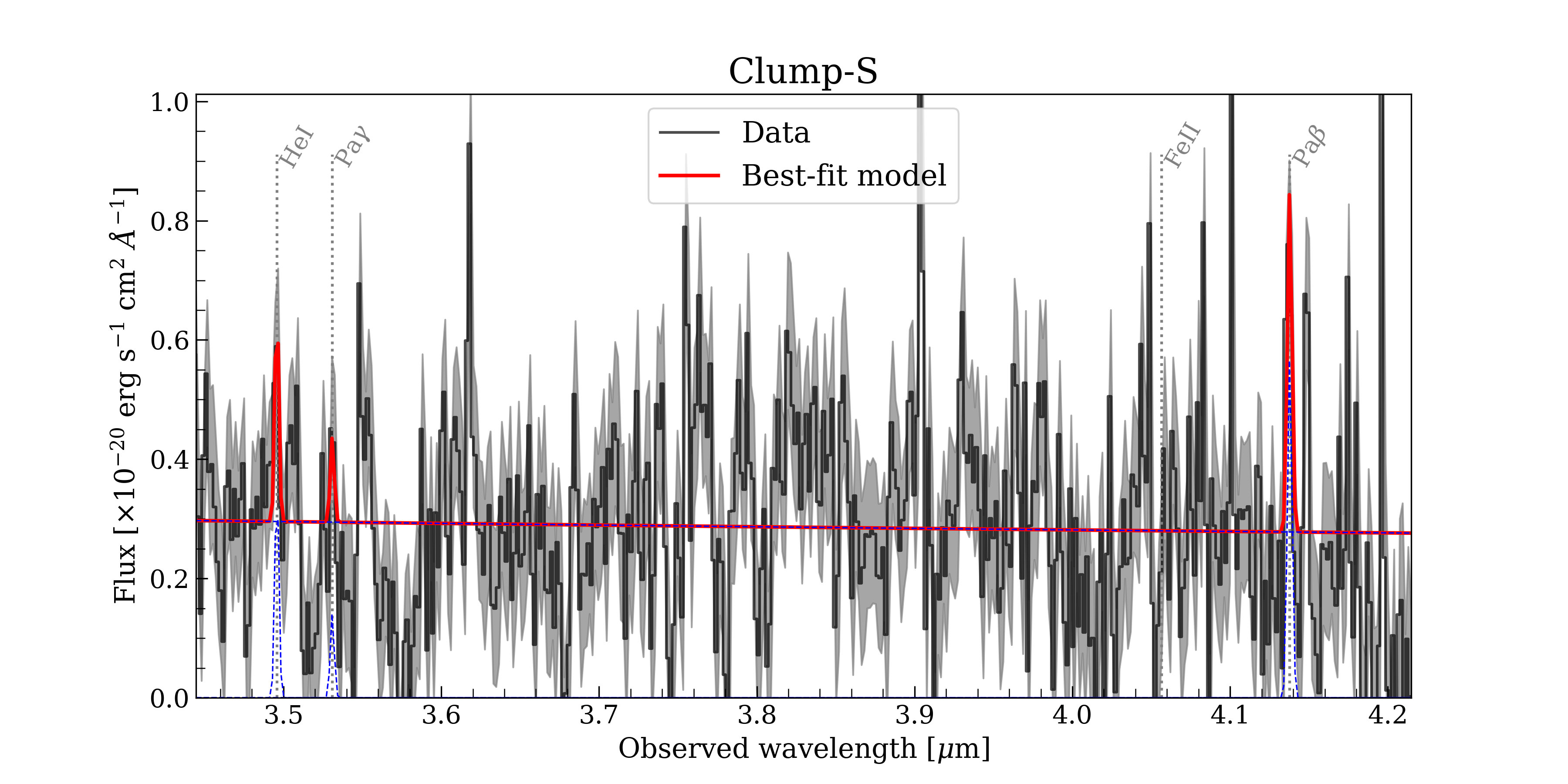}
    \includegraphics[width=0.45\linewidth]{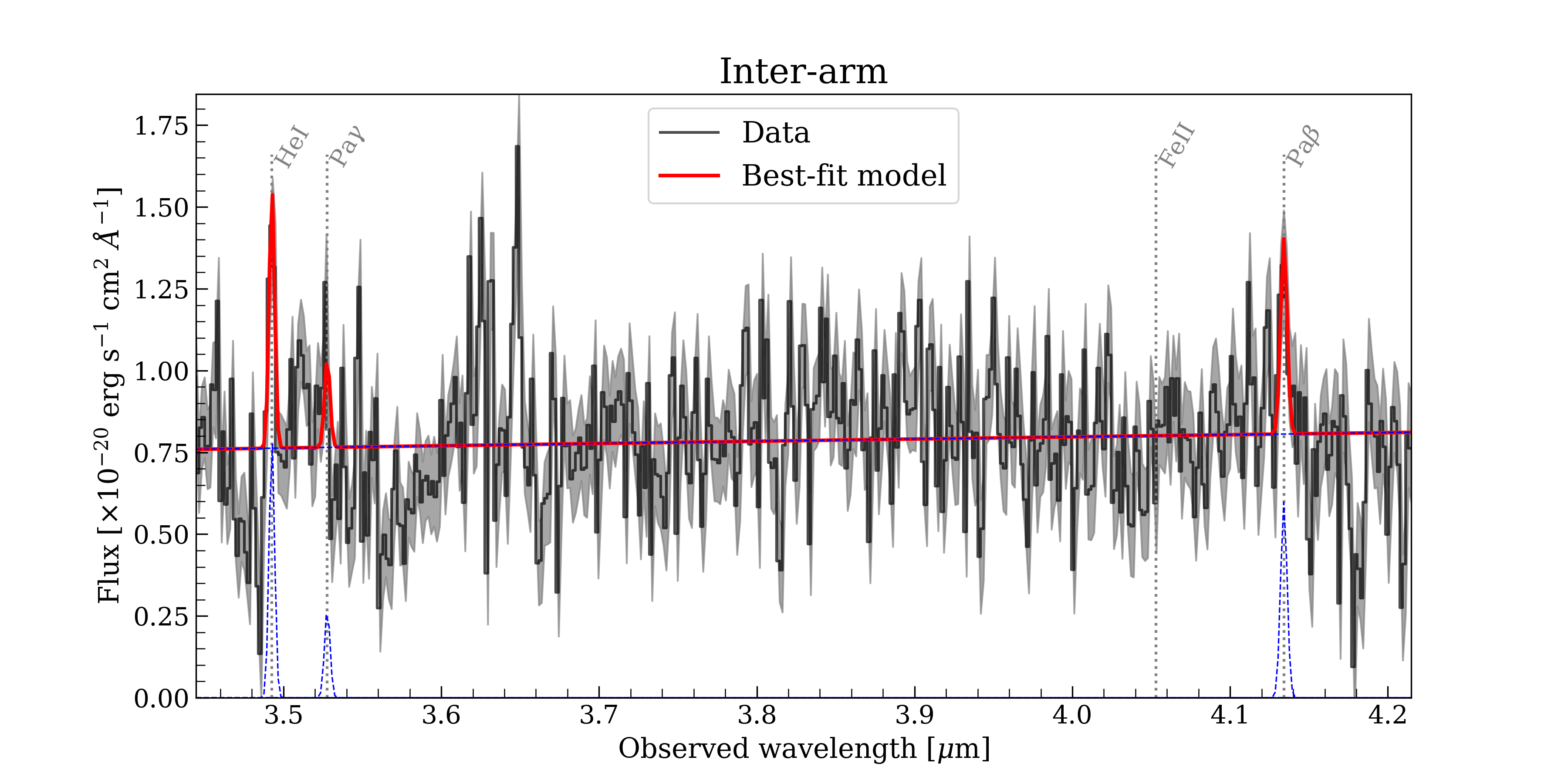}

    \caption{NIRSpec MSA high-resolution spectra ($R$\,=\,1000) of \target, obtained with the G395M/F290LP, for the shutters where we detect emission lines at S/N>3. Data are shown in black, with the best-fit model overplotted in red.}
    \label{fig:bestfit_g395m}
\end{figure*}

\begin{figure*}
    \centering
    \includegraphics[width=0.45\linewidth]{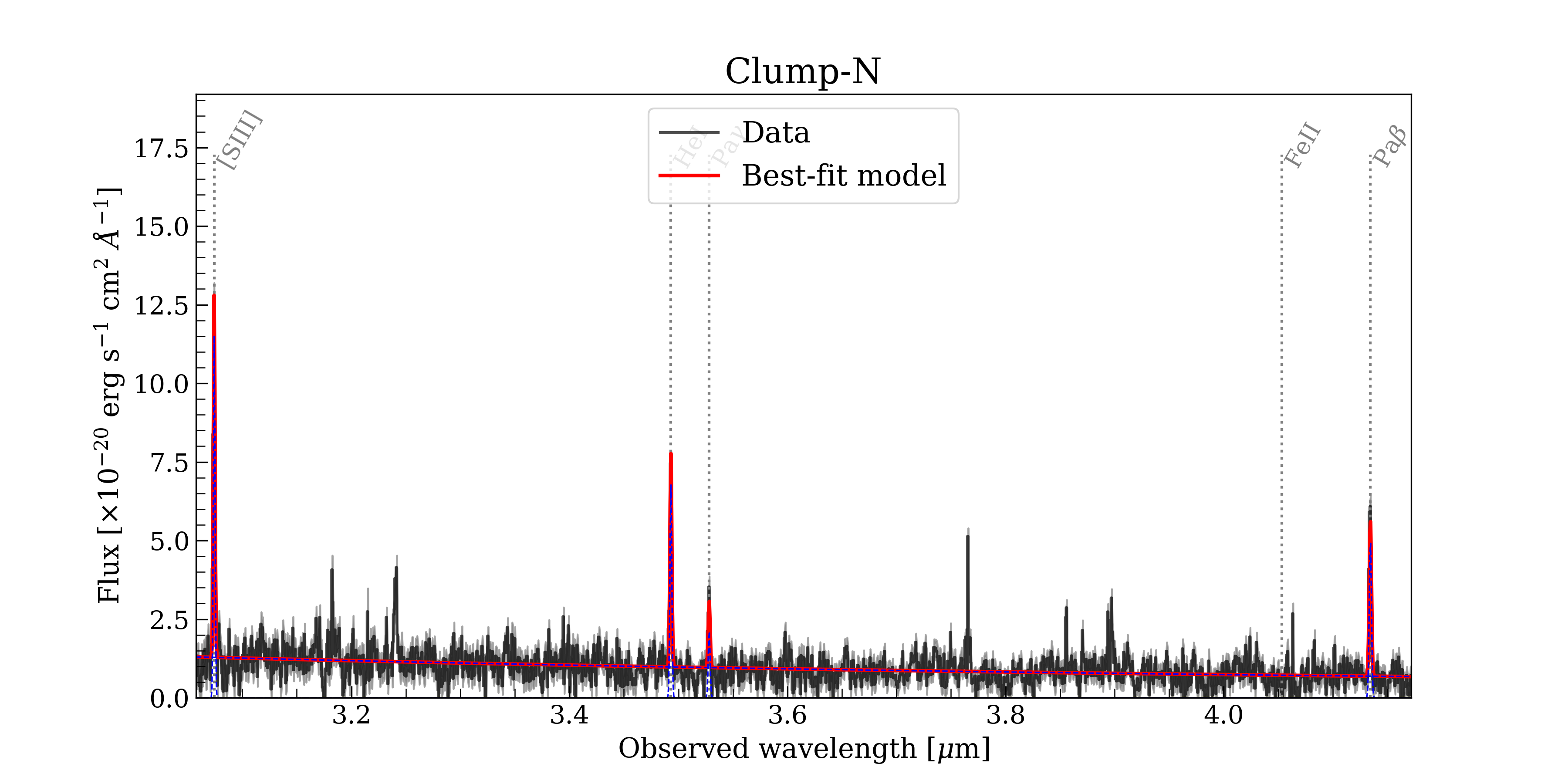}
    \includegraphics[width=0.45\linewidth]{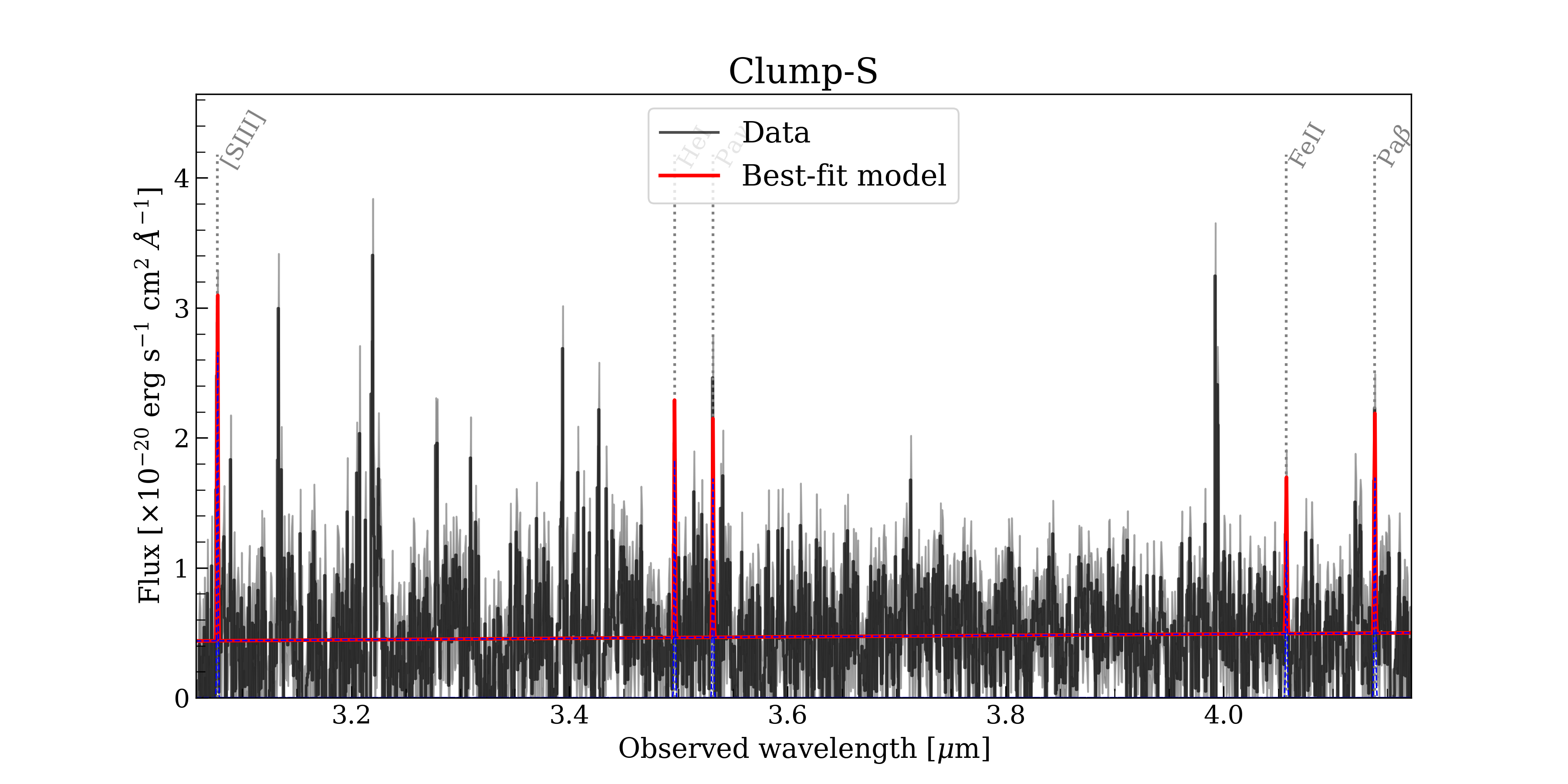}
    \includegraphics[width=0.45\linewidth]{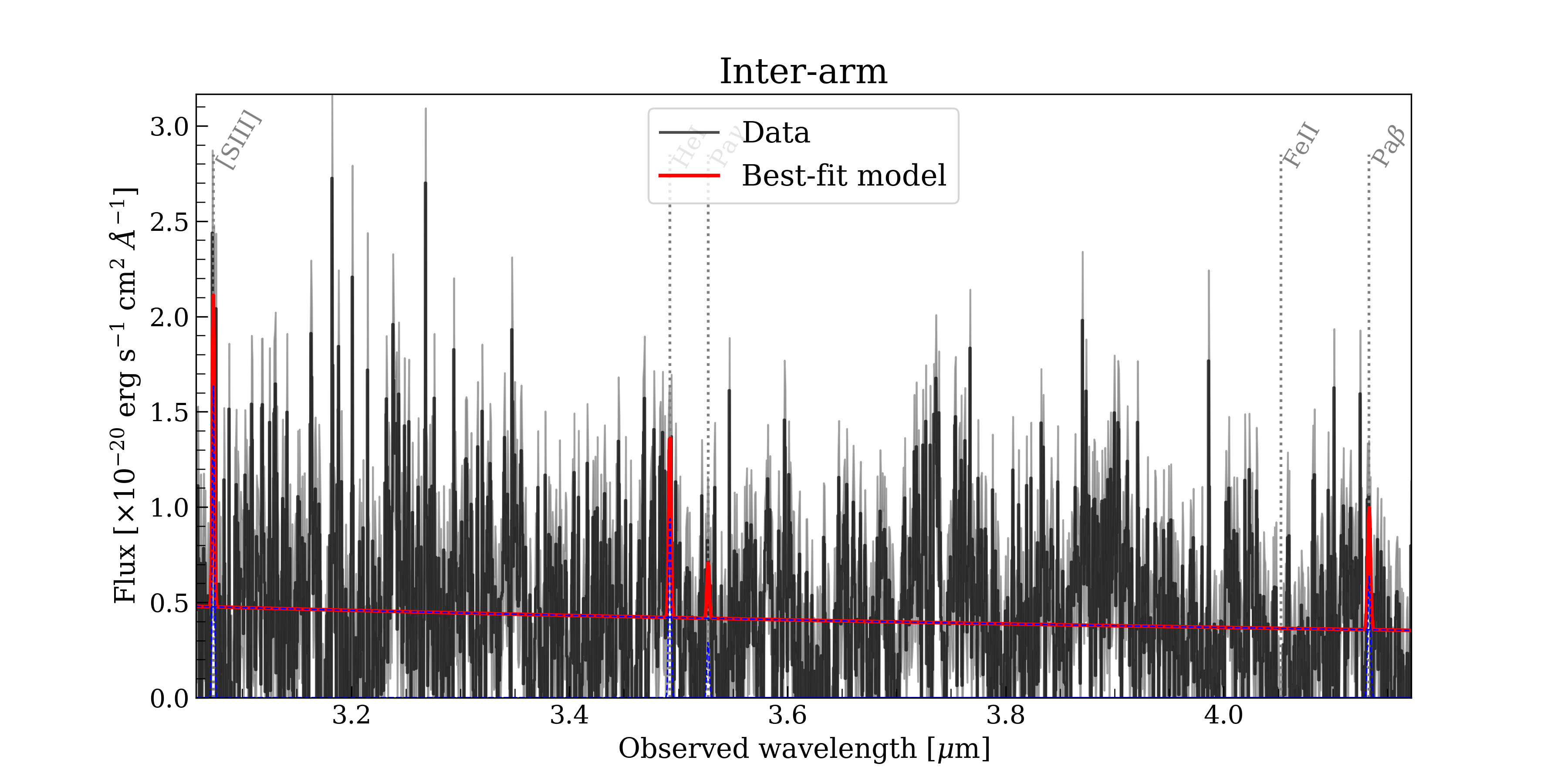}

    \caption{NIRSpec MSA high-resolution spectra ($R$\,=\,2700) of \target, obtained with the G395H/F290LP, for the shutters where we detect emission lines at S/N>3. Data are shown in black, with the best-fit model overplotted in red.}
    \label{fig:bestfit_g395h}
\end{figure*}

\begin{table*}[]
\caption{Best-fit results for emission lines detected at S/N\,$>$\,3 in the NIRSpec MSA spectra of \target at different spectral resolution.}
\centering
\footnotesize
\label{tab:results_line_fitting}
\resizebox{0.95\textwidth}{!}{
\begin{tabular}{c|cccccc|ccccc}
\hline\hline

\multirow{3}{*}{Line}             & \multicolumn{6}{c|}{PRISM/CLEAR}                                                                                                                             & \multicolumn{5}{c}{G140M/F070LP, G235M/F170LP, G395M/F290LP}            \\ 
     & \multicolumn{6}{c|}{}                                                                                                                             & \multicolumn{5}{c}{(G395H/F290LP)}            \\
\cline{2-12} 
 & Arm & Inter-Arm & Clump-N & Ext-N & Clump-S & Ext-SW & Arm & Inter-Arm & Clump-N & Clump-S & Ext-SW \\
\hline\hline
\hline
\oiid & -  & 85 $\pm$ 33 $^{(*)}$ & 917 $\pm$ 43 $^{(*)}$ & - & 184 $\pm$ 34 $^{(*)}$ & - & -  & - & - & - & - \\
\hd & - &  - & - & - & - & -  & -  & - & 111 $\pm$ 14 & - & - \\
\hg & - &  - & 20 $\pm$ 1 & - & -  & - & - & - & 191 $\pm$ 16 & - & - \\
\hb & - &  - & 380 $\pm$ 31  & - & -  & - & - & - & 525 $\pm$ 18 & 102 $\pm$ 15  & - \\
\oiii$\lambda$5007 & - & - & 997 $\pm$ 35  & - & 192 $\pm$ 24  & - & - & 112 $\pm$ 23 & 972 $\pm$ 18 & 155 $\pm$ 17 & - \\
\ha & 606 $\pm$ 190 & 350 $\pm$ 16 & 1782 $\pm$ 22 & 83 $\pm$ 14 & 355 $\pm$ 13 & 65 $\pm$ 10 & 332 $\pm$ 45 & 293 $\pm$ 13 & 1792 $\pm$ 12 & 281 $\pm$ 10 & 41 $\pm$ 10\\
\nii & - & 70 $\pm$ 4 & 357 $\pm$ 4 & - & 71 $\pm$ 3 & - & 136 $\pm$ 40 & 82 $\pm$ 10 & 351 $\pm$ 9 & 40 $\pm$ 9 & - \\
\siileft & - & 122 $\pm$ 19 $^{(*)}$  & 424 $\pm$ 20 $^{(*)}$ & - & 108 $\pm$ 15 $^{(*)}$ & - & - & 37 $\pm$ 11 & 171 $\pm$ 9 & 47 $\pm$ 10 & - \\
\siiright & - & - & - & - & - & - &  - & 24 $\pm$ 11 & 148 $\pm$ 10 & 20 $\pm$ 9 & - \\
\siiileft & - & - & 153 $\pm$ 9 & - & 30 $\pm$ 7 & - & -& - & 76 $\pm$ 8 & - & - \\
\hline
\multirow{2}{*}{\siiiright}  & - & 34 $\pm$ 12 & 325 $\pm$ 10 & - & 61 $\pm$ 9 & - & - & - & 249 $\pm$ 8 & 65 $\pm$ 15 & - \\
 & - & - & - & - & - & - & - & (39 $\pm$ 5) & (218 $\pm$ 6) & (33 $\pm$ 4) & - \\
\hline
\multirow{2}{*}{\hei} & - & 41 $\pm$ 10 & 175 $\pm$ 7 & - & 40 $\pm$ 6 & - & -& 33 $\pm$ 6 & 146 $\pm$ 5 & 11 $\pm$ 4 & - \\
 & - & - & - & - & - & - & - & (24 $\pm$ 5) & (144 $\pm$ 6) & (24 $\pm$ 6) & - \\
\hline
\multirow{2}{*}{\Pag} & - & - & 60 $\pm$ 6 & - & 20 $\pm$ 4 & - & - & - & 59 $\pm$ 4 & - & -  \\
 & - & - & - & - & - & - & - & - & (46 $\pm$ 5) & (22 $\pm$ 3) & - \\
\hline
\multirow{2}{*}{\Pab} & -  & 35 $\pm$ 7 & 151 $\pm$ 7 & - & 29 $\pm$ 5 & - & - & 30 $\pm$ 6 & 112 $\pm$ 4 & 22 $\pm$ 5 & - \\
 & - & - & - & - & - & - & - & (20 $\pm$ 5) & (127 $\pm$ 5) & (26 $\pm$ 4) & - \\
\hline
 $\sigma_{\rm gas}$ & - & - & - & - & - & - & - & (79 $\pm$ 10) & (55 $\pm$ 2) & (44 $\pm$ 3) & - \\
\hline
\end{tabular}
}
\tablefoot{All fluxes are in units of $10^{-20}$ \ergscm. For the reddest G395M and G395H gratings, we provide line fluxes measured at both medium and high spectral resolution (the latter in brackets). The last row reports the intrinsic velocity dispersion $\sigma_{\rm gas}$ (i.e., corrected for instrumental broadening) in units of \kms, as measured in high-resolution G395H data. $^{(*)}$ Total fluxes computed as the sum of the flux of the two \oiid and \siid doublet components, which are unresolved in PRISM/CLEAR spectra ($R$\,=\,100).}
\end{table*}

\section{Resolved SED best models}\label{sec:sed_fitting_models}

\begin{figure*}
    \centering
    \includegraphics[width=0.95\textwidth]{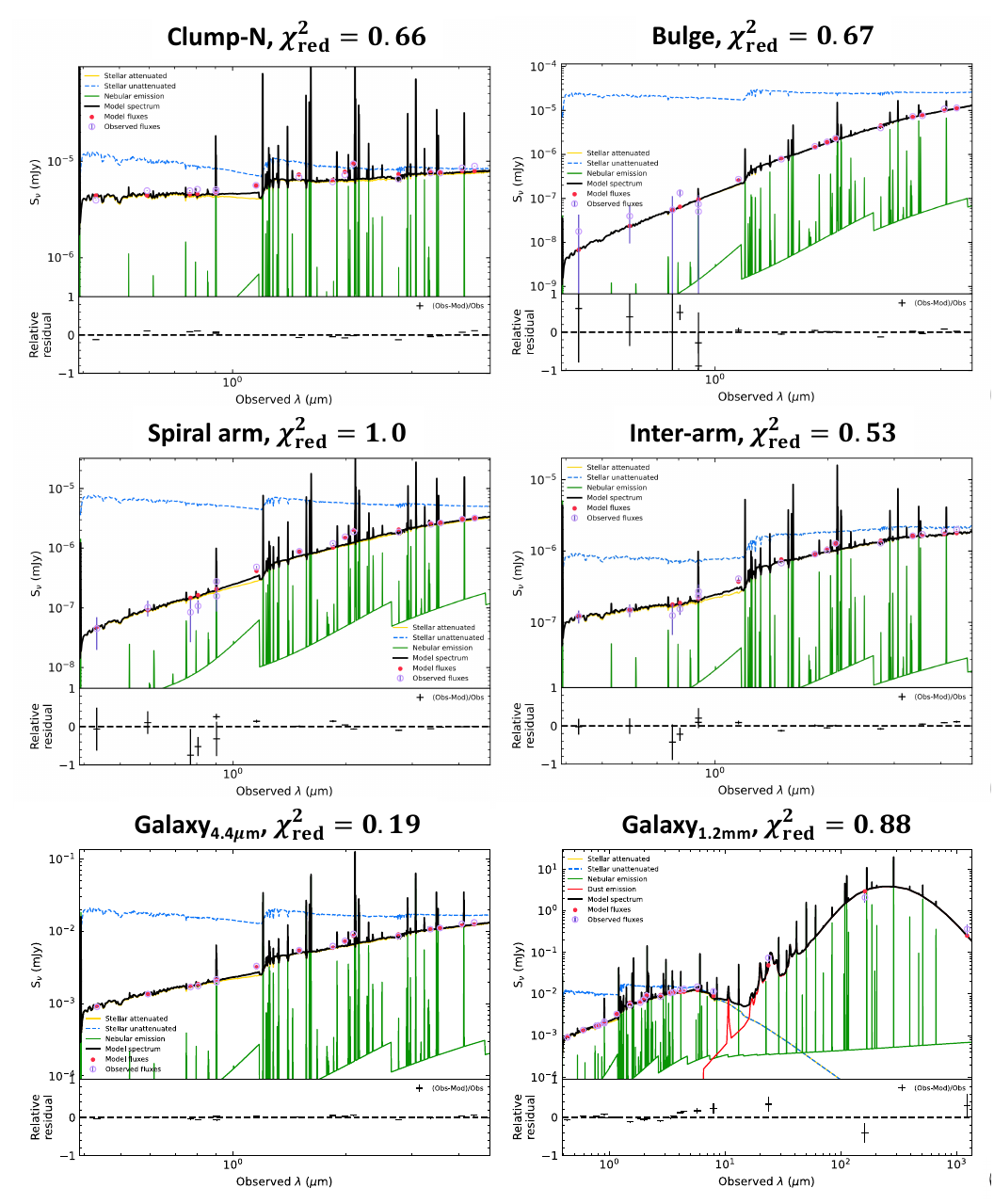}
    \caption{Best-fit models and observed SEDs as obtained from our resolved (top and middle panels) and integrated (bottom) SED modeling with \texttt{CIGALE}. \textit{Top and middle.} Resolved results are shown for four representative pixels from four different regions, namely: Clump-N, the bulge, spiral arm, and inter-arm regions. The difference in the SED shape and in the relative intensities between distinct spectral components highlights the need for spatially resolved SED fitting to accurately account for the different physical properties of distinct galaxy regions. \textit{Bottom}. Results from our integrated SED modeling of photometry up to 4.4\,$\mu$m (left) and 1.2\,mm (right), respectively.}
    \label{fig:sed_models}
\end{figure*}

Figure \ref{fig:sed_models} shows the observed SEDs and the best-fit models, resulting from our resolved SED fitting with \texttt{CIGALE} (performed in Sect. \ref{sec:sed_analysis}), for four individual pixels representative of distinct galaxy regions (top and middle panels), namely: Clump-N (or Clump A), the bulge, the spiral arm and the inter-arm region southeast and north of the bulge, respectively. Bottom panels show the best fits of the integrated SED modeling (obtained in Sect. \ref{sec:sed_integ_results}), including photometry up to NIRCam F444W (left) and ALMA Band 6 1.2\,mm (right).

\section{Nebular $A_V$ from NIRCam imaging}\label{sec:av_imaging}

\begin{figure*}
    \centering
    \includegraphics[width=0.95\linewidth]{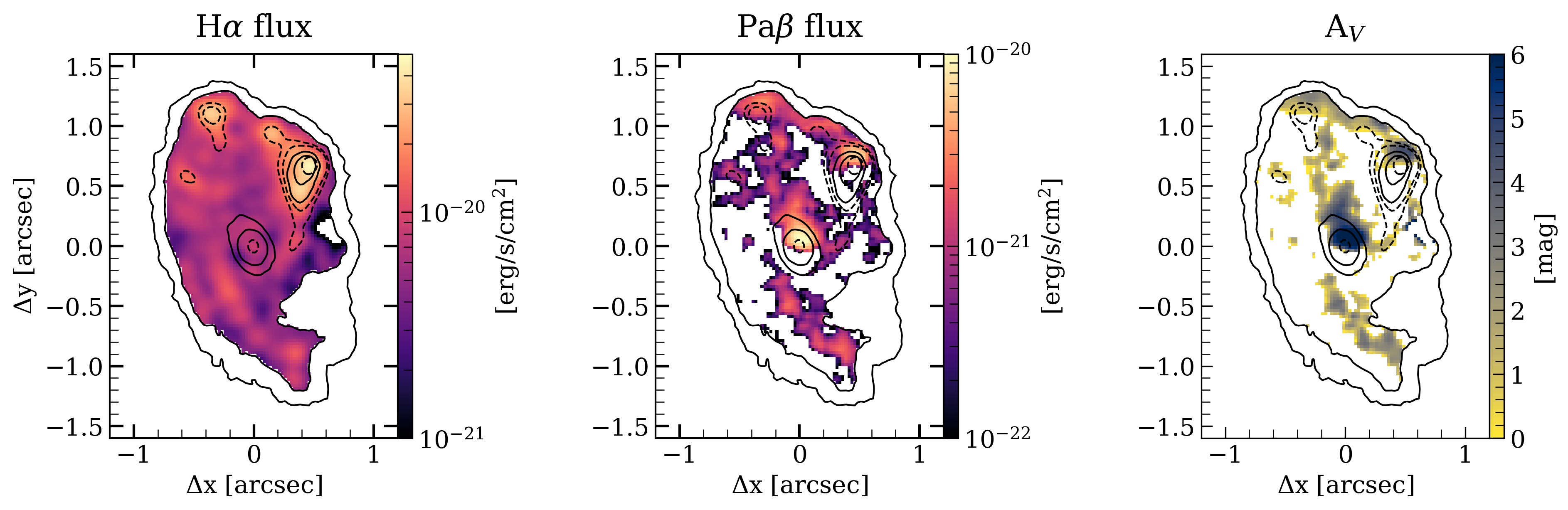}
    \caption{Line emission from medium-band F210M and F410M NIRCam imaging. From the left, we show maps of recovered \ha\ and \Pab line flux (not corrected for dust attenuation), and finally of nebular $A_V$, as inferred from \Pab/\ha\ line ratios. The solid and dashed black contours are the same as in Fig. \ref{fig:SED_fitting_maps}.}
    \label{fig:maps_from_photometry}
\end{figure*}

 In addition to the stellar $A_{V,{\rm star}}$ map inferred from SED fitting in Sect. \ref{sec:sed_cigale}, we derive an independent map of nebular $A_{V,{\rm neb}}$ from \Pab/\ha\ ratios, as inferred by combining available medium-band NIRCam imaging containing \ha\ and \Pab (i.e., F210M and F410M, respectively), and broadband NIRCam filters (see also \citealt{Lorenz:2025}). We first fit a third-degree polynomial to the broadband photometry of each pixel, considering only filters at longer wavelengths than the Balmer break (i.e., F150W, F200W, F277W, F356W, and F444W), to derive the continuum emission. In the polynomial fitting, we assigned a small uncertainty (i.e., 1\%) to the flux of filters with no significant line emission. For the F200W and F444W filters, we assigned larger uncertainties of 10\% and 5\%, respectively, to account for the contribution of \ha\ and \Pab emission lines to the total broadband emission. Then, we convolve the continuum best-fit model of each pixel with the F210M and F410M filter transmission, respectively, and integrate the convolved continuum model over the wavelength range of the corresponding filter, thus separately obtaining the contribution of continuum emission to the F210M and F410M bands. By removing the inferred continuum level from the total observed F210M and F410M flux in each pixel, we end up with residual line emission, including also contributions from emission lines other than \ha\ and \Pab. In particular, line emission of \target in the F210M and F410M filters is mostly due to \ha+\nii\ and \Pab+[Fe{\sc{ii}}], respectively. Therefore, to obtain an estimate of the \ha\ and \Pab\ fluxes only, we correct the resulting total line fluxes by adopting \ha/\nii\ and \Pab/[Fe{\sc{ii}}] line ratios, as measured from the MSA spectra. In particular, for the northern clumps (i.e., Clump-N and the smaller northernmost clump) we adopt the flux measurements obtained for Clump-N (\ha/\nii\,=\,5.1 and \Pab/[Fe{\sc{ii}}]\,=\,14); while for all the rest of the galaxy (i.e., spiral arms, inter-arm regions, and bulge), we rely on the line ratios inferred for the Inter-Arm and/or Arm region, when available (i.e., \Pab/[Fe{\sc{ii}}]\,=\,8.4 from Inter-Arm spectra; \ha/\nii\,=\,3.0, computed as average between line ratios measured from the Inter-Arm and Arm regions).

In Fig. \ref{fig:maps_from_photometry}, we show the resulting maps of observed (i.e., not dust corrected) \ha\ and \Pab\ flux, along with the nebular $A_{V,{\rm neb}}$ map, as obtained from pixel-per-pixel \Pab/\ha\ line ratios, according to a \citet{Calzetti:2000} dust attenuation law ($R_V$\,=\,4.05). Unfortunately, due to faintness of the \Pab line emission compared to the continuum, this approximate procedure allows us to recover only \Pab line flux in part of the most actively star-forming regions (i.e., clumps and spiral arms). Apparently \Pab is also detected in the bulge, as opposed to the NIRCam WFSS \Pab map (shown in Fig. \ref{fig:Pa_beta_reconstructed}). For this reason, we caution on the reliability of the \Pab detection in the inner central regions. The right-hand panel of Fig. \ref{fig:maps_from_photometry} displays the resulting $A_{V,{\rm neb}}$ map inferred from \Pab/\ha\ line ratios, after masking pixels with (nonphysical) negative $A_{V,{\rm neb}}$ values. Due to the sparse $A_{V,{\rm neb}}$ map, we estimated the average $A_{V,{\rm neb}}$ value after masking pixels within the 0.25$''$-diameter bulge aperture (subjected to the large uncertainties in the \Pab line flux). We adopted this value as the global dust attenuation of the galaxy, assuming an uncertainty of 30\% (i.e., $A_{V,{\rm neb}}$\,=\,2.0\,$\pm$\,0.6).

\section{Validating resolved SED fitting results}\label{sec:sed_fitting_check}

\begin{figure*}
    \centering
    \includegraphics[width=0.70\linewidth]{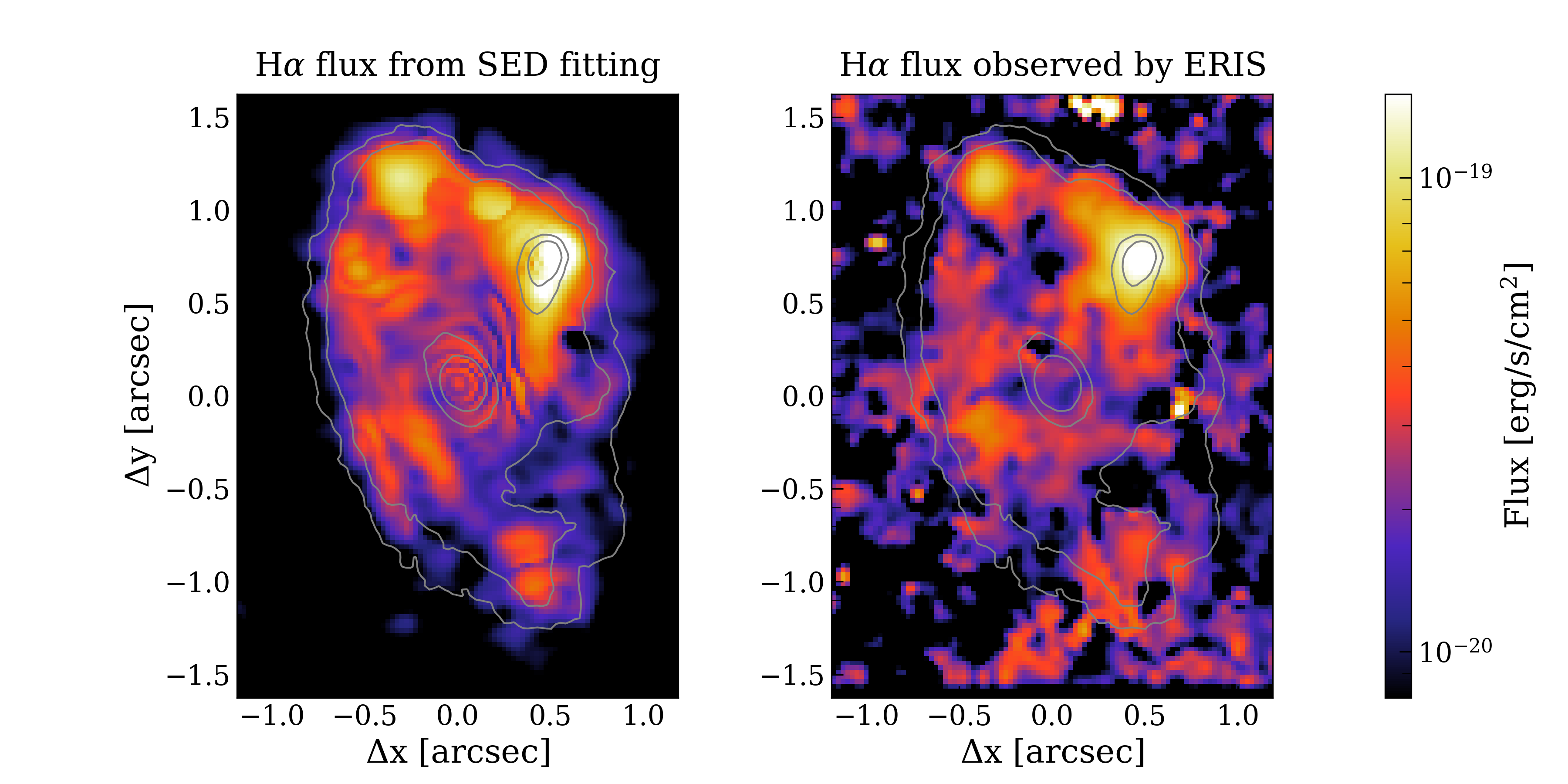}
    \caption{Expected observed \ha\ flux (left), as recovered from the SFR10 and $A_{V,{\rm star}}$ maps resulting from our spatially resolved SED fitting (see Fig. \ref{fig:SED_fitting_maps}), and the real observed \ha\ flux from VLT/ERIS observations (right). The gray contours trace NIRCam F444W emission at 5\%, 10\%, 35\%, 50\% of the emission peak. Both maps have the same color scale and a 0.025$''$ pixel scale.}
    \label{fig:ha_comparison}
\end{figure*}

To check on the reliability of the results of our spatially resolved SED fitting, we recover the expected observed \ha\ flux by combining the SFR10 and $A_{V,{\rm star}}$ maps. As we discussed in Sects. \ref{sec:pab}-\ref{sec:discussion}, nebular line emission is expected to be subject to extra attenuation compared to the stellar populations in regions of intense star formation (e.g., \citealt{Calzetti:2000,Price:2014,Tacchella:2018}). Therefore, as a first approximation, we used the $A_{V,{\rm star}}$ map -- the only full $A_V$ map available -- while acknowledging that it may locally underestimate the true attenuation of the \ha\ line emission.
In particular, we first convert the SFR10 into \ha\ intrinsic luminosity according to the \citet{Kennicutt:2012} relation, which is in turn converted into the observed \ha\ flux by (de-)correcting for dust attenuation \citep{Calzetti:2000} law, using the $A_{V,{\rm star}}$ values previously inferred for each single pixel.
In Fig. \ref{fig:ha_comparison} we show the comparison of our recovered observed \ha\ flux (left-hand panel) with the real observed one from ERIS observations (right-hand panel, same color scale; Forster Schreiber The two \ha\ maps overall show a good agreement, especially in correspondence of the two northern \ha-bright clumps (i.e., Clumps A and B), while the similarity is less evident in the regions of fainter \ha\ emission, due to the lower S/N of ERIS data. However, the ERIS \ha\ map displays increased emission at the southernmost clump (Clump C), and along the spiral arm east of the bulge. The overall good similarity between the two \ha\ maps points to the reliability of our SFR10 map, although some degeneracy might still be present between SFR10 and $A_{V,{\rm star}}$ values for a given observed \ha\ flux. By summing over all pixels, we obtain a total recovered \ha\ flux of 
$2.0 \times 10^{-16}$ \ergscm, which is consistent with the total \ha\ flux measured from ERIS and SINFONI data \citep{Forster:2009}.
We note that, although in principle it is possible to insert as input in \texttt{CIGALE} also emission-line fluxes, we do not do it because of the difficulties in correctly aligning (on a pixel level of 0.025$''$) the NIRCam, ACS, and ERIS data. This test highlights that we are able not only to correctly recover the total \ha\ flux, but also its spatial distribution, showing the reliability of our spatially resolved SED analysis.

\end{appendix}

\end{document}